\documentclass[twocolumn]{aastex63}

\usepackage{amsmath}	% Advanced maths commands
\usepackage{enumitem}

\definecolor{darkgreen}{HTML}{449900}

\def\Kepler{\textit{Kepler}} %italicized 'Kepler'
\def\Gaia{\textit{Gaia}} %italicized 'Gaia'

\defcitealias{HFR2019}{Paper I}
\defcitealias{HFR2020}{Paper II}

\received{28 July 2020}
\revised{17 September 2020}
\accepted{23 November 2020}
\shorttitle{AMD-stable Clustered Planetary Systems}
\shortauthors{He et al.}
\graphicspath{{./}{Figures/}}
\begin{document}

\title{Architectures of Exoplanetary Systems. III: Eccentricity and Mutual Inclination Distributions of AMD--stable Planetary Systems}

\correspondingauthor{Matthias Yang He}
\email{myh7@psu.edu}

\author[0000-0002-5223-7945]{Matthias Y. He}
\affiliation{Department of Astronomy \& Astrophysics, 525 Davey Laboratory, The Pennsylvania State University, University Park, PA 16802, USA}
\affiliation{Center for Exoplanets \& Habitable Worlds, 525 Davey Laboratory, The Pennsylvania State University, University Park, PA 16802, USA}
\affiliation{Center for Astrostatistics, 525 Davey Laboratory, The Pennsylvania State University, University Park, PA 16802, USA}
\affiliation{Institute for Computational \& Data Sciences, 525 Davey Laboratory, The Pennsylvania State University, University Park, PA 16802, USA}

\author[0000-0001-6545-639X]{Eric B. Ford}
\affiliation{Department of Astronomy \& Astrophysics, 525 Davey Laboratory, The Pennsylvania State University, University Park, PA 16802, USA}
\affiliation{Center for Exoplanets \& Habitable Worlds, 525 Davey Laboratory, The Pennsylvania State University, University Park, PA 16802, USA}
\affiliation{Center for Astrostatistics, 525 Davey Laboratory, The Pennsylvania State University, University Park, PA 16802, USA}
\affiliation{Institute for Computational \& Data Sciences, 525 Davey Laboratory, The Pennsylvania State University, University Park, PA 16802, USA}

\author[0000-0003-1080-9770]{Darin Ragozzine}
\affiliation{Department of Physics \& Astronomy, N283 ESC, Brigham Young University, Provo, UT 84602, USA}

\author[0000-0001-6259-3575]{Daniel Carrera}
\affiliation{Department of Physics \& Astronomy, 101 Physics Hall, Iowa State University, Ames, IA 50011, USA}



\begin{abstract}

The angular momentum deficit (AMD) of a planetary system is a measure of its orbital excitation and a predictor of long--term stability.
We adopt the AMD--stability criteria to constrain the orbital architectures for exoplanetary systems.
Previously, \citet{HFR2019} showed through forward modeling (SysSim) that the observed multiplicity distribution can be well reproduced by two populations consisting of a low and a high mutual inclination component.
Here, we show that a broad distribution of mutual inclinations arising from systems at the AMD--stability limit can also match the observed \Kepler{} population.
We show that distributing a planetary system's maximum AMD amongst its planets results in a multiplicity--dependent distribution of eccentricities and mutual inclinations. 
Systems with intrinsically more planets have lower median eccentricities and mutual inclinations, and this trend is well described by power--law functions of the intrinsic planet multiplicity ($n$): $\tilde{\mu}_{e,n} \propto n^{-1.74_{-0.07}^{+0.11}}$ and $\tilde{\mu}_{i,n} \propto n^{-1.73_{-0.08}^{+0.09}}$, where $\tilde{\mu}_{e,n}$ and $\tilde{\mu}_{i,n}$ are the medians of the eccentricity and inclination distributions.
We also find that intrinsic single planets have higher eccentricities ($\sigma_{e,1} \sim 0.25$) than multi-planet systems, and that the trends with multiplicity appear in the observed distributions of period--normalized transit duration ratios.
We show that the observed preferences for planet size orderings and uniform spacings are more extreme than what can be produced by the detection biases of the \Kepler{} mission alone.
Finally, we find that for systems with detected transiting planets between 5 and 10 days, there is another planet with a greater radial velocity signal $\simeq~53\%$ of the time.

\end{abstract}

\keywords{Exoplanet systems (484); Exoplanet detection methods (489); Exoplanet catalogs (488); Exoplanets (498); Extrasolar rocky planets (511); Astrostatistics distributions (1884); Computational methods (1965); Radial velocity (1332); Planet hosting stars (1242); Exoplanet dynamics (490); Planetary system formation (1257)}


\section{Introduction} \label{sec:intro}

While NASA's \Kepler{} \textit{Space Telescope} \citep{B2010, B2011a, B2011b, B2013} was launched over a decade ago and has since been decommissioned, the ensemble of exoplanet candidates it discovered during its primary mission continues to serve as the single largest and most uniformly vetted exoplanet catalog known to date. The abundance of relatively short period ($P \lesssim 1$ yr) transiting planets in the super--Earth to sub--Neptune size regime ($R_p \lesssim 4 R_\oplus$) observed by \Kepler{} around FGKM dwarf stars continues to advance our understanding of exoplanetary systems in the inner regions of main sequence stellar environments \citep{La2011, Li2011a, Li2011b, Li2014, R2014}. Beyond the sheer number of exoplanet detections, the \Kepler{} population also includes a wealth of systems with multiple transiting planets, sometimes called Systems with Tightly-spaced Inner Planets (STIPs). These multi-transiting systems are incredibly informative because they also provide information about the architectures of their intrinsic systems beyond simply the occurrence rates, insights which are not possible from systems with only a single planet \citep{RH2010, F2014, WF2015, HFR2019}.

Numerous studies have attempted to explore the mutual inclination distribution of the multi-planet systems \citep{La2011, Li2011b, FM2012, J2012, TD2012, WSS2012, F2014}. A similarly large number of studies have also focused on the eccentricity distribution, showing that most \Kepler{} planets tend to have relatively low eccentricities \citep{M2011, WL2013, HL2014, F2014, SDC2015, X2016, vE2019, M2019}. These studies have largely contributed to the picture that most planets in multi-transiting systems have near coplanar orbits, consistent with planet formation theories involving gaseous discs, as needed to explain the frequency of systems with many transiting planets. However, these studies typically had fewer detections with which to constrain their results, more simplistic treatments of the \Kepler{} detection efficiency, and a more limited understanding of the stellar properties. They also had a narrower focus on certain elements of the multi-planet systems instead of attempting to simultaneously model all of the architectural properties at once, including the distributions of periods, period ratios, planet sizes, orbital eccentricities and mutual inclinations, multiplicities, and the fraction of stars with planets.

As shown in \citet{HFR2019} (hereafter \citetalias{HFR2019}), a detailed model that can simultaneously reproduce all of these features is especially powerful for probing the underlying correlations in multi-planet systems. A full forward model for the \Kepler{} primary mission has been enabled only recently by advancements in both our understanding of the data and the methodology. For example, the Exoplanets Systems Simulator (``SysSim''; \citealt{H2018, H2019, HFR2019, HFR2020}) makes use of multiple \Kepler{} data products \citep{C2017, BC2017a, BC2017b, BC2017c, Co2017} to provide a sophisticated simulator for the \Kepler{} detection pipeline. We also adopt the final, \Kepler{} DR25 catalog of exoplanet candidates \citep{T2018}, which was uniformly vetted in a fully automated manner with the \Kepler{} \textit{Robovetter} \citep{Co2017}. With the aid of improved stellar properties thanks to \Gaia{} DR2 \citep{Gaia2018} and consistent isochrone fitting \citep{B2020}, forward models have become more powerful than ever for constraining the true properties of the planetary systems and their distributions.

\subsection{\Kepler{} Dichotomy}

One trend that has emerged from even the earliest studies of the \Kepler--observed planet multiplicity distribution is an apparent excess of single transiting systems that cannot be easily explained together with the low mutual inclination, high multiplicity systems \citep{Li2011b, J2012, HM2013, BJ2016}. 
Perhaps the simplest potential solution would be to invoke a high fraction of intrinsic single--planet systems \citep{FM2012, SKC2019}.  However, \citetalias{HFR2019} showed that a large population of intrinsically single--planet systems was not a viable explanation for the abundance of single transiting systems making use of the high occurrence rate of planetary systems.  
A second potential solution would be to invoke two populations of planetary systems, one characterized by low--mutual inclinations and a second with planets with substantial mutual inclinations.
Using this model, \citet{Mu2018} and \citet{HFR2019} provided constraints on the architectures of planetary systems.
A more creative solution has been posited by \citet{Z2018}, which involves a strong anti--correlation between the mutual inclination scale and the multiplicity of each system. 
While including a higher mutual inclination population can fit the observed multiplicity and transit duration ratio distributions \citepalias{HFR2019}, the extent of their high inclinations is difficult to constrain with \Kepler{} data (due to their nature of being observed as single--transiting systems) and raises concerns about their long--term stability.
Statistical studies of stellar obliquity measurements (i.e. the misalignment of the stellar spin axis compared to the planets' orbits) can also shed light on the high mutual inclination planets, including their formation pathways and dynamical histories \citep{FT2007, NIB2008, MJ2011, MP2018}.

Architectural models of multi-planet systems usually adopt simple, approximate conditions for stability such as requiring a minimum separation between adjacent planets of several (mutual) Hill radii \citep{G1993, CWB1996, PW2015}. While these stability criteria are physically motivated for the two--planet case, it is unclear how well they generalize to higher multiplicity systems. Furthermore, the mutual Hill stability criteria does not consider the mutual inclinations between planet orbits, which are known to significantly influence the orbital stability and evolution of planetary systems. A more sophisticated and general approach to stability is to consider the angular momentum deficit (AMD) of a system, which is a conserved quantity derived in the secular approximation of planetary orbits that can be used to predict long--term stability \citep{L1997, L2000, LP2017, PLB2017}. The AMD of an orbit is a measure of its excitation compared to the circular and coplanar case \citep{L1997, LP2017}. Thus, it naturally accounts for both the eccentricity and mutual inclination (relative to the system invariant plane) of a given planet. Additionally, this quantity extends easily to multi-planet systems with any number of planets, as the AMD of a planetary system is simply the sum over the AMD of each planet. Finally, the AMD stability criteria is also relatively computationally efficient to evaluate, and has been extended to treat the cases of first--order mean motion resonance (MMR) overlap \citep{W1980, DPH2013, PLB2017}. In this paper, we therefore use the AMD stability criteria as a physically motivated view of the orbital eccentricity and mutual inclination distributions of multi-planet systems, with a focus on providing additional constraints on solutions to the \Kepler{} dichotomy problem.

\subsection{Correlations of Periods and Sizes in Multi-planet Systems}

Recent studies of \Kepler{} exoplanetary systems have identified additional patterns in their observed architectures, with the three most prominent being the apparent similar sizes of planets in the same system, their preference for an increasing size ordering, and their correlated spacings in systems with three or more planets \citep{C2013, MWL2017, W2018a, WP2019, GF2020}. Due to the complex detection biases to be considered, the physical nature of these so--called ``peas in a pod'' patterns have also been hotly debated \citep{W2018a, Z2019, WP2019, MT2020}. A proper treatment of the detection biases, such as a detailed forward model \citep{Mu2018, HFR2019, HFR2020}, is necessary to disentangle real correlations in the intrinsic planetary systems from observational artifacts. While our forward model in \citetalias{HFR2019} was used to show that the observed similarities in orbital periods and in planet sizes are indicative of real clustering in the underlying systems, our analysis was driven by fits to distributions of pair--wise statistics (i.e. period ratios, transit depth ratios, and transit duration ratios of adjacent planet pairs), which do not fully capture more complex patterns. Luckily, the recent study by \citet{GF2020} developed several key metrics with roots in complexity theory to better capture the global structures of multi-planet systems. Thus, we also adopt (slightly modified versions of) these metrics to further constrain the intrinsic architectures of planetary systems in this paper.

% Summarize the structure of this paper
\bigskip
We organize this paper as follows.
In \S\ref{Methods} we describe our forward modelling procedure (summarized from \citetalias{HFR2019}) and how we modify our previous model. This involves describing our updated stellar catalog (\S\ref{secStars}), summarizing our previous clustered Poisson point process model from \citet{HFR2019, HFR2020} (\S\ref{secOldModel}), defining the AMD stability criteria and how we use it in our new model (\S\ref{secMaxAMDmodel}), and detailing our observational constraints including the new terms from \citet{GF2020} (\S\ref{Obs}).
In \S\ref{Results} we present the key results of our new model along with a side-by-side comparison with the old model.
A discussion of the implications and limitations of our new model is provided in \S\ref{Discussion}, including a discussion of the \Kepler{} dichotomy (\S\ref{Kepler_dichotomy}), inferences about the ``peas in a pod'' trends (\S\ref{Peas_in_a_pod}), and implications for radial velocity (RV) surveys (\S\ref{RVs}).
Finally, we summarize all of our main conclusions in \S\ref{Conclusions}.

\section{Methods} \label{Methods}

We develop our models as an extension of the Exoplanets Systems Simulator (``SysSim'') codebase, which can be installed as the ExoplanetsSysSim.jl package \citep{F2018b}. This package provides the core SysSim functions as well as detailed models of the \Kepler{} detection efficiency and vetting pipeline. Specific details about the detection model, as well as the broader SysSim project with applications to planet occurrence rates, are described in \citet{H2018, H2019, HFT2020}. The clustered models are provided in the \url{https://github.com/ExoJulia/SysSimExClusters} repository, which are described in \citetalias{HFR2019}; we provide a separate code branch for each paper. We also provide step-by-step instructions on how to download our simulated catalogs or generate new catalogs.

In \citetalias{HFR2019}, we defined the following multi-stage procedure for studying the intrinsic architectures of planetary systems, which constitutes a \textit{full forward model}:
\begin{itemize}[leftmargin=*, label={}]
 \item \textbf{Step 0: Define a statistical description for the intrinsic distribution of exoplanetary systems.}
 \item \textbf{Step 1: Generate an underlying population of exoplanetary systems (\textit{physical catalog}).}
 \item \textbf{Step 2: Generate an observed population (\textit{observed catalog}) from the \textit{physical catalog}.}
 \item \textbf{Step 3: Compare the simulated \textit{observed catalog} with the \Kepler{} data.}
 \item \textbf{Step 4: Optimize a distance function to find the best-fit model parameters.}
 \item \textbf{Step 5: Explore the posterior distribution of model parameters using a Gaussian Process (GP) emulator.}
 \item \textbf{Step 6: Compute credible intervals for model parameters and simulated catalogs using Approximate Bayesian Computing (ABC).}
\end{itemize}

In this study, we retain the above framework and most elements of the full forward model. The most important updates for this paper are in how we assign eccentricities and inclinations in ``Step 1'', as described in \S\ref{secNewProc}. While in \citetalias{HFR2019} we drew eccentricities and inclinations directly and independently, our new model assigns eccentricities and inclinations so as to create dynamically ``packed'' multiple planet systems.

In \S\ref{secStars}, we first describe our stellar input catalog, which includes updated parameters from the \Gaia--\Kepler{} Stellar Properties catalog \citep{B2020}. In \S\ref{secOldModel}, we provide an overview of our previous model.  We describe the concept of ``AMD stability'' in \S\ref{secMaxAMDmodel}  and provide details for the updated process for generating planetary systems in \S\ref{secNewProc}.

\subsection{Stellar catalog} \label{secStars}

Our ability to characterize planetary properties, which then affects our inferences of their system architectures, is limited by our knowledge of the stellar properties. In an effort to mitigate the effect of stellar uncertainties in our analyses, we purposefully defined a set of summary statistics in \citetalias{HFR2019} that minimize the impact of uncertainties in the stellar radii, by fitting to the \Kepler{} distributions of measured transit depths instead of planet radii, and of ratios of observables (i.e. transit depth ratios and transit duration ratios) where the stellar radii cancel out. Nevertheless, the stellar properties (and thus their error bars) propagate through our forward model when simulating physical and observed catalogs. Moreover, some key observables such as the transit durations and circular--normalized transit durations are sensitive to the underlying distribution of eccentricities but rely on having well characterized and consistent stellar radii and masses in order to provide meaningful constraints (e.g., \citealt{M2011, PBC2014, vEA2015, X2016}).

\emph{A clean sample of FGK dwarfs:} We adopt a very similar stellar catalog as the one defined in \citetalias{HFR2019} and \citet{H2019}, which involves a series of cuts on the \Kepler{} DR25 target list (see \S3.1 therein). To summarize, this list of cuts includes requiring: consistent values between the \Kepler{} magnitude and the \Gaia{} G magnitude; a good astrometric fit (\Gaia{} GOF\_AL $\leq 20$ and astrometric excess noise $\leq 5$); and a precise parallax (fractional parallax error within 10\% of the parallax). These cuts are primarily made to filter out likely close--in binary stars and stars with poorly measured radii. We also select for stars on the main sequence by requiring $0.5 \leq b_p-r_p \leq 1.7$ and $L \leq 1.75 L_{\rm MS}(b_p-r_p)$ where $L_{\rm MS}(b_p-r_p)$ is derived from iteratively fitting to the main sequence.

\emph{Revised stellar radii and masses:} While we adopted revised stellar radii from \Gaia{} DR2 in \citetalias{HFR2019}, we had kept the stellar masses from \Kepler{} DR25 since our analyses in those studies were relatively insensitive to stellar mass. Here, we take advantage of the new \Gaia--\Kepler{} Stellar Properties Catalog \citep{B2020}, which provides a homogeneous set of stellar properties derived from isochrone fitting using \Gaia{} DR2 inputs. This yields a self--consistent set of stellar mean densities, crucial to the calculation of circular--normalized transit durations, which we adopt as a summary statistic in this paper (see \S\ref{SummaryStats}).

\emph{Reddening correction:} We retain an explicit model dependence on host star spectral type by adopting the \Gaia{} DR2 $b_p-r_p$ colors for each star and correcting for reddening. We account for differential reddening by constructing a simple model for $E(b_p-r_p)$ as a smooth function of $b_p-r_p$.
The remaining stars are binned into 20 quantiles by $b_p-r_p$ and the median distance--normalized reddening, $E(b_p-r_p)/d$ where $d = 1/\pi$ is the distance computed from the parallax $\pi$, is computed for each bin. We then compute the interpolated reddening for each target, $E^* = E^*(b_p-r_p)$, by interpolating $E(b_p-r_p)/d$ as a function of $b_p-r_p$ and multiplying by $d$. Finally, we apply the reddening correction derived this way for all targets, and re-cut and re-fit the FGK main sequence using the corrected colors, with $0.5 \leq b_p-r_p-E^* \leq 1.7$.

Our final stellar catalog contains 86,760 targets. The median corrected color is $b_p-r_p-E^* \simeq 0.81$ mag, which is close to the Solar value.

\subsection{Previous clustered model} \label{secOldModel}

Our clustered model with a host star dependence consists of the following features:

\begin{itemize}[leftmargin=*, label={}]
 \item \textbf{Fraction of stars with planets:} Each star has a probability of hosting a planetary system (between $3-300$ d and $0.5-10 R_\oplus$), that is a linear function of its \Gaia{} intrinsic color ($c \equiv b_p-r_p-E^*$):
 \begin{eqnarray}
 %\begin{split}
  && f_{\rm swpa}(c) = \\
  && \quad \max \Big\{ 0,\min \Big[ m \Big(c - c_{\rm med} \Big) + f_{\rm swpa,med}, 1 \Big] \Big\} \nonumber \label{eq_fswp_bprp}
 %\end{split}
 \end{eqnarray}
 where $m = d{f_{\rm swpa}}/d(c)$ is the slope and $f_{\rm swpa,med} = f_{\rm swpa}(c_{\rm med})$ is the normalization (at the median color, $c_{\rm med} \simeq 0.81$ for our sample of FGK dwarfs). The value of $f_{\rm swpa}$ is always bounded between 0 and 1, since the fraction of stars with planets cannot be negative or greater than 1.
 \item \textbf{Planet clusters:} For stars assigned a non-empty planetary system, each system is composed of ``clusters'' of planets. We attempt to assign both the number of clusters and planets per cluster by drawing from a zero-truncated Poisson (ZTP) distribution, $N_c \sim {\rm ZTP}(\lambda_c)$ and $N_p \sim {\rm ZTP}(\lambda_p)$, respectively. We note that some clusters may be rejected due to failing our stability criteria (see below), so the true distributions may not exactly match a ZTP, especially for rather large values.
 \item \textbf{Orbital periods:} A power-law describes the distribution of cluster period scales $P_c$.  The period of each planet in a cluster is drawn from a log-normal distribution with cluster width $N_p \sigma_P$ (where $N_p$ is the number of planets in the cluster and $\sigma_P$ is a width scale parameter), between $P_{\rm min} = 3$ and $P_{\rm max} = 300$ d:
 \begin{eqnarray}
  f(P_c) &\propto& {P_c}^{\alpha_P} \label{eq_Pc} \\
  P'_i &\sim& {\rm Lognormal}(0, N_p\sigma_P) \label{eq_P_unscaled} \\
  P_i &=&  P_c P'_i,\quad P_{\rm min} \leq P_i \leq P_{\rm max}
 \end{eqnarray}
 where $P_i$ are true periods and $P'_i$ are unscaled periods (i.e. before multiplying by the period scale).
 \item \textbf{Planet radii:} A broken power-law describes the distribution of cluster radius scales $R_{p,c}$.  The radius of each planet in a cluster is drawn from a log-normal distribution centred on $R_{p,c}$ with cluster width $\sigma_R$, between $R_{p,\rm min} = 0.5$ and $R_{p,\rm max} = 10 R_\oplus$.
\begin{eqnarray}
 f(R_{p,c}) &\propto& \left\{
 \begin{array}{ll}
  {R_{p,c}}^{\alpha_{R1}}, & R_{p,\rm min} \leq R_{p,c} \leq R_{p,\rm break} \label{eq_Rc} \\
  {R_{p,c}}^{\alpha_{R2}}, & R_{p,\rm break} < R_{p,c} \leq R_{p,\rm max}
 \end{array} \right., \\
 R_{p,i} &\sim& {\rm Lognormal}(R_{p,c}, \sigma_R) \label{eq_Rp}
\end{eqnarray}
where $\alpha_{R1}$ and $\alpha_{R2}$ are power-law indices and $R_{p,\rm break} = 3 R_\oplus$ is the break radius.
\item \textbf{Planet masses:} A non-parametric, probabilistic mass--radius relation from \citet{NWG2018} is used to draw the masses of the planets conditioned on their radii.

 \item \textbf{Eccentricities:} The orbital eccentricities for all planets are drawn from a Rayleigh distribution, $e \sim {\rm Rayleigh}(\sigma_e)$.
\item \textbf{Mutual inclinations:} Two Rayleigh distributions for the mutual inclinations are used, corresponding to a high and a low mutual inclination population (with scales $\sigma_{i,\rm high}$ and $\sigma_{i,\rm low}$, respectively, such that $\sigma_{i,\rm high} \geq \sigma_{i,\rm low}$), where the fraction of systems belonging to the high inclination population is $f_{\sigma_{i,\rm high}}$:
\begin{equation}
 i_m \sim \left\{
 \begin{array}{ll}
  {\rm Rayleigh}(\sigma_{i,\rm high}), & u < f_{\sigma_{i,\rm high}} \label{eq_incl} \\
  {\rm Rayleigh}(\sigma_{i,\rm low}), & u \geq f_{\sigma_{i,\rm high}}
 \end{array} \right.,
\end{equation}
where $u \sim {\rm Unif}(0,1)$.
 \item \textbf{Planets near resonance:} Peaks near the first-order mean motion resonances (MMRs) in the observed period ratio distribution are produced by drawing low mutual inclinations for the planets ``near an MMR'' with another planet (which we define as cases where the period ratio is in the range $[\mathcal{P}_{\rm mmr}, 1.05 \mathcal{P}_{\rm mmr}]$ for any $\mathcal{P}_{\rm mmr}$ in \{2:1, 3:2, 4:3, 5:4\}), such that these planets have mutual inclinations drawn from the Rayleigh distribution with $\sigma_{i,\rm low}$ regardless of which mutual inclination population the system belongs to.
\item \textbf{Stability criteria:} Adjacent planets are separated by at least $\Delta_c = 8$ mutual Hill radii ($R_H$), and orbital periods are resampled until this criteria is met:
\begin{eqnarray}
 \Delta &=& \frac{a_{\rm out}(1-e_{\rm out}) - a_{\rm in}(1+e_{\rm in})}{R_H} > \Delta_c, \label{eq_Nhill} \\
 R_H &=& \bigg(\frac{a_{\rm in} + a_{\rm out}}{2}\bigg)\bigg[\frac{m_{\rm in} + m_{\rm out}}{3 M_\star}\bigg]^{1/3}. \label{eq_mhill}
\end{eqnarray}
For clusters where a maximum number of resampling attempts has been met, the entire cluster is discarded.
\end{itemize}

Hereafter, we will refer to this previous model as the ``two--Rayleigh'' model, due to the parameterization of the mutual inclinations as a mixture of two Rayleigh distributions.

Although this model does closely fit many of the marginal distributions for the \Kepler{} catalog of exoplanet candidates \citep{HFR2019, HFR2020}, and provides meaningful constraints on many of its model parameters, there are some limitations worth addressing. First, our stability criteria, while simple, is likely an inadequate requirement for some planetary systems, especially those with many planets.  The strict cutoff at $\Delta_c = 8$ is abrupt and only treats adjacent planet pairs. Furthermore, this stability metric also ignores the inclinations, allowing $\sigma_{i,\rm high}$ to reach arbitrarily large values which generate extreme ($>45^\circ$) mutual inclinations that are unlikely and probably unstable due to dynamical evolution through secular interactions. Our mutual inclination distribution is also limited to a mixture of two Rayleigh distributions; while this parametrization is well motivated by previous studies and evidence of the \Kepler{} dichotomy \citep{Li2011b, J2012, HM2013, BJ2016, ZCH2019, HFR2019, SKC2019}, the number of free parameters ($\sigma_{i,\rm high}$, $\sigma_{i,\rm low}$, and $f_{\sigma_{i,\rm high}}$) is high and does not allow for a smooth transition from the low to high mutual inclination regime. 
Additionally, the eccentricity distribution in our original model is limited to a single Rayleigh distribution, for which we find a small scale ($\sigma_e \simeq 0.02$).  While most planets do have near circular orbits, there are some confirmed exoplanets with larger eccentricities which this model does not have the flexibility of producing. 
These parameterizations for the eccentricity and inclinations also imply that we assume that orbital eccentricities and mutual inclinations are independent, which is likely a poor assumption when considering the role of dynamical interactions. 
Finally, the assumed radius broken power-law is not adequate for a detailed description for the true radius distribution, which is bimodal and sculpted by photoevaporation \citep{OW2013, F2017, OW2017, vE2017, C2018} and heating mechanisms (e.g. core-powered mass loss; \citealt{GSS2016, GSS2018, GS2019}).

While the radius distribution is a key component to understanding the nature of the planet radius valley and the correlations of planet sizes, both with orbital period and with each other \citep{C2013, W2018a, Z2019, HFR2019, WP2019, MT2020}, we do not address this topic in this work. In this paper, we address all the other concerns previously described: we adopt a more sophisticated, dynamically motivated view of stability in multi-planet systems using the angular momentum deficit (AMD) stability criterion \citep{LP2017, PLB2017}. We develop a new forward model that makes use of AMD stability to generate planetary systems from a more realistic joint eccentricity and mutual inclination distribution.

\subsection{New clustered model: maximum AMD model} \label{secMaxAMDmodel}

\subsubsection{AMD stability}

The angular momentum deficit (AMD) is 
the difference between the total angular momentum of a planetary system and what the total angular momentum would be if all orbits were circular and coplanar (with the same semi-major axes and masses).  
First described by \citet{L1997, L2000}, the AMD of a planetary system is a conserved quantity in the secular theory of orbital motion (i.e., ignoring resonant interactions). 
A comprehensive discussion of AMD stability including derivations of the AMD from the Hamiltonian and conditions for stability against collisions (in the absence of MMRs) is presented in \citet{LP2017}. The conditions for stability against MMR overlap are derived in \citet{PLB2017}. While the proofs in those works are outside the scope of this paper, we restate the main equations required to evaluate the AMD-stability condition (considering both collisions and MMR overlap) for a planetary system. Following the notation of \citet{LP2017} and \citet{PLB2017}, for a system of $N$ planets, the total AMD is simply the sum of the AMD of the planets:
\begin{eqnarray}
 {\rm AMD}_{\rm tot} &=& \sum_{k=1}^{N} {\rm AMD}_k \label{eq_amd_tot} \\
  &=& \sum_{k=1}^{N} \Lambda_k \Big(1 - \sqrt{1-e_k^2}\cos{i_{m,k}} \Big), \label{eq_amd} \\
 \Lambda_k &=& \mu_k\sqrt{a_k} \label{eq_amd_circ}
\end{eqnarray}
where $\mu_k = M_{p,k}/M_\star$ is the planet--star mass ratio (here we work in units of $GM_\star \equiv 1$), $a_k$ is the semi-major axis, $e_k$ is the orbital eccentricity, and $i_{m,k}$ is the mutual inclination relative to the system invariant plane, for the $k^{\rm th}$ planet ($k=1,\dots,N$).

\bigskip
\emph{Stability against collisions:} In intuitive terms, the AMD stability criteria (against collisions) requires that any pair of planets in the system must not have crossing orbits if all the AMD of \textit{the entire system} were assigned to just those two planets.   While ${\rm AMD}_{\rm tot}$ is conserved, the AMD is exchanged between the orbits due to secular gravitational interactions of the planets. Following \citet{LP2017}, we consider pairs of adjacent planets (where 1 denotes the inner planet and 2 denotes the outer planet) and define the planet mass ratio ($\gamma$) and semi-major axis ratio ($\alpha$):
\begin{eqnarray}
 \gamma &=& \mu_1/\mu_2 = M_{p,1}/M_{p,2}, \label{eq_mass_ratio} \\
 \alpha &=& a_1/a_2. \label{eq_a_ratio}
\end{eqnarray}
A quantity called the ``critical relative AMD for collision'', $\mathcal{C}_{\rm coll}$, is then given by \citep{LP2017}:
\begin{eqnarray}
 \mathcal{C}_{\rm coll} &=& \gamma \sqrt{\alpha} \Big(1 - \sqrt{1 - e_1^2}\Big) + \Big(1 - \sqrt{1-e_2^2}\Big), \label{eq_amd_coll} \\
 e_1 &=& e_{\rm crit}(\gamma,\alpha), \label{eq_e1} \\
 e_2 &=& 1 - \alpha - \alpha{e_1} \label{eq_e2}
\end{eqnarray}
where $e_{\rm crit}(\gamma,\alpha)$ is the solution for $e$ in the following equation, that can be solved numerically:
\begin{equation}
 \alpha{e} + \frac{\gamma{e}}{\sqrt{\alpha(1-e^2) + \gamma^2 {e}^2}} -1 + \alpha = 0. \label{eq_e_crit}
\end{equation}

The AMD stability criteria is then simply given by comparing $\mathcal{C}_{\rm coll}$ to the ``relative AMD'' of each planet:
\begin{equation}
 \mathcal{C}_{x,k} = \frac{\rm AMD_{\rm tot}}{\Lambda_k}, \label{eq_amd_rel}
\end{equation}
where $\mathcal{C}_{x,k}$ is the relative AMD of the $k^{\rm th}$ planet (a measure of its orbital excitation), to yield the AMD stability condition against collisions:
\begin{eqnarray}
 && \mathcal{C}_{x,k} < \mathcal{C}_{\rm coll} \label{eq_amd_stability_coll1} \\
 \implies && {\rm AMD}_{\rm tot} < \Lambda_k \mathcal{C}_{\rm coll}, \quad k = 2,\dots,N \label{eq_amd_stability_coll2}
\end{eqnarray}
where $\mathcal{C}_{\rm coll}$ is evaluated for the $(k-1, k)$ planet pair. For $k = 1$, we consider the case where the total AMD must also not be enough to allow the innermost planet to collide with the star, i.e. ${\rm AMD}_{\rm tot} < \Lambda_1$.

\bigskip
\emph{Stability against MMR overlap:} The AMD stability criteria against MMR overlap follows a similar logic, by also considering pairs of planets and deriving a ``critical relative AMD for MMR overlap'' \citep{PLB2017}. Two cases must be considered: circular orbits and eccentric orbits. First, for circular orbits, the following criteria must be satisfied by all planet pairs:
\begin{eqnarray}
 \alpha &<& \alpha_{\rm crit} \simeq 1 - 1.46\epsilon^{2/7}, \label{eq_circ_overlap} \\
 \epsilon &=& \mu_1 + \mu_2, \label{eq_eps}
\end{eqnarray}
i.e. it is simply a function of the semi-major axes and masses. Similar results were found in \citet{W1980, DPH2013}. For eccentric orbits, the ``critical relative AMD for MMR overlap'', $\mathcal{C}_{\rm mmr}$, is given by \citep{PLB2017}:
\begin{eqnarray}
 \mathcal{C}_{\rm mmr} &=& \frac{g^2 \gamma \sqrt{\alpha}}{2 + 2\gamma \sqrt{\alpha}}, \label{eq_amd_mmr} \\
 g &=& \frac{3^4 (1-\alpha)^5}{2^9 r \epsilon} - \frac{32r \epsilon}{9(1-\alpha)^2}, \label{eq_g} \\
 r &\simeq& 0.80199.
\end{eqnarray}

Likewise, the AMD stability criteria in this case (only considering MMR overlap) is given by:
\begin{eqnarray}
 && \mathcal{C}_{x,k} < \mathcal{C}_{\rm mmr} \label{eq_amd_stability_mmr1} \\
 \implies && {\rm AMD}_{\rm tot} < \Lambda_k \mathcal{C}_{\rm mmr}, \quad k = 2,\dots,N \label{eq_amd_stability_mmr2}
\end{eqnarray}
where $\mathcal{C}_{\rm mmr}$ again is evaluated for the $(k-1, k)$ planet pair.

\bigskip
In summary, the full condition for AMD stability (against both collisions and MMR overlap) is that the criteria in equations \ref{eq_amd_stability_coll2}, \ref{eq_circ_overlap}, \& \ref{eq_amd_stability_mmr2} must be satisfied. If the condition in equation \ref{eq_circ_overlap} is true, we can define a limit on the total system AMD by combining equations \ref{eq_amd_stability_coll2} \& \ref{eq_amd_stability_mmr2} (along with the requirement that planet $k=1$ does not collide with the star):
\begin{eqnarray}
 {\rm AMD}_{\rm tot} &<& \Lambda_k {\rm min}(\mathcal{C}_{\rm coll}, \mathcal{C}_{\rm mmr}), \quad k = 2,\dots,N \label{eq_amd_stability} \\
 {\rm AMD}_{\rm tot} &<& \Lambda_1
\end{eqnarray}
In other words, the total system AMD must be less than the \textit{critical AMD}:
\begin{eqnarray}
 {\rm AMD}_{\rm tot} &<& {\rm AMD}_{\rm crit} \nonumber \\
 &=& {\rm min}\bigg[ \Big\{ \Lambda_k {\rm min}\big( \mathcal{C}_{\rm coll}, \mathcal{C}_{\rm mmr} \big) : k=2,\dots,N \Big\} \nonumber \\
 && \quad\quad \cup \Big\{\Lambda_1 \Big\} \bigg]. \label{eq_amd_crit}
\end{eqnarray}
Following this formalism, we can define a maximum amount of AMD for a given set of planet masses and semi-major axes, such that if this AMD were distributed in any way between the planets, the conditions against collisions and MMR overlap would still hold (and the inner-most planet would not collide with the star).

\subsubsection{Distributing maximum system AMD} \label{Max_AMD_model}

\begin{figure}
\includegraphics[scale=0.24,trim={0.5cm 0.3cm 0.5cm 0.5cm},clip]{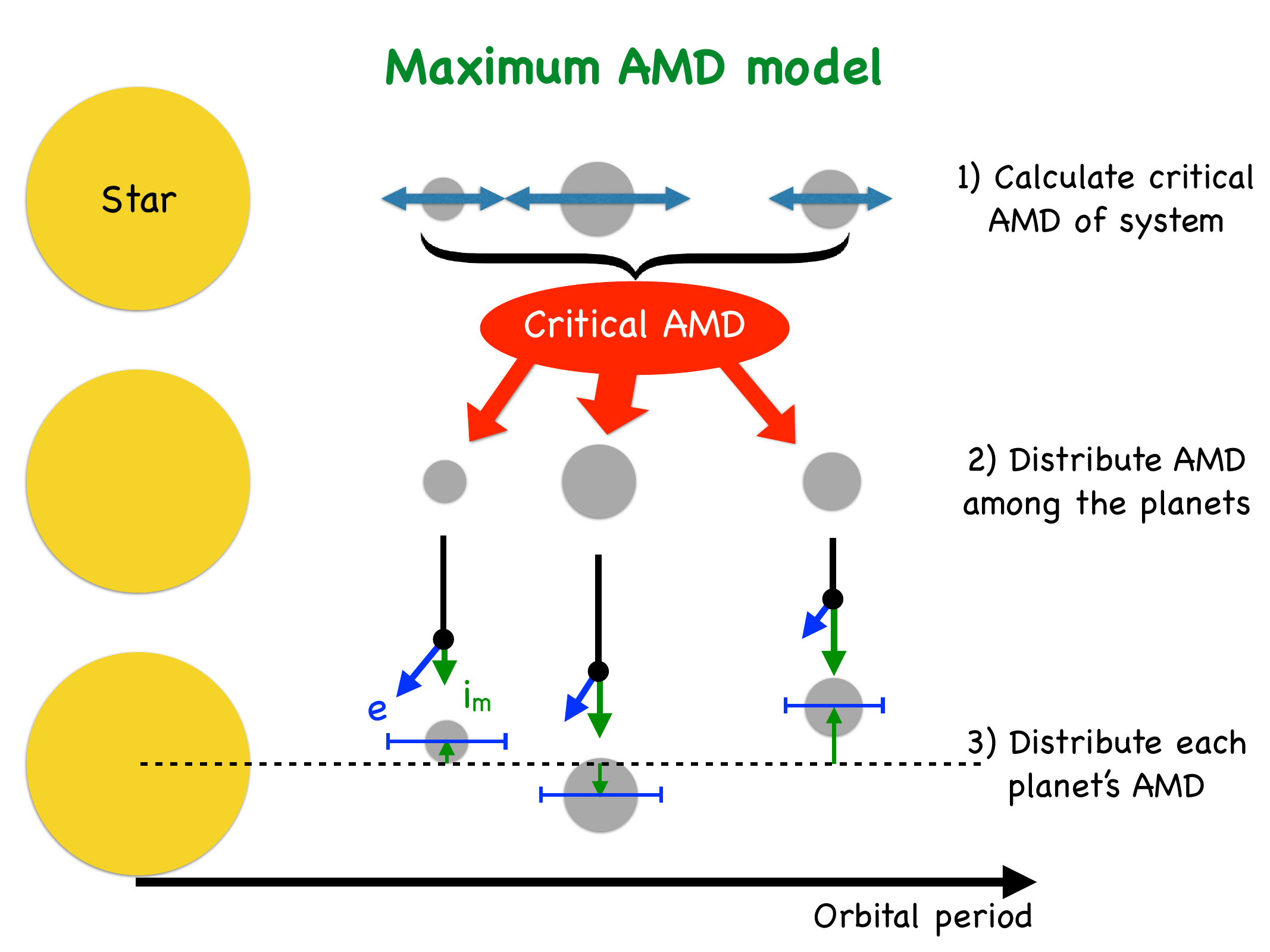}
\caption{Cartoon illustration of our maximum AMD model. Given a set of planet masses and orbital periods (satisfying our mutual Hill stability criteria for circular orbits) in a system, we first compute the critical AMD of the system in (1) from the collision and MMR overlap conditions (equation \ref{eq_amd_crit}). We then distribute the critical AMD amongst the planets, per unit mass, in (2). Finally, we distribute the AMD of each planet amongst their eccentricity and mutual inclination components in (3). The details are described in \S\ref{Max_AMD_model}.}
\label{fig:model_cartoon}
\end{figure}

\citet{LP2017} showed that during collisions of planets, the total AMD of a system always decreases. In this way, collisional events act to stabilize a system, and it is sensible to imagine that many planetary systems evolved from outside the stability limit to inside the limit after a sequence of collisions. Such systems would likely result in having a total AMD just below the critical value, as their final stable configuration prevents further loss of AMD. However, even inner planetary systems that are apparently AMD stable could be perturbed by the presence of giant planets or binary stellar companions at long periods (e.g., \citealt{TR2005}). While large masses at wide orbits can provide a significant amount of AMD to the system (equation \ref{eq_amd_circ}), the timescale for AMD transfer between widely separated planets is also slow.
We note that AMD--unstable systems do exist \citep{LP2017}, as some systems could exceed the critical AMD and still be long--lived; we discuss this further in \S\ref{AMD_parameter}. Nevertheless, we explore a conservative model in which all systems are formally AMD--stable, by replacing our two--Rayleigh model for inclinations with the assumption that all planetary systems have the critical (i.e., maximum) AMD. Hence, we will refer to our new model as the ``maximum AMD model'' for the remainder of this paper. We also relax this assumption of maximum AMD in \S\ref{AMD_parameter} and test the model when systems are below (or above) the maximum AMD. We distribute this total AMD ``budget'' amongst the individual planets as follows, providing a natural constraint on their orbital eccentricities and mutual inclinations. A cartoon illustration of this process is summarized in Figure \ref{fig:model_cartoon}.

We keep all other aspects of our clustered (two--Rayleigh) model the same, only replacing the Rayleigh$(\sigma_e)$ distribution of eccentricities and two-population (also Rayleighs with $\sigma_{i,\rm low}$ and $\sigma_{i,\rm high}$) distribution of mutual inclinations.\footnote{Since our treatment of the planets near resonance in \citetalias{HFR2019} also rely on drawing their mutual inclinations from the low scale ($\sigma_{i,\rm low}$), we do not retain our MMR features but discuss this further in \S\ref{MMRs}.} As such, we still draw a number of clusters $N_c \sim {\rm ZTP}(\lambda_c)$ and planets per cluster $N_p \sim {\rm ZTP}(\lambda_p)$ for a fraction of planet hosting stars ($f_{\rm swpa}(b_p-r_p-E^*)$). To retain the correlations in planet sizes and orbital periods, the planet radii within each cluster are still drawn from a lognormal distribution, where the cluster scale is drawn from a broken power-law; the periods in each cluster are also drawn from a lognormal distribution, where the period scale follows a single power-law.
Finally, we continue to use the mutual Hill stability criteria as a precondition for stability ($\Delta > \Delta_c$ for adjacent planet pairs), but assuming circular orbits at this stage of drawing the periods since the eccentricities have not been set, and additionally checking that equation \ref{eq_circ_overlap} is met.
The mutual Hill stability criteria is necessary to set the semi-major axes.

For a given planetary system of $N > 1$ planets with drawn planet radii $\{R_{p,k}\}$, masses $\{M_{p,k}\}$, and orbital periods $\{P_k\}$, we first compute ${\rm AMD}_{\rm crit}$ from equation \ref{eq_amd_crit}. We then distribute ${\rm AMD}_{\rm tot} = {\rm AMD}_{\rm crit}$ amongst the planets per unit mass, so the $k^{\rm th}$ planet gets:
\begin{equation}
 {\rm AMD}_k = \bigg(\frac{M_{p,k}}{\sum_{k^\prime=1}^{N} M_{p,k^\prime}} \bigg) {\rm AMD}_{\rm tot}. \label{eq_amd_per_pl}
\end{equation}
Since the AMD of a planet's orbit is proportional to its mass, this choice provides the same degree of dynamical ``excitation'' for all the planets in a given system.

For each planet, we then further distribute its AMD$_k$ randomly amongst the three orbital excitation components: $x = e\sin{\omega}$, $y = e\cos{\omega}$, and $z = \sin{i_m}$, as follows.
In order to do this, we must constrain the sum of their squares; it can be shown from equation \ref{eq_amd} (just the term inside the summation) that the constraint is:
\begin{equation}
 x^2 + y^2 + z^2 = \frac{{\rm AMD}_k}{\Lambda_k} \bigg(2 - \frac{{\rm AMD}_k}{\Lambda_k}\bigg). \label{eq_xyz}
\end{equation}
First, we draw two random numbers partitioning the unit interval, e.g. $a, b \sim {\rm Unif}(0,1)$. Re--labeling $a$ and $b$ such that $a < b$, we then assign $x^2 = a$, $y^2 = b-a$, and $z^2 = 1-b$ so that $x^2+y^2+z^2 = 1$. Each component is then multiplied by the total sum in equation \ref{eq_xyz} to yield the constraint.\footnote{This procedure is equivalent to drawing from a (symmetric) Dirichlet distribution with $\bm{\alpha} = (1,1,1)$.} 

Physically, the $x$ and $y$ components can be interpreted as kicks in the system plane, while $z$ represents a kick out of the plane. Once the values of $x,y,z$ are drawn satisfying the above equation, it is easy to compute the eccentricity, argument of pericenter ($\omega$), and mutual inclination.
%via: $e = \sqrt{x^2+y^2}$, $\omega = {\rm atan}(x, y)$, and $i_m = \sin^{-1}{z}$, respectively. 
Drawing the eccentricities $\{e_k\}$ and mutual inclinations $\{i_{m,k}\}$ of the planets in this way ensures that the system is AMD stable. %, by definition of the total system AMD.

For intrinsic single planet ($N = 1$) systems, the ``critical AMD'' is simply $\Lambda_1$ (equation \ref{eq_amd_circ}). Since the orbits of single planets define the system invariant plane and thus do not have a ``mutual inclination'' relative to it, allowing these planets to have the critical AMD would force their eccentricities to unity. Instead, we draw their eccentricities from a separate distribution, Rayleigh$(\sigma_{e,1})$.

\emph{Summary of free parameters:} Altogether, our maximum AMD model has 11 free parameters:
\begin{itemize}[leftmargin=*]
 \item $f_{\rm swpa,med}$: the fraction of stars with planets at the median color ($b_p-r_p-E^* = 0.81$ mag),
 \item $\frac{df_{\rm swpa}}{d(b_p-r_p-E^*)}$: the rate of change of $f_{\rm swpa}$ with color,
 \item $\lambda_c$: the mean number of clusters per system$^\dagger$,
 \item $\lambda_p$: the mean number of planets per cluster$^\dagger$,
 \item $\Delta_c$: the minimum separation in mutual Hill radii for adjacent planets,
 \item $\alpha_P$: the power--law index of the period distribution,
 \item $\alpha_{R1}$: the radius power--law index below $R_{p,\rm break}$,
 \item $\alpha_{R2}$: the radius power--law index above $R_{p,\rm break}$,
 \item $\sigma_{e,1}$: the Rayleigh scale for the eccentricities of true singles,
 \item $\sigma_R$: the standard deviation in log--radius for planets in the same cluster, and
 \item $\sigma_P$: the standard deviation in log--period, per planet, for planets in the same cluster.
\end{itemize}
$^\dagger$The ``mean'' is before zero--truncating and rejection sampling due to the stability criteria.

This is fewer than that of our two--Rayleigh model (even after allowing $\Delta_c$ to vary, which was not the case in \citetalias{HFR2019}), while maintaining the same features and providing a better match to the \Kepler{} data as we will show in \S\ref{Results}.

\subsubsection{Mass--radius relation} \label{secMR}

Numerous mass--radius (M-R) relations for exoplanets exist in the literature, including probabilistic models (e.g., \citealt{WRF2016, CK2017}). These models are necessary to account for the significant scatter in mass as a function of radius due to the diversity of planet compositions. For example, \citet{WRF2016} assumed a power-law relation with a scatter in mass that is normally distributed. \citet{CK2017} extended a similar probabilistic model to a much wider range of radii and masses, with power-laws broken into four regimes from moons ($\sim 0.1 R_\oplus$) to stars ($\sim 100 R_\oplus$). In \citetalias{HFR2019}, we used the M-R relationship from \citet{NWG2018} (hereafter NWG18). This M-R relation is both non--parametric (it does not assume a functional form with fixed number of parameters, but rather is defined by a set of basis functions with many weights) and probabilistic (there is a distribution of planet masses at any given radius). It is defined using a series of Bernstein polynomials for the joint M-R distribution, fit to a sample of 127 \Kepler{} exoplanets with masses measured from RVs or transit timing variations (TTVs). Specifically, this M-R model involves 55 degrees of freedom in each dimension, for a total of 3025 weights that describe the structure of the joint distribution.

%\textbf{While numerous other M-R relations exist (including probabilistic models, e.g. \citealt{CK2017}), the NWG18 M-R relation is very flexible and offers benefits over simpler or parametric models. For example, \citet{CK2017} provide a probabilistic broken power-law model for the M-R relation with four regimes, covering an extremely wide range of radii and masses from moons ($\sim 0.1 R_\oplus$) to stars ($\sim 100 R_\oplus$). For our narrower range of planet sizes ($0.5-10 R_\oplus$), the NWG18 relation is more flexible, as it provides a similar scatter while not being confined to power-law forms (and does not involve sharp transitions/breaks). However, there are a few drawbacks which prompt us to revise the M-R relation.}
While the NWG18 model is very flexible and offers many benefits over simpler, parametric models, there are a few drawbacks that prompt us to revise the M-R relation. First, there are only a few planets with mass and radius measurements informing the lower limit of the model; only three data points are below $1 R_\oplus$, where our radius power--law distribution peaks. Moreover, we find that this relation produces a bimodal distribution of planet mass towards the lower mass limit, due to a single data point at the lowest end (see Figure 3 of \citealt{NWG2018}; there is a jump at $\sim 0.7 R_\oplus$). Given our range of planet radii between 0.5 and 10$R_\oplus$, we find that this yields a sharp peak of planet masses just above $0.1 M_\oplus$. Second, this M-R relation produces a significant scatter in planet masses for sizes larger than $\sim 0.7 R_\oplus$. While the large scatter is reasonable and driven by data for larger radii, it leads to extreme densities at smaller radii with a majority of planets denser than pure iron. While our previous models are only weakly dependent on the planet mass distribution through the mutual Hill stability criteria, the new model considered in this paper is affected to a greater extent by the assumed planet masses due to the direct calculation of each system's critical AMD and its subsequent distribution amongst the planets. Thus, we revise our M-R relation by adopting a more physically plausible relation for small sizes.

For planets above a certain transition radius ($R_p \geq R_{p,\rm trans}$), we still use the NWG18 relation; the large scatter in planet mass as a function of planet radius is consistent with previous findings that most planets above $\sim 1.6 R_\oplus$ are not rocky \citep{Rogers2015}. For planets with $R_p < R_{p,\rm trans}$, we switch to a different M-R relation based on the more physical, ``Earth--like rocky'' model from \citet{Zeng2019}. We choose $R_{p,\rm trans} = 1.472 R_\oplus$ as the transition radius because it is where the mean prediction for $M_p | R_p$ from NWG18 intersects the Earth--like rocky relation.\footnote{There are two additional intersection points, both below $0.7 R_\oplus$, but setting $R_{p,\rm trans}$ to either of them would not resolve our concerns regarding the bimodal mass distribution due to the sharp jump at $0.7 R_\oplus$ or the prevalence of planets with densities greater than that of pure iron planets.} We interpolate the Earth--like rocky table from \citet{Zeng2019} for $M_p$ as a function of $R_p$, which we denote as $M_{p,\rm ELR}(R_p)$, and use it as the mean prediction for a lognormal distribution of $M_p | R_p$ with a standard deviation ($\sigma$) that also scales with $R_p$:
\begin{eqnarray}
 \log_{10}\bigg(\frac{M_p}{M_\oplus}\bigg) &\sim& \mathcal{N}(\mu, \sigma), \nonumber \\
 \mu &=& \log_{10}\Big(\frac{M_{p,\rm ELR}(R_p)}{M_\oplus}\Big), \nonumber \\
 \sigma &=& m(R_p - R_{p,\rm trans}) + \sigma(R_{p,\rm trans}) \label{eq_MRsmall}
\end{eqnarray}
where $m = (\sigma(R_{p,\rm trans}) - \sigma(R_{p,\rm min}))/(R_{p,\rm trans} - R_{p,\rm min})$ is the slope of the linear relation for $\sigma(R_p)$.
The choice of a lognormal distribution is motivated by the symmetric scatter in $\log{M_p}$ from the NWG18 relation. We parametrize $\sigma$ as a linear function of $R_p$ for simplicity, where $\sigma(R_{p,\rm min}) = 0.04$ (corresponding to about a factor of $\sim 10\%$) and $\sigma(R_{p,\rm trans}) = 0.3$ (corresponding to a factor of $\sim 2$), chosen to match the scatter in the NWG18 relation at the transition radius. Thus, our M-R relation is approximately continuous in both the median prediction and the scatter in $\log_{10}M_p$ at all radii considered, including at $R_p = R_{p,\rm trans}$.

We also caution that the NWG18 relation is not strictly appropriate for drawing planet masses in a physical catalog because it is fit to a set of observed masses and radii and therefore does not account for the relevant detection biases. \citet{NR2020} developed a model for the underlying mass--radius--period distribution which would be more appropriate, but their work was more focused on describing the methodology rather than producing the best M-R relationship for a wide range of radii. Thus, we employ a combination of the NWG18 M-R relation for large planets (where the distribution is well-constrained by observations) and our simple physical model for small planets to address the issues discussed above.

\subsubsection{New procedure for generating a \textit{physical catalog}}
\label{secNewProc}
Here, we provide a step-by-step procedure for generating a \textit{physical catalog} from the maximum AMD model, by adapting the procedure outlined in \citetalias{HFR2019} for the old clustered model (\S2.2 therein). First, we set a number of target stars and a value for each model parameter. For each target:
\begin{enumerate}[leftmargin=*]
 \item Assign a random star from the \Kepler{} stellar catalog crossmatched with \Gaia{} DR2 and the \citet{B2020} catalog (see \S\ref{Kepler_data}).
 \item Compute the fraction of stars with planets ($f_{\rm swpa}$) for this star's $b_p-r_p-E^*$ color using equation \ref{eq_fswp_bprp}. Draw a number $u \sim {\rm Unif}(0,1)$. If $u > f_{\rm swpa}$, return the star with no planets; otherwise, continue.
 \item Draw a number of clusters in the system, $N_c \sim {\rm ZTP}(\lambda_c)$, to attempt. Re-sample until $N_c \leq N_{c,\rm max}$.
 \item For each cluster:
 \begin{enumerate}
  \item Draw a number of planets in the cluster, $N_p \sim {\rm ZTP}(\lambda_p)$. Re-sample until $N_p \leq N_{p,\rm max}$.
  \item Draw a characteristic radius, $R_{p,c}$. If $N_p = 1$, the radius of the one planet in this cluster is also $R_p = R_{p,c}$. If $N_p > 1$, draw a radius for each of the cluster's planets, $R_{p,k} \sim {\rm Lognormal}(R_{p,c}, \sigma_R)$, where $k = 1,\dots,N_p$ (the log is base-$e$).
  \item Draw the planet masses conditioned on their radii using the mass--radius relations described in \S\ref{secMR}.
  \item Draw unscaled periods for the planets in the cluster. If $N_p = 1$, assign an unscaled period of $P' = 1$. If $N_p > 1$, draw their unscaled periods $P'_k \sim {\rm Lognormal}(0, N_p\sigma_P)$, where $k = 1,\dots,N_p$ (the log is base-$e$), and sort them in increasing order. Check if $\Delta \geq \Delta_c$ and if equation \ref{eq_circ_overlap} are satisfied for all pairs in the cluster. Re-sample the unscaled periods $P'_k$ until this condition is satisfied or the maximum number of attempts (100) is reached. If the latter case occurs, discard the cluster.
  \item Draw a period scale factor $P_c$ (days) and multiply each planet's unscaled periods by the period scale for its parent cluster: $P_k = {P'_k}P_c$, where $k = 1,\dots,N_p$. Check if $\Delta \geq \Delta_c$ and if equation \ref{eq_circ_overlap} are satisfied for all adjacent planet pairs in the entire system, including planets from previously drawn clusters. Re-sample $P_c$ for the current cluster until this condition is satisfied or until the maximum number of attempts (100) is reached. If the latter case occurs, discard the cluster.
 \end{enumerate}
 \item If the total number of (successfully attempted) planets in the system is $N = 1$, draw an eccentricity $e \sim {\rm Rayleigh}(\sigma_{e,1})$ and argument of pericenter $\omega \sim {\rm Unif}(0,2\pi)$, and skip to step \ref{step_orbit}.
 \item Compute ${\rm AMD}_{\rm crit}$ for the system using equation \ref{eq_amd_crit}.
 \item Distribute ${\rm AMD}_{\rm crit}$ amongst the planets using equation \ref{eq_amd_per_pl}.
 \item Distribute the AMD of each planet randomly amongst the $x = e\sin{\omega}$, $y = e\cos{\omega}$, and $z=\sin{i_m}$ components subject to equation \ref{eq_xyz}. Compute $e = \sqrt{x^2+y^2}$, $\omega = {\rm atan}(x,y)$, and $i_m = \sin^{-1}{z}$ (mutual inclination relative to the system invariant plane) for each planet.
 \item Draw an angle of ascending node, $\Omega \sim {\rm Unif}(0,2\pi)$, and mean anomaly, $M \sim {\rm Unif}(0,2\pi)$, (relative to the system invariant plane) for each planet. \label{step_orbit}
 \item Specify the system invariant plane by drawing a random normal vector relative to the observer sky ($z$) axis.
 \item Compute the inclination angle $i$ (relative to the plane of the sky) for each planet's orbit, using rotations and dot products relative to the system invariant plane.
\end{enumerate}

\subsection{Observational comparisons} \label{Obs}

We define an expanded set of summary statistics and several distance functions that accounts for these summary statistics.

\subsubsection{Summary statistics} \label{SummaryStats}

We divide the stellar sample into two halves based on their $b_p-r_p-E^*$ colors (a ``bluer'' half and a ``redder'' half), in order to constrain the occurrence of planetary systems as a function of spectral type.
We also further expand on our set of summary statistics. For each observed catalog, we compute each of the following three times; once for the full sample and once for each half:
\begin{enumerate}[leftmargin=*]
 \item the total number of observed planets $N_{p,\rm tot}$ relative to the number of target stars $N_{\rm stars}$, $f = N_{p,\rm tot}/N_{\rm stars}$,
 \item the observed multiplicity distribution, $\{N_m\}$, where $N_m$ is the number of systems with $m$ observed planets and $m = 1,2,3,...$,
 \item the observed orbital period distribution, $\{P\}$,
 \item the observed period ratio distribution, $\{\mathcal{P} = P_{i+1}/P_i\}$,
 \item the observed transit depth distribution, $\{\delta\}$,
 \item the observed transit depth ratio distribution, $\{\delta_{i+1}/\delta_i\}$,
 \item the observed transit duration distribution, $\{t_{\rm dur}\}$,
 \item the observed circular-normalized transit duration distribution, $\{t_{\rm dur}/t_{\rm circ}\}$ where $t_{\rm circ} = \frac{R_\star{P}}{\pi{a}}$, of observed singles ($\{t_{\rm dur}/t_{\rm circ}\}_{1}$) and observed multis ($\{t_{\rm dur}/t_{\rm circ}\}_{2+}$),
 \item the observed period-normalized transit duration ratio distribution of adjacent planets apparently near an MMR, $\{\xi\}_{\rm res}$, and not near an MMR, $\{\xi\}_{\rm non-res}$. The normalized transit duration ratio is given by $\xi = (t_{\rm dur,in}/t_{\rm dur,out})(P_{\rm out}/P_{\rm in})^{1/3}$ \citep{S2010, F2014}.
\end{enumerate}

In addition to the list above, we also compute a few system--level summary statistics adapted from the metrics defined in \citet{GF2020}. In that study, several measures drawn from information theory are used to capture the global architectures of planetary systems. Specifically, \citet{GF2020} applied the concepts of ``Shannon entropy'' \citep{Sh1948}, ``disequilibrium'' (qualitatively opposite to entropy), and ``convex complexity'' (a product of entropy and disequilibrium; \citealt{LMC1995, LMC2010}) to planetary systems. With a focus on quantifying the correlations within systems (e.g. uniforming spacing and ``peas in a pod''; \citealt{MWL2017, W2018a, Z2019}), they defined metrics including \textit{mass partitioning} (to quantify the similarities in planet masses), \textit{monotonicity} (to quantify the mass ordering of planets), and \textit{gap complexity} (to quantify the uniformity of spacings between planets), amongst other statistics. \citet{GF2020} tested EPOS \citep{Mu2018} and SysSim (the clustered periods and sizes model from \citetalias{HFR2019}) using these metrics and found that while our clustered models performed well in many ways (including \textit{mass partitioning}, due to our clustering in planet sizes), there were statistically significant differences in both \textit{monotonicity} and \textit{gap complexity} between our models and the \Kepler{} data. In particular, our clustered model from \citetalias{HFR2019} tends to produce too many systems with negative monotonicity (i.e. planet sizes decreasing with increasing period) and too much gap complexity (i.e. too much variation between spacings of adjacent planets). However, these statistics were not included in our distance functions in \citetalias{HFR2019}, so it is unclear how well our models could perform in these metrics, or what sort of model constraints are provided by these metrics.

Here, we define analogous metrics to those mentioned above, using planet radius instead of planet mass as the relevant quantity. We choose to work with radius rather than mass because it is an observable quantity from the \Kepler{} mission (or any other transit survey), provided the stellar radii are well characterized. Indeed, few planets in the \Kepler{} catalog have measured masses, which would also be subject to other intractable detection biases; the alternative (and what \citealt{GF2020} opted to do) is to rely on a mass-radius relation, which is highly model dependent. Following \citet{GF2020}, we define \textit{radius partitioning} (analogous to their \textit{mass partitioning}; equations 7 and 8 therein) as:
\begin{eqnarray}
 \mathcal{Q}_R &\equiv& \bigg(\frac{m}{m - 1}\bigg) \Bigg(\sum_{k=1}^{m}\Big(R_{p,k}^{*} - \frac{1}{m}\Big)^2 \Bigg), \label{eq_radius_partitioning} \\
 R_{p,k}^{*} &=& \frac{R_{p,k}}{\sum_{i=1}^{m}R_{p,i}}, \label{eq_norm_R}
\end{eqnarray}
where $m$ is the observed multiplicity (number of planets in a system) and $R_{p,k}^{*}$ is the normalized planet radius for the $k^{\rm th}$ planet. We also define \textit{radius monotonicity} (to differentiate from the \textit{monotonicity} in \citealt{GF2020}; equation 9 therein) as:
\begin{equation}
 \mathcal{M}_R \equiv \rho_S{\mathcal{Q}_R}^{1/m}, \label{eq_radius_monotonicity}
\end{equation}
where $\rho_S$ is the Spearman's rank correlation coefficient of the planet radii (i.e. vs. their indices when sorting by their periods) in a system. The value of $\rho_S$ ranges from $-1$ (strictly decreasing order) to 1 (strictly increasing order), but does not encapsulate the magnitude of any monotonic trend, which is achieved by the inclusion of the ${\mathcal{Q}_R}^{1/m}$ factor (see \S3.3 of \citealt{GF2020} for a further explanation). Finally, we use the same definition for \textit{gap complexity} (equations 13 and 14 in \citealt{GF2020}):
\begin{eqnarray}
 \mathcal{C} &\equiv& -K \Bigg(\sum_{i=1}^n {p_i^{*}\log{p_i^{*}}} \Bigg) \cdot \Bigg(\sum_{i=1}^n \Big(p_i^{*} - \frac{1}{n}\Big)^2 \Bigg), \label{eq_gap_complexity} \\
 p_i^{*} &=& \frac{\log{\mathcal{P}_i}}{\log(P_{\rm max}/P_{\rm min})}, \label{eq_norm_P}
\end{eqnarray}
where $n = m-1$ is the number of adjacent planet pairs (i.e. gaps) in a system, $\mathcal{P}_i$ are their period ratios, and $K$ is a normalization constant such that $\mathcal{C}$ is always in the range $(0,1)$. The exact value of $K = 1/\mathcal{C}_{\rm max}$ is a function of $n$ that must be computed numerically \citep{AP1996}; \citet{GF2020} provide a table of $\mathcal{C}_{\rm max}$ for $n = 2,\dots,9$ along with an empirical relation for $\mathcal{C}_{\rm max}(n)$ fit to these values.

As with our other summary statistics, we compute the distributions of these system-level metrics, $\{\mathcal{Q}_R\}$, $\{\mathcal{M}_R\}$, and $\{\mathcal{C}\}$, for the full catalog as well as for the bluer and redder halves. The \textit{radius partitioning} and \textit{monotonicity} can be computed for all systems with $m \geq 2$ observed planets, while the \textit{gap complexity} can only be computed for systems with $m \geq 3$ planets (since at least two gaps are needed).

\subsubsection{Distance function} \label{Distance_functions}

\begin{deluxetable}{lcccccc}
\centering
\tablecaption{Weights for the individual distance terms computed from a reference clustered periods and sizes model (\citetalias{HFR2019}).}
\tablehead{
 \colhead{Distance term} & \multicolumn2c{All} & \multicolumn2c{Bluer} & \multicolumn2c{Redder} \\
 & \colhead{$\hat{\sigma}(\mathcal{D})$} & \colhead{$w$} & \colhead{$\hat{\sigma}(\mathcal{D})$} & \colhead{$w$} & \colhead{$\hat{\sigma}(\mathcal{D})$} & \colhead{$w$}
}
\decimalcolnumbers
\startdata
 $D_f$ & 0.00103 & 971 & 0.00146 & 683 & 0.00154 & 649 \\
 $D_{\rm mult}$ & 0.00593 & 169 & 0.01150 & 87 & 0.01373 & 73 \\
 \hline
 $\mathcal{D}$ (KS): & & & & & & \\
 $\{P\}$ & 0.02616 & 38 & 0.03544 & 28 & 0.03805 & 26 \\
 $\{\mathcal{P}\}$ & 0.04836 & 21 & 0.06441 & 16 & 0.07167 & 14 \\
 $\{\delta\}$ & 0.02907 & 34 & 0.03988 & 25 & 0.04121 & 24 \\
 $\{\delta_{i+1}/\delta_i\}$ & 0.05106 & 20 & 0.06821 & 15 & 0.07437 & 13 \\
 $\{t_{\rm dur}\}$ & 0.02831 & 35 & 0.03928 & 25 & 0.03995 & 25  \\
 $\{t_{\rm dur}/t_{\rm circ}\}_{1}$ & 0.03554 & 28 & 0.05019 & 20 & 0.05066 & 20 \\
 $\{t_{\rm dur}/t_{\rm circ}\}_{2+}$ & 0.04054 & 25 & 0.05673 & 18 & 0.05785 & 17 \\
 $\{\xi_{\rm res}\}$ & 0.11572 & 9 & 0.16131 & 7 & 0.17897 & 6 \\
 $\{\xi_{\rm non-res}\}$ & 0.05607 & 18 & 0.07361 & 14 & 0.08078 & 12 \\
 $\{\mathcal{Q}_R\}$ & 0.06078 & 16 & 0.08128 & 12 & 0.09019 & 11 \\
 $\{\mathcal{M}_R\}$ & 0.06558 & 15 & 0.08828 & 11 & 0.09546 & 10 \\
 $\{\mathcal{C}\}$ & 0.10404 & 10 & 0.13641 & 7 & 0.15676 & 6 \\
 \hline
 $\mathcal{D}$ (AD$'$): & & & & & & \\
 $\{P\}$ & 0.00113 & 882 & 0.00218 & 459 & 0.00233 & 429 \\
 $\{\mathcal{P}\}$ & 0.00329 & 304 & 0.00602 & 166 & 0.00736 & 136 \\
 $\{\delta\}$ & 0.00138 & 723 & 0.00263 & 380 & 0.00276 & 362 \\
 $\{\delta_{i+1}/\delta_i\}$ & 0.00392 & 255 & 0.00698 & 143 & 0.00862 & 116 \\
 $\{t_{\rm dur}\}$ & 0.00145 & 691 & 0.00291 & 344 & 0.00302 & 331 \\
 $\{t_{\rm dur}/t_{\rm circ}\}_{1}$ & 0.00221 & 453 & 0.00421 & 237 & 0.0043 & 233 \\
 $\{t_{\rm dur}/t_{\rm circ}\}_{2+}$ & 0.00267 & 374 & 0.00563 & 178 & 0.00533 & 188 \\
 $\{\xi_{\rm res}\}$ & 0.02098 & 48 & 0.04515 & 22 & 0.05154 & 19 \\
 $\{\xi_{\rm non-res}\}$ & 0.00479 & 209 & 0.00808 & 124 & 0.00982 & 102 \\
 $\{\mathcal{Q}_R\}$ & 0.00612 & 163 & 0.01045 & 96 & 0.01303 & 77 \\
 $\{\mathcal{M}_R\}$ & 0.00700 & 143 & 0.01310 & 76 & 0.01533 & 65 \\
 $\{\mathcal{C}\}$ & 0.01701 & 59 & 0.03099 & 32 & 0.03942 & 25 \\
\enddata
\tablecomments{Each weight $w$ is computed as the inverse of the root mean square of the distances $\hat{\sigma}(\mathcal{D})$ between repeated realizations of the same (i.e. ``perfect'') model, $w = 1/\hat{\sigma}(\mathcal{D})$, using the same number of target stars as our \Kepler{} sample. The weights are shown here as rounded whole numbers for guidance purposes only.}
\label{tab:weights}
\end{deluxetable}

In \citetalias{HFR2019}, we used a linear weighted sum of individual distance terms to combine the fits to each summary statistic into a single distance function. Two separate distance functions were used, with one adopting the two-sample Kolmogorov--Smirnov (KS; \citealt{K1933, S1948}) distance for each marginal distribution and the other adopting a modified version of the two-sample Anderson--Darling (AD; \citealt{AD1952, P1976}; see equations 23--24 in \citetalias{HFR2019} for our modification) statistic. Each distance function includes a term for the overall rate of planets, $D_f$, and the observed multiplicity distribution, $D_{\rm mult}$:
\begin{eqnarray}
 D_f &=& | f_{\rm sim} - f_{\rm Kepler}|, \label{eq_dist_rate} \\
 D_{\rm mult} &=& \rho_{\rm CRPD} = \frac{9}{5}\sum_j O_j \bigg[{\bigg(\frac{O_j}{E_j}\bigg)}^{2/3} - 1\bigg]. \label{eq_dist_mult}
\end{eqnarray}
The term for fitting the rate of planets is simply the absolute difference in the ratios of observed planets to target stars, where $f_{\rm sim} = N_{p,\rm tot}/N_{\rm stars}$ (and likewise for \Kepler{}). For the term in equation \ref{eq_dist_mult}, we adopt the ``Cressie--Read power divergence'' \citep{CR1984}, where $O_j$ are the numbers of ``observed'' systems in our models, and $E_j$ are the numbers of expected systems from the \Kepler{} data, for multiplicity bins $j = 1,2,3,4,5+$ (see the discussion surrounding equation 19 in \citetalias{HFR2019}).

In this paper, we define three different distance functions (with KS and AD versions for each, totaling six separate analyses). We start with the exact same distance function as in \citet{HFR2020}:
\begin{eqnarray}
 \mathcal{D}_{W,1} &=& \sum_{\rm samples} \sum_{i'} w_{i'} \mathcal{D}_{i'} \label{eq_dist_general} \\
 &=& \sum_{\rm samples} \bigg[ \frac{D_f}{\hat{\sigma}(D_f)} + \frac{D_{\rm mult}}{\hat{\sigma}(D_{\rm mult})} + \sum_{i \in S_1} \frac{\mathcal{D}_i}{\hat{\sigma}(\mathcal{D}_i)} \bigg], \label{eq_dist1}
\end{eqnarray}
where $w_{i'} = 1/\hat{\sigma}(\mathcal{D}_{i'})$ are the weights for each individual distance term (listed in Table \ref{tab:weights}) and everything within the outer summation refer to the distances computed using the summary statistics in a given sample only. The distances $\mathcal{D}_i$ within the inner summation are either KS or AD distances, where the summation is over the indices labeling the summary statistics in the set $S_1 = \{3,\dots,7, 9\}$. The purpose of applying the same distance function (including the weights $w_{i'}$) to our new model is to enable a direct comparison between the two models.

For the second distance function, we swap out the term for the $\{t_{\rm dur}\}$ distribution with terms for the distributions of circular-normalized transit durations of observed singles and multis, $\{t_{\rm dur}/t_{\rm circ}\}_{1}$ and $\{t_{\rm dur}/t_{\rm circ}\}_{2+}$, respectively:
\begin{equation}
 \mathcal{D}_{W,2} = \sum_{\rm samples} \bigg[ \frac{D_f}{\hat{\sigma}(D_f)} + \frac{D_{\rm mult}}{\hat{\sigma}(D_{\rm mult})} + \sum_{i \in S_2} \frac{\mathcal{D}_i}{\hat{\sigma}(\mathcal{D}_i)} \bigg], \label{eq_dist2}
\end{equation}
where $S_2 = \{3,\dots,6,8,9\}$. We use the circular-normalized transit durations because they are more sensitive to the distribution of eccentricities, which is a key feature of our new model for planetary system architectures. The motivation for separating the observed singles from the observed multis is due to our separate treatment of the intrinsic single planet systems; in particular, we aim to constrain the eccentricity scale ($\sigma_{e,1}$) of these systems.

Finally, we test a third distance function to also incorporate the system--level metrics inspired or taken from \citet{GF2020}, as we denote here by $S_{\rm GF}$ and as we defined in \S\ref{SummaryStats}, by adding weighted terms to our previous distance function:
\begin{eqnarray}
 \mathcal{D}_{W,3} &=& \mathcal{D}_{W,2} +  \sum_{\rm samples} \sum_{i \in S_{\rm GF}} \frac{\mathcal{D}_i}{\hat{\sigma}(\mathcal{D}_i)} \\
 &=& \sum_{\rm samples} \bigg[ \frac{D_f}{\hat{\sigma}(D_f)} + \frac{D_{\rm mult}}{\hat{\sigma}(D_{\rm mult})} \nonumber \\
 && \quad\quad\quad + \sum_{i \in S_2} \frac{\mathcal{D}_i}{\hat{\sigma}(\mathcal{D}_i)} + \sum_{i \in S_{\rm GF}} \frac{\mathcal{D}_i}{\hat{\sigma}(\mathcal{D}_i)} \bigg]. \label{eq_dist3}
\end{eqnarray}
In other words, the terms in the summation over $S_{\rm GF}$ are also either KS or AD distances computed between the observed distributions (of $\{\mathcal{Q}_R\}$, $\{\mathcal{M}_R\}$, and $\{\mathcal{C}\}$) of our model and of the \Kepler{} data.

\subsubsection{The Kepler planet catalog} \label{Kepler_data}

Our stellar catalog is described in \S\ref{secStars}. To constrain our models, we use a planet catalog derived from the \Kepler{} DR25 KOI table (only keeping planet candidates around stars in our stellar catalog), where we also:
\begin{enumerate}[leftmargin=*]
 \item replace the transit depths and durations with the median values from the posterior samples in \citet{R2015},
 \item replace the planet radii based on the transit depths and the updated \citet{B2020} stellar radii, and
 \item only keep planets in the period range $[3, 300]$ days and planet radii range $[0.5, 10] R_\oplus$.
\end{enumerate}

\subsection{Model optimization}

The full details for performing inference using approximate Bayesian computation (ABC) on our model parameters (i.e. steps 4--6 as listed at the beginning of \S\ref{Methods}) are described in \citetalias{HFR2019} (\S2.5-2.6 therein). Here, we further summarize our method.

\begin{deluxetable}{lccc}
\centering
\tablecaption{Optimizer bounds, GP length scales $\lambda_i$, and emulator bounds for the parameters of the maximum AMD model.}
\tablehead{\colhead{Parameter} & \colhead{Optimizer bounds} & \colhead{$\lambda_i$} & \colhead{Emulator bounds}}
\decimalcolnumbers
\startdata
 $f_{\rm swpa,med}$ & $(0, 1)$ & 0.2 & $(0.6, 1)$ \\
 $\frac{df_{\rm swpa}}{d(b_p-r_p-E^*)}$ & $(-1, 2)$ & 1 & $(-0.6, 2)$ \\
 $\ln{(\lambda_c)}$ & $(\ln(0.2), \ln(10))$ & - & - \\
 $\ln{(\lambda_p)}$ & $(\ln(0.2), \ln(10))$ & - & - \\
 $\ln{(\lambda_c \lambda_p)}$ & - & 1 & $(-1, 2)$ \\
 $\ln{(\frac{\lambda_p}{\lambda_c})}$ & - & 1.5 & $(-1.5, 3)$ \\
 $\Delta_c$ & $(3, 20)$ & 3 & $(6, 15)$ \\
 $\alpha_P$ & $(-2, 2)$ & 1 & $(-0.8, 1.6)$ \\
 $\alpha_{R1}$ & $(-4, 2)$ & 0.5 & $(-2, -0.5)$ \\
 $\alpha_{R2}$ & $(-6, 0)$ & 1 & $(-6, -3)$ \\
 $\sigma_{e,1}$ & $(0, 0.5)$ & 0.2 & $(0, 0.5)$ \\
 $\sigma_R$ & $(0, 0.5)$ & 0.15 & $(0.1, 0.5)$ \\
 $\sigma_P$ & $(0, 0.5)$ & 0.15 & $(0.1, 0.5)$ \\
\enddata
\tablecomments{The same values are used for all analyses (all distance functions, including KS and AD terms). We varied the parameters $\ln(\lambda_c)$ and $\ln(\lambda_p)$ separately in the optimization stage, while we trained and predicted on $\ln(\lambda_c \lambda_p)$ and $\ln(\lambda_p/\lambda_c)$ during the emulator stage.}
\label{tab:optim_GP}
\end{deluxetable}

\subsubsection{Optimization stage}

For each distance function (e.g. $\mathcal{D}_{W,1}$, $\mathcal{D}_{W,2}$, and $\mathcal{D}_{W,3}$, each involving KS or AD terms), we attempt to find the minimum of the function using a Differential Evolution optimizer (in the ``BlackBoxOptim.jl'' Julia package). This optimizer implements a population-based genetic algorithm, where ``individuals'' of a starting population are evolved such that ones with better ``fitness'' are more likely to survive and pass on their properties to future generations; for our problem, each ``individual'' in the population is a set of model parameters, and its ``fitness'' is the distance evaluated at that point in parameter space. We choose a population size of four times the number of free model parameters ($4 \times 11 = 44$) and evolve for 5000 model evaluations, saving the model parameters and distances at each evaluation. We then repeat this optimization process 50 times with a different random seed each time. Thus, this results in a collection of $5000 \times 50 = 2.5 \times 10^5$ points (model evaluations) for each distance function. The optimizer bounds for each parameter are listed in Table \ref{tab:optim_GP}.

\subsubsection{GP emulator stage}

Simulating a full \textit{physical} and \textit{observed catalog} is computationally expensive and the genetic algorithm must evaluate the distance function in series, causing the optimization stage to be limited to a few thousand model evaluations per optimizer run. The distance function is also noisy due to the stochasticity of our model. An adequate exploration of the 11-dimensional parameter space requires tens to hundreds of millions of points; thus, we train a Gaussian process (GP) emulator \citep{RW2006} for each distance function, which is described by a prior mean function $m(\bm{x})$ and a covariance (i.e. kernel) function $k(\bm{x},\bm{x'};\bm{\phi})$:
\begin{eqnarray}
 f(\bm{x}) &\sim& \mathcal{GP}\big(m(\bm{x}), k(\bm{x},\bm{x'};\bm{\phi})\big), \label{eq_GP} \\
 k(\bm{x},\bm{x'};\bm{\phi}) &=& \sigma_f^2 {\rm exp} \Bigg[-\frac{1}{2} \sum_i \frac{(x_i - {x_i}')^2}{\lambda_i^2} \Bigg], \label{eq_kernel}
\end{eqnarray}
where $f(\bm{x}) = \mathcal{D}_W$ is the distance function we wish to model, $\bm{x}$ and $\bm{x^\prime}$ are sets of model parameters, and $\bm{\phi} = (\sigma_f, \lambda_1, \lambda_2,..., \lambda_d)$ are the relevant hyperparameters of this ``squared exponential'' kernel.

For a given mean function, kernel function, and set of training points, the GP emulator is fully defined and can ``predict'' the outputs (i.e. distance function evaluations) given inputs (i.e. model parameters). The training points are a subset of the points from the optimization stage. For inputs far away from any training points, the emulator will return values distributed close to the mean function; we choose a constant mean function that is set to a large value relative to the majority of our training points, so that emulated distances at such points will be significantly worse than the best model evaluations. The value of the mean function for each distance function is listed in Table \ref{tab:d_thres}; as in \citetalias{HFR2019}, they are significantly larger when involving AD distances because we find that our AD distance is more sensitive to deviations from a perfect fit than the KS distance.

\subsubsection{ABC inference stage}

We compute the ABC posterior distributions for the model parameters, for each distance function, by using the emulator to evaluate each distance function at a large number of points. We draw these points from our prior, which we assume is a uniform distribution for each parameter (with bounds listed in Table \ref{tab:optim_GP}), and keep points passing a certain distance threshold ($\mathcal{D}_{\rm thres}$). The distance threshold is chosen based on the best distances achieved during the optimization stage. In this paper, we used three pairs of distance functions: KS and AD for $\mathcal{D}_{W,1}$, $\mathcal{D}_{W,2}$, and $\mathcal{D}_{W,3}$. Each of these is a weighted sum of individual terms that are normalized (weighted) such that a perfect model would contribute a distance of $\sim 1$ for each term. The number of individual terms and the distance threshold for each distance function are shown in Table \ref{tab:d_thres}.

\begin{deluxetable}{lcccc}
\centering
\tablecaption{Best distances, mean functions, and distance thresholds for the maximum AMD model, for each distance function.}
\tablehead{\colhead{Distance function} & \colhead{\# of terms} & \colhead{Best dist.} & \colhead{$m(\bm{x})$} & \colhead{$\mathcal{D}_{\rm thres}$}}
\decimalcolnumbers
\startdata
 $\mathcal{D}_{W,1}$ (KS) & $9 \times 3$ & $\sim 33$ & 75 & 45 \\
 $\mathcal{D}_{W,2}$ (KS) & $10 \times 3$ & $\sim 35$ & 75 & 45 \\
 $\mathcal{D}_{W,3}$ (KS) & $13 \times 3$ & $\sim 50$ & 100 & 65 \\
 \hline
 $\mathcal{D}_{W,1}$ (AD$'$) & $9 \times 3$ & $\sim 50$ & 150 & 80 \\
 $\mathcal{D}_{W,2}$ (AD$'$) & $10 \times 3$ & $\sim 50$ & 150 & 80 \\
 $\mathcal{D}_{W,3}$ (AD$'$) & $13 \times 3$ & $\sim 75$ & 250 & 120 \\
\enddata
\tablecomments{In column (2), the number of terms for each distance function is a multiple of three because we compute distances for the full sample as well as the bluer and redder halves, and is equivalent to the typical total distance for a perfect model (as each term is weighted to one).}
\label{tab:d_thres}
\end{deluxetable}

\section{Results} \label{Results}

We organize the main results as follows.
First, we briefly report how the new ``maximum AMD model'' compares to the ``two-Rayleigh model'' in terms of fitting the \Kepler{} data, in \S\ref{Results_old_new}. 
In \S\ref{Results_params}, we present and discuss the best--fit parameters of the maximum AMD model and the underlying distributions of planetary systems resulting from it. 
Next, we explore the primary new features of the maximum AMD model, the eccentricity and mutual inclination distributions, in \S\ref{Results_ecc_incl}.  
In particular, we show that the maximum AMD model naturally: 
(1) produces correlations in the distribution of eccentricities and mutual inclinations with intrinsic multiplicity,
(2) leads to trends in the observed $\xi$ distribution with observed multiplicity that match the patterns seen in the \Kepler{} data, and
(3) generates a physically plausible joint distribution of orbital eccentricities and mutual inclinations.
Finally, we discuss the eccentricity distribution of intrinsically single--planet inner planetary systems (\S\ref{secEccSingles}) and correlations of eccentricity and mutual inclination with the minimum ratio of orbital periods (\S\ref{secCorMinPeriodRatio}).

\begin{figure*}
\centering
\begin{tabular}{cc}
 \includegraphics[scale=0.425,trim={0 0.5cm 0 0.2cm},clip]{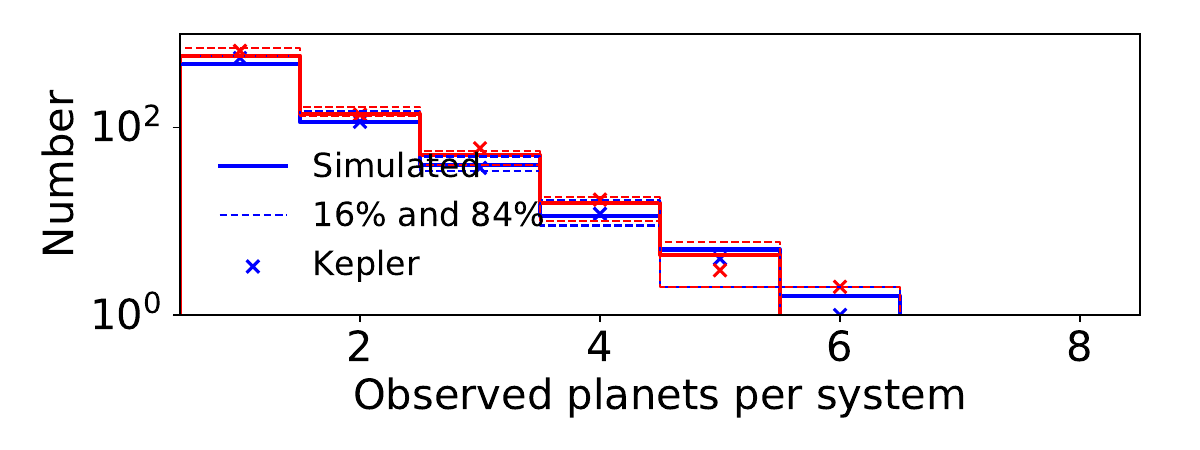} &
 \includegraphics[scale=0.425,trim={0 0.5cm 0 0.2cm},clip]{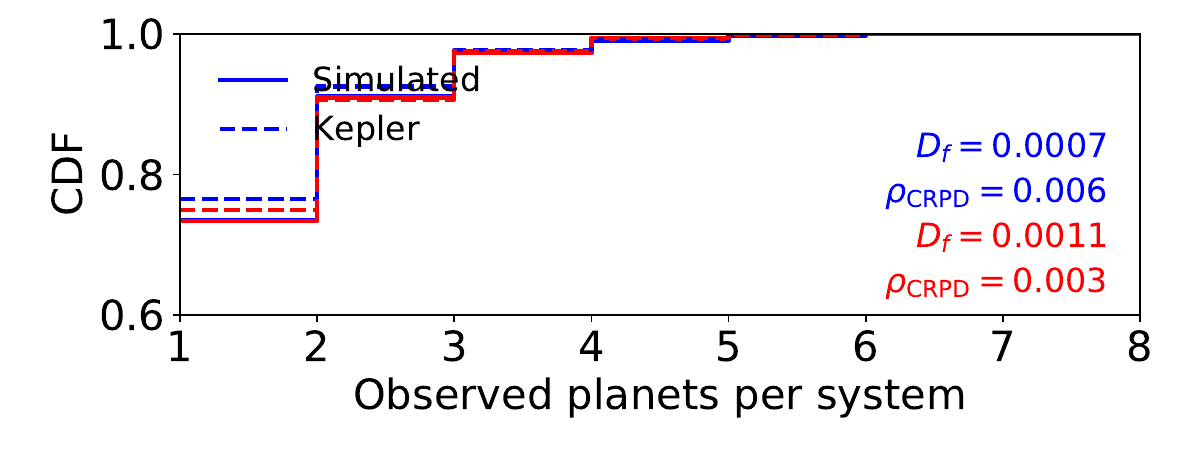} \\
 \includegraphics[scale=0.425,trim={0 0.5cm 0 0.2cm},clip]{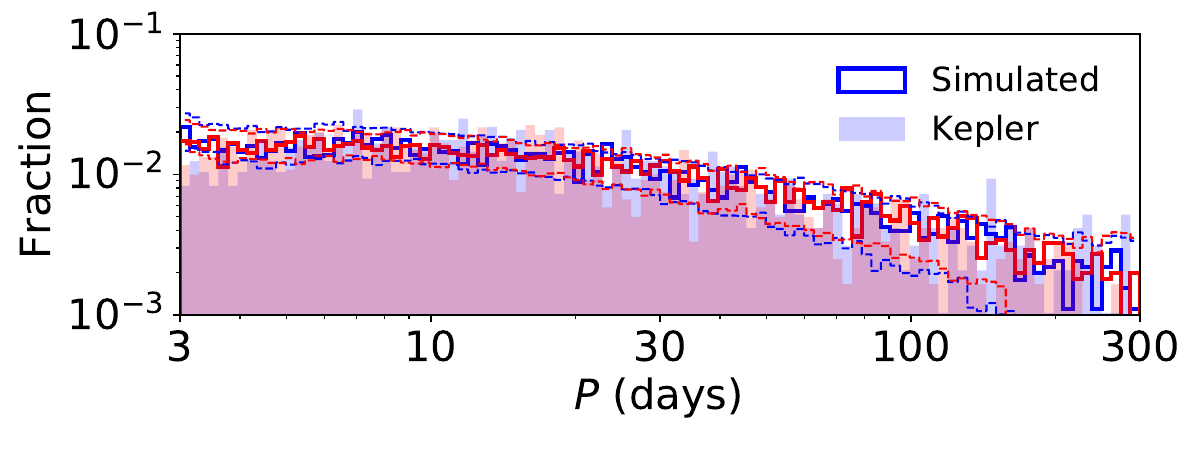} &
 \includegraphics[scale=0.425,trim={0 0.5cm 0 0.2cm},clip]{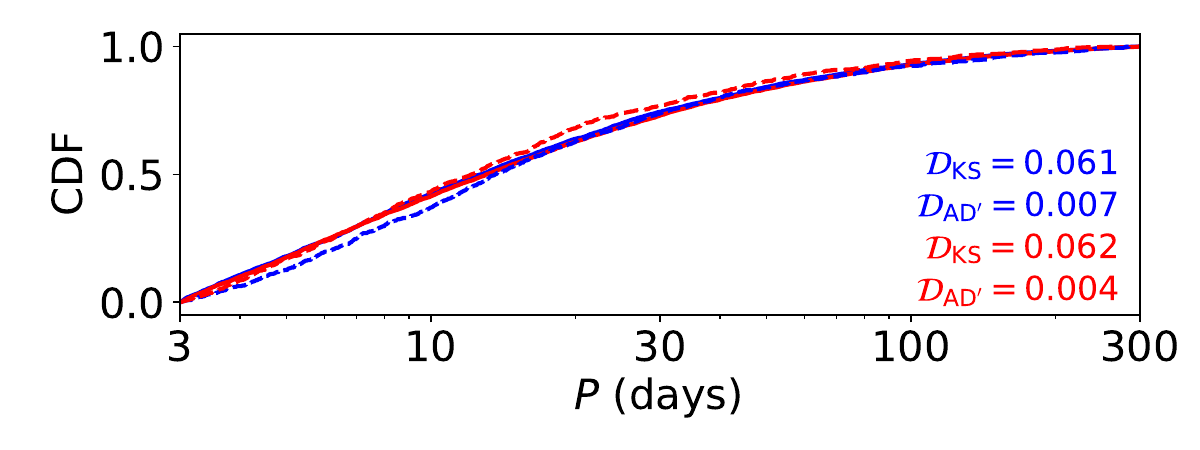} \\
 \includegraphics[scale=0.425,trim={0 0.5cm 0 0.2cm},clip]{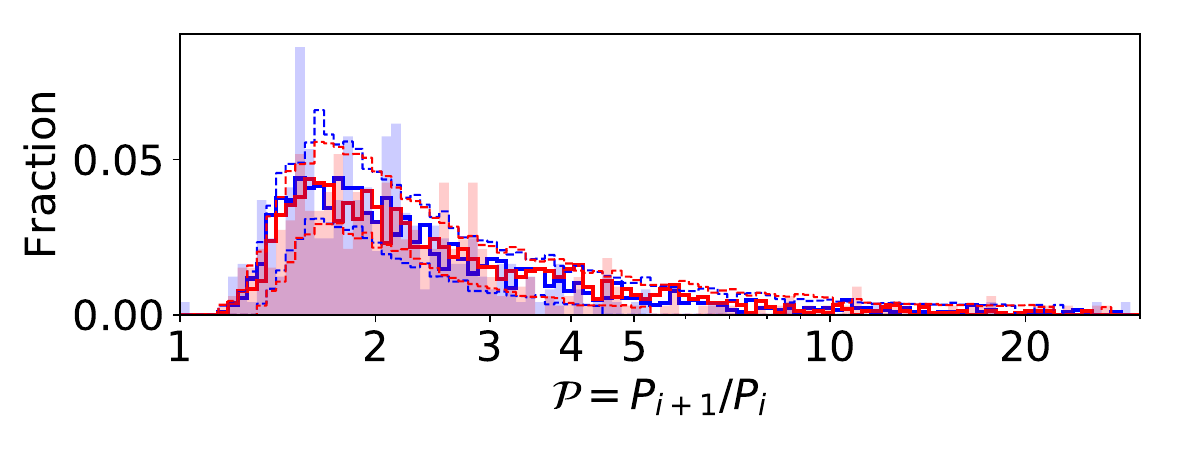} &
 \includegraphics[scale=0.425,trim={0 0.5cm 0 0.2cm},clip]{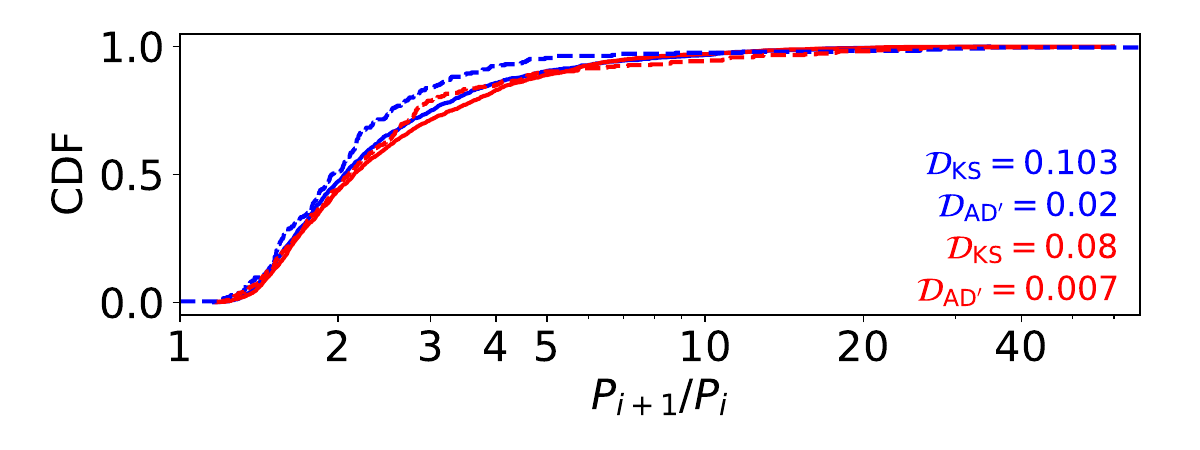} \\
 \includegraphics[scale=0.425,trim={0 0.5cm 0 0.2cm},clip]{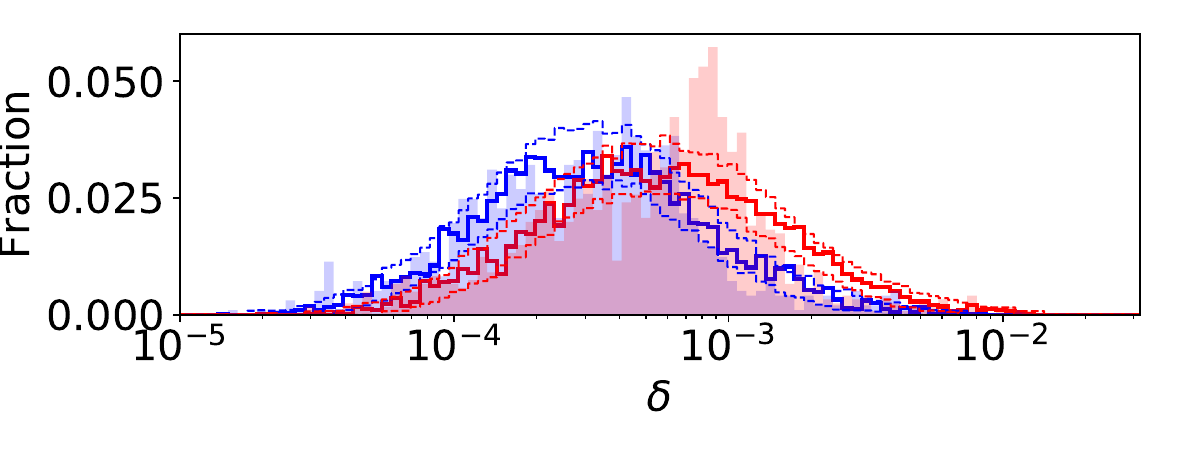} &
 \includegraphics[scale=0.425,trim={0 0.5cm 0 0.2cm},clip]{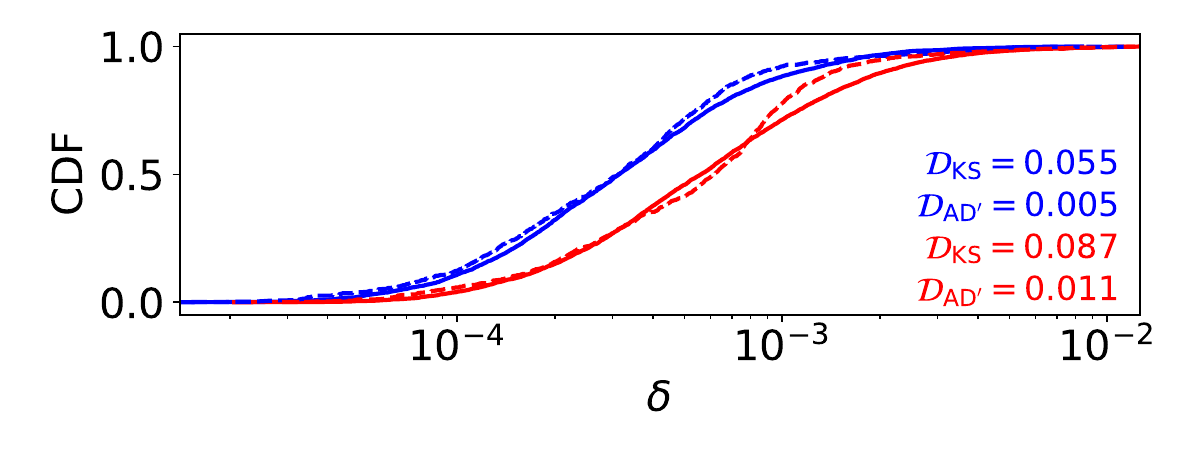} \\
 \includegraphics[scale=0.425,trim={0 0.5cm 0 0.2cm},clip]{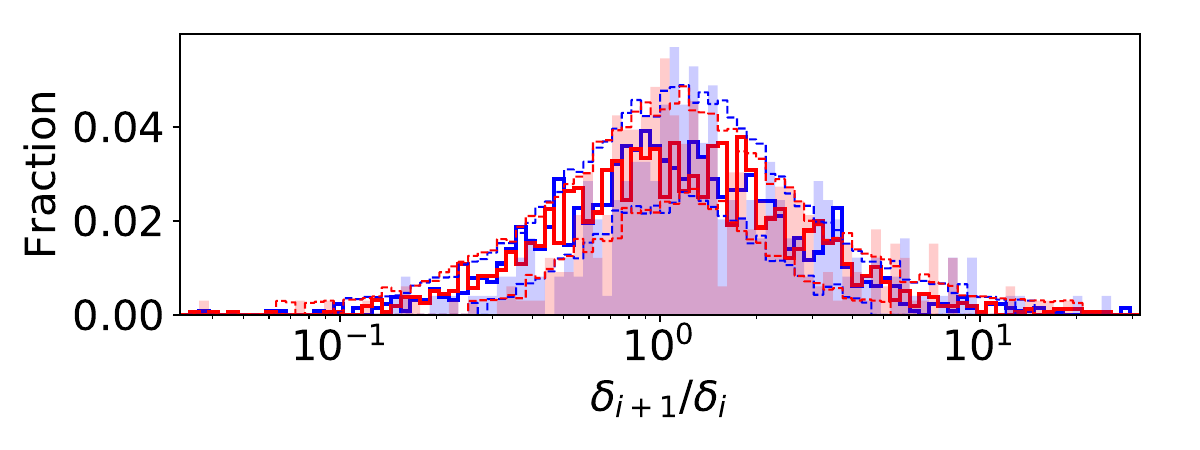} &
 \includegraphics[scale=0.425,trim={0 0.5cm 0 0.2cm},clip]{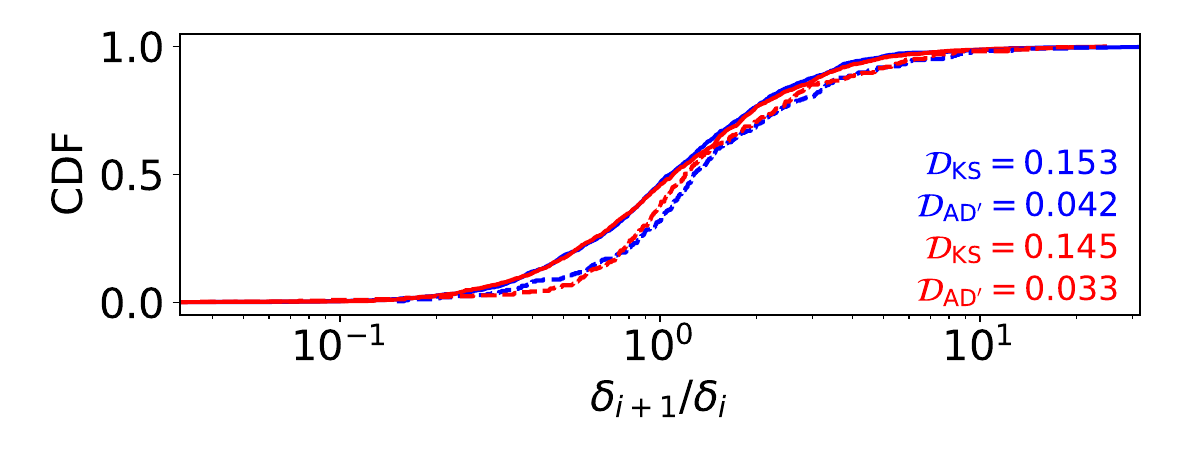} \\
 \includegraphics[scale=0.425,trim={0 0.5cm 0 0.2cm},clip]{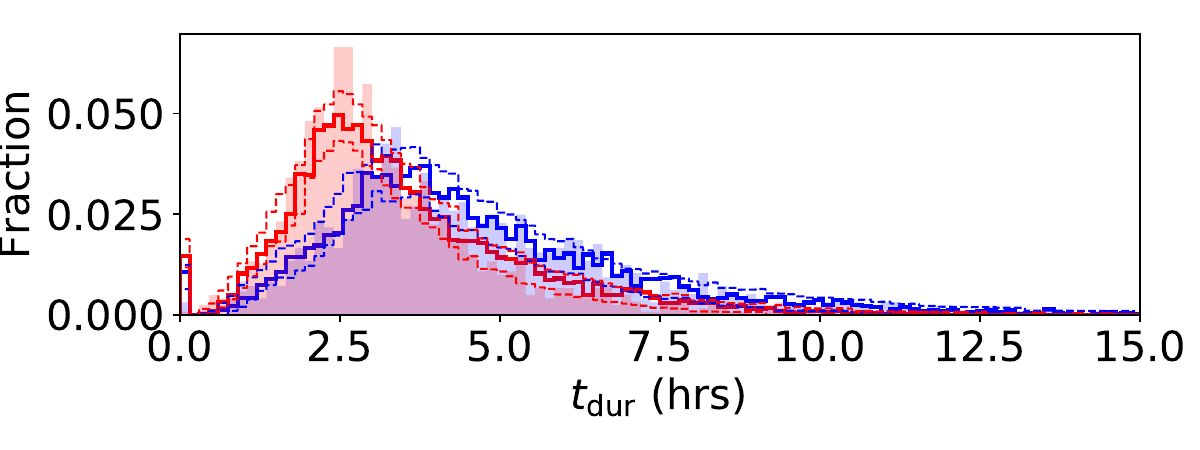} &
 \includegraphics[scale=0.425,trim={0 0.5cm 0 0.2cm},clip]{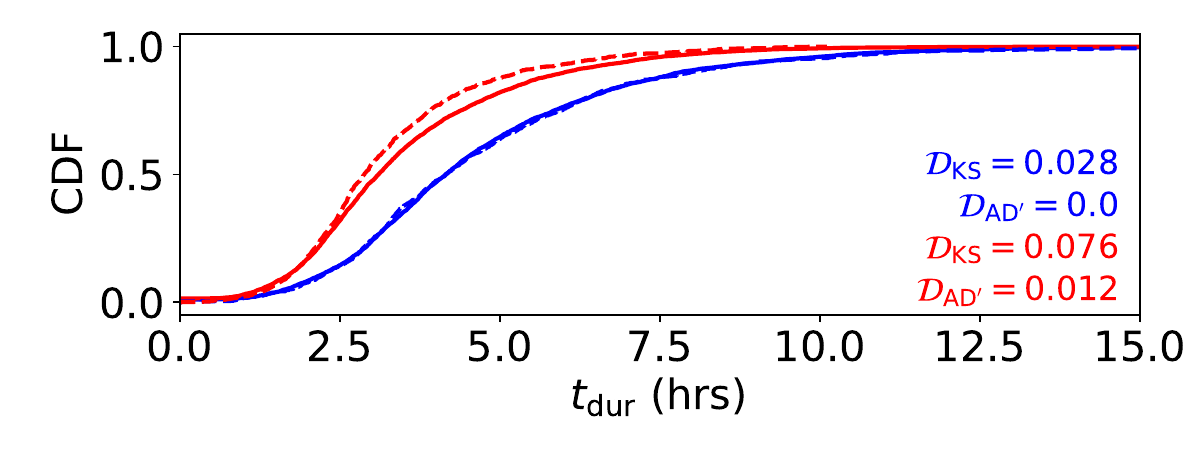} \\
 \includegraphics[scale=0.425,trim={0 0.5cm 0 0.2cm},clip]{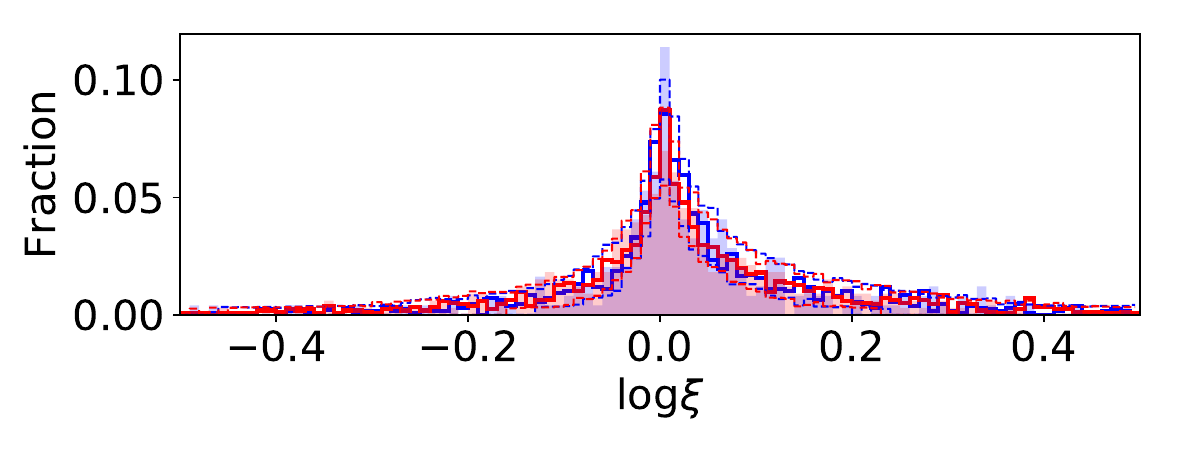} &
 \includegraphics[scale=0.425,trim={0 0.5cm 0 0.2cm},clip]{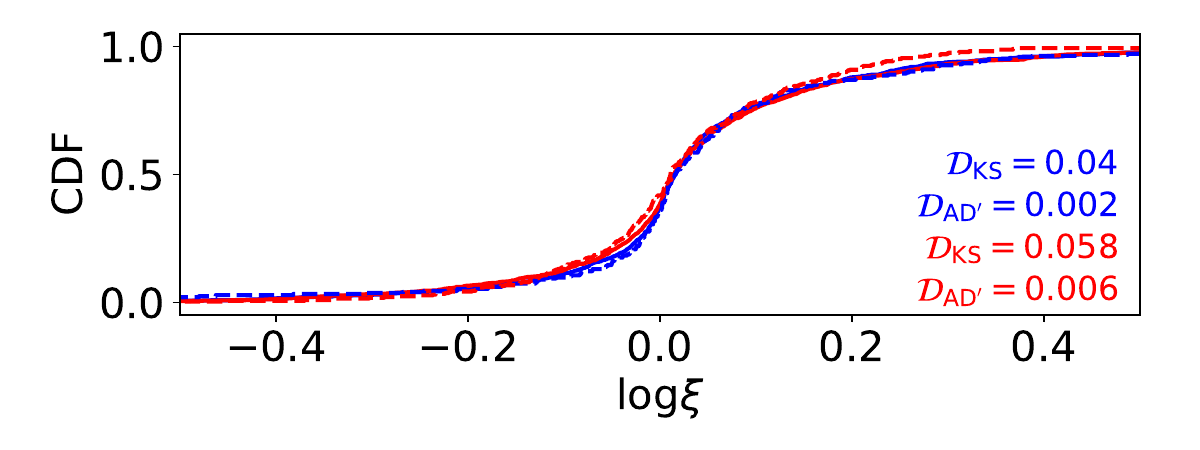} \\
\end{tabular}
\caption{Marginal distributions of observable properties for our maximum AMD model as compared to the \Kepler{} data, split into bluer and redder halves as likewise colored. \textbf{Left-hand panels:} histograms of these observables, as labeled. The solid bold lines show one simulated observed catalog from this model (with parameter values listed in Table \ref{tab:param_fits}), while the \Kepler{} DR25 exoplanets are plotted as shaded, filled histograms for comparison.
The dashed lines show the 16th and 84th percentiles of each bin based on 100 simulated catalogs with parameters drawn from our emulator with $\mathcal{D}_{W,3} \leq 65$ (KS).
\textbf{Right-hand panels:} the corresponding CDFs to the left-hand panels. The solid bold lines show the one simulated catalog, while the dashed lines show the \Kepler{} distributions. The relevant KS and AD distances (unweighted) are shown in each panel.}
\label{fig:model_split_bprp}
\end{figure*}

\subsection{Comparison of old and new models} \label{Results_old_new}

Before we describe all the new results using our maximum AMD model, we first show how well this model fits the \Kepler{} data in comparison to the previous two--Rayleigh model.
To facilitate direct comparison, we use the same summary statistics and distance function from \citet{HFR2020} (i.e., $\mathcal{D}_{W,1}$). In Figure \ref{fig:model_split_bprp}, we plot the marginal distributions of a simulated \textit{observed catalog} from our maximum AMD model (bold blue and red histograms), with the \Kepler{} DR25 catalog over-plotted (shaded blue and red histograms) for comparison. We split the observed catalogs (both simulated and real) into two halves at the median stellar $b_p-r_p-E^*$ color, as we fit to the marginal distributions of each of the bluer and redder samples simultaneously (\S\ref{SummaryStats}). The parameters used to generate this catalog are listed in Table \ref{tab:param_fits}.

Overall, the maximum AMD model performs very well for reproducing the observed \Kepler{} data in terms of these marginal distributions. The marginal distributions of observables from the best--fitting observed catalogs generated from this model are almost indistinguishable by eye from those generated from the two--Rayleigh model. In Appendix Figures \ref{fig:dists1_KS} \& \ref{fig:dists1_AD}, we show histograms of the individual weighted distance terms for each of our summary statistics, for KS and AD versions of $\mathcal{D}_{W,1}$, respectively. While we are able to choose a smaller distance threshold (for both KS and AD) for the new model compared to the old model to achieve a similar efficiency in the rate of accepted points, this is at least partially due to the fact that the old model involves more free parameters. The best distances achieved are similar for the two models. Thus, while it is unclear if the maximum AMD model provides a significantly better fit to the \Kepler{} data as the two--Rayleigh model, it is at least as good of a description.

While the two-Rayleigh model provides a slightly better fit to the observed multiplicity, period, and period ratio distributions, both models reproduce the observed distributions well.
In contrast, the maximum AMD model performs better for the transit duration (bottom left panel in Figures \ref{fig:dists1_KS} \& \ref{fig:dists1_AD}) and slightly better for the period--normalized transit duration ratio distributions (for both near--resonant and non--resonant pairs, but only in AD distances; bottom middle and right panels in Figure \ref{fig:dists1_AD}).
Indeed, the better agreement for the transit duration and transit duration ratio distributions was one of the motivations for developing the maximum AMD model.
The main difference between the two models is how the eccentricities and mutual inclinations are drawn, and these most directly affect the observed distributions for the transit durations and duration ratios. 
Interestingly, there is also a noticeable improvement to the transit depth distribution but a worse fit to the transit depth ratio distribution (especially in AD distances) for the maximum AMD model, which was not anticipated.

While both models fit the \Kepler{} catalog near equally well, the maximum AMD model is appealing for several reasons, as previously motivated in \S\ref{secOldModel}. The main advantage is that it incorporates a more sophisticated criteria for long-term orbital stability. The two--Rayleigh model shows a strong preference for a high mutual inclination population \citepalias{HFR2019}, characterized by a Rayleigh scale of $\sigma_{i,\rm high} \sim 45^\circ$. This results in some planets with extremely high orbital inclinations (including retrograde, $i_m \gtrsim 90^\circ$).
Secular interactions between highly--inclined planets within a system are very likely to lead to orbital instabilities and planets colliding or being ejected from the system.
The maximum AMD model produces systems that are AMD-stable by design, 
so secular interactions in the resulting systems are unlikely to result in close encounters, making it a more physically reasonable model. 
A second reason to prefer the maximum AMD model is that it uses several fewer free parameters, yet it can explain the observed data equally well. As described in \S\ref{Max_AMD_model}, the new model removes several parameters we previously used to characterize the distribution of eccentricities ($\sigma_e$; which was replaced by a parameter for the eccentricity scale of single planets, $\sigma_{e,1}$) and the distribution of mutual inclinations ($\sigma_{i,\rm low}$, $\sigma_{i,\rm high}$, and $f_{\sigma_{i,\rm high}}$).
Finally, in \S\ref{Results_ecc_incl}, we will show additional features in the \Kepler{} data that match the predictions of the maximum AMD model resulting from the improved eccentricity and inclination distributions.

\begin{figure*}
\centering
\includegraphics[scale=0.29,trim={0 0 0 0},clip]{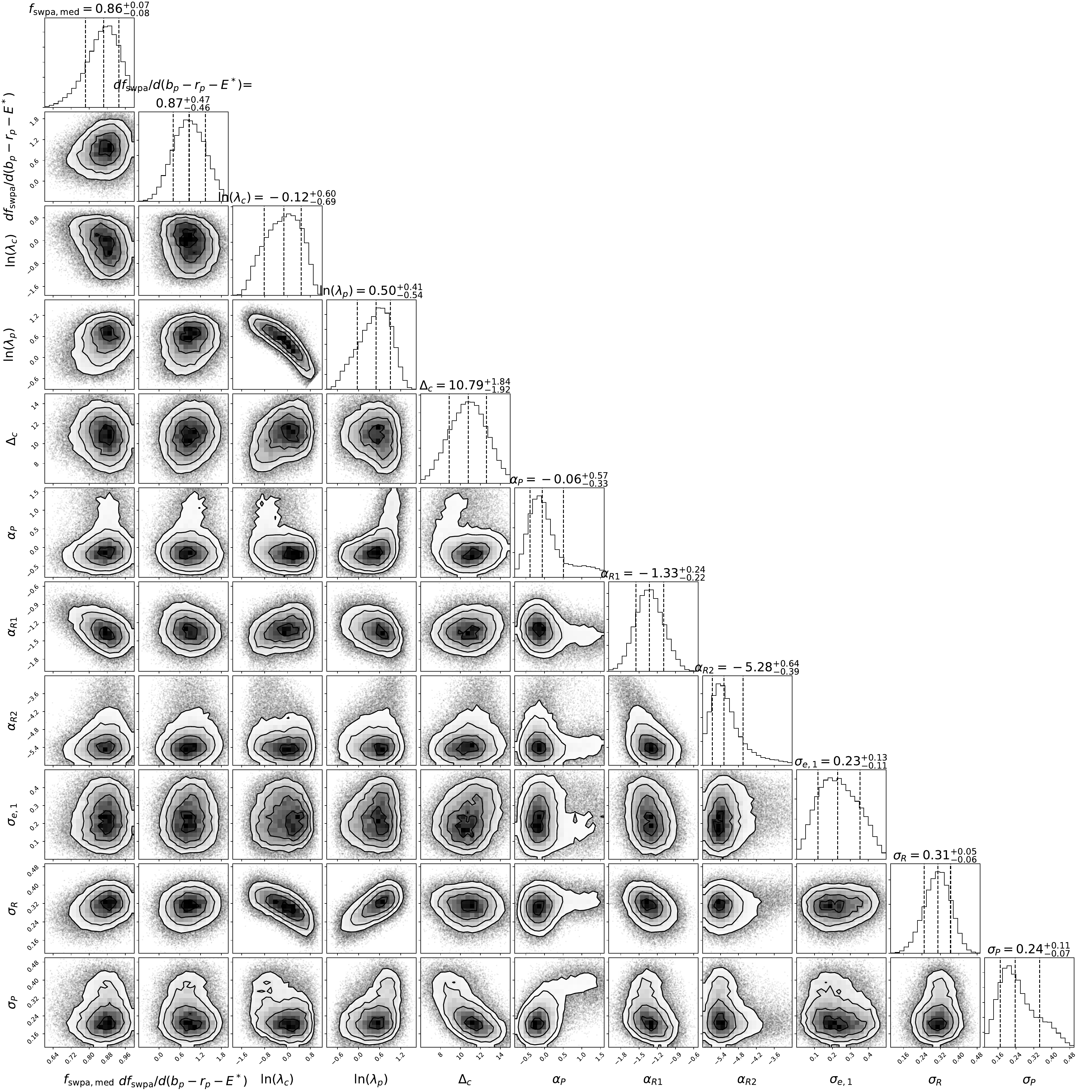}
\caption{ABC posterior distributions of the free model parameters for the maximum AMD model, using our $\mathcal{D}_{W,3}$ distance function (with KS distance terms). A total of $5 \times 10^4$ points passing a distance threshold (65) evaluated using the GP emulator are plotted. The 68.3\% credible intervals (as listed for each parameter) are also presented in Table \ref{tab:param_fits}, and analogous figures resulting from using the other distance functions are shown in the Appendix.}
\label{fig:d3_KS_corner}
\end{figure*}

\subsection{The distribution of planetary systems and their architectures} \label{Results_params}

In this section, we re--examine the constraints on the remaining free parameters of our new model, which retain their interpretations, as well as correlations between the parameters. Since our new model involves both AMD stability and mutual Hill stability, we allowed the $\Delta_c$ parameter (minimum spacing in mutual Hill radii) to vary, which had been kept fixed at $\Delta_c = 8$ in previous papers.

\begin{deluxetable*}{lccccccccc}
\centering
\tablecaption{Best-fitting values for the free parameters of each clustered model.}
\tablehead{
 \colhead{Parameter} & \multicolumn2c{Two--Rayleigh model} & \multicolumn7c{Maximum AMD model} \\
 & \colhead{$\mathcal{D}_{W,1}$ (KS)} & \colhead{$\mathcal{D}_{W,1}$ (AD)} & \colhead{Fig. \ref{fig:model_split_bprp}} & \colhead{$\mathcal{D}_{W,1}$ (KS)} & \colhead{$\mathcal{D}_{W,1}$ (AD)} & \colhead{$\mathcal{D}_{W,2}$ (KS)} & \colhead{$\mathcal{D}_{W,2}$ (AD)} & \colhead{$\mathcal{D}_{W,3}$ (KS)} & \colhead{$\mathcal{D}_{W,3}$ (AD)}
}
\decimalcolnumbers
\startdata
 $f_{\sigma_{i,\rm high}}$                 & $0.43_{-0.09}^{+0.09}$ & $0.44_{-0.09}^{+0.10}$ & - & - & - & - & - & - & - \\[5pt]
 $f_{\rm swpa,med}$                        & $0.60_{-0.12}^{+0.13}$ & $0.57_{-0.11}^{+0.12}$ & 0.88 & $0.87_{-0.08}^{+0.07}$ & $0.87_{-0.08}^{+0.07}$ & $0.88_{-0.07}^{+0.06}$ & $0.90_{-0.07}^{+0.06}$ & $0.86_{-0.08}^{+0.07}$ & $0.89_{-0.08}^{+0.06}$ \\[5pt]
 $\frac{df_{\rm swpa}}{d(b_p-r_p-E^*)}$ & $0.84_{-0.35}^{+0.37}$ & $1.15_{-0.36}^{+0.35}$ & 0.9 & $1.02_{-0.47}^{+0.45}$ & $0.80_{-0.50}^{+0.50}$ & $0.90_{-0.41}^{+0.39}$ & $0.66_{-0.59}^{+0.56}$ & $0.87_{-0.46}^{+0.47}$ & $0.67_{-0.72}^{+0.61}$ \\[5pt]
 $\ln{(\lambda_c)}$                           & $0.18_{-0.73}^{+0.80}$ & $0.99_{-0.84}^{+0.60}$ & 0 & $-0.30_{-0.65}^{+0.72}$ & $0.09_{-0.88}^{+0.53}$ & $-0.33_{-0.66}^{+0.73}$ & $-0.11_{-0.64}^{+0.62}$ & $-0.12_{-0.69}^{+0.60}$ & $0.06_{-0.71}^{+0.56}$ \\[5pt]
 $\lambda_c$                                   & $1.20_{-0.62}^{+1.46}$ & $2.68_{-1.52}^{+2.23}$ & 1 & $0.74_{-0.36}^{+0.79}$ & $1.09_{-0.64}^{+0.76}$ & $0.72_{-0.35}^{+0.77}$ & $0.90_{-0.42}^{+0.77}$ & $0.88_{-0.44}^{+0.73}$ & $1.06_{-0.54}^{+0.80}$ \\[5pt]
 $\ln{(\lambda_p)}$                          & $1.17_{-0.40}^{+0.36}$ & $0.77_{-0.55}^{+0.54}$ & 0.47 & $0.48_{-0.55}^{+0.43}$ & $0.31_{-0.51}^{+0.55}$ & $0.50_{-0.56}^{+0.41}$ & $0.50_{-0.55}^{+0.41}$ & $0.50_{-0.54}^{+0.41}$ & $0.54_{-0.61}^{+0.43}$ \\[5pt]
 $\lambda_p$                                   & $3.22_{-1.05}^{+1.41}$ & $2.15_{-0.91}^{+1.55}$ & 1.6 & $1.62_{-0.68}^{+0.86}$ & $1.37_{-0.55}^{+1.01}$ & $1.65_{-0.71}^{+0.85}$ & $1.65_{-0.70}^{+0.83}$ & $1.65_{-0.68}^{+0.84}$ & $1.72_{-0.78}^{+0.91}$ \\[5pt]
 $\Delta_c$                                       & 8 (fixed)                        & 8 (fixed)                          & 10 & $9.36_{-1.68}^{+1.85}$ & $9.28_{-1.19}^{+1.24}$ & $9.23_{-1.56}^{+1.73}$ & $9.38_{-1.20}^{+1.45}$ & $10.79_{-1.92}^{+1.84}$ & $11.27_{-1.35}^{+1.34}$ \\[5pt]
 $\alpha_P$                                      & $0.64_{-0.58}^{+0.56}$ & $0.81_{-0.44}^{+0.43}$ & 0 & $-0.12_{-0.32}^{+0.60}$ & $-0.15_{-0.26}^{+0.29}$ & $-0.07_{-0.32}^{+0.91}$ & $-0.12_{-0.28}^{+0.35}$ & $-0.06_{-0.33}^{+0.57}$ & $-0.05_{-0.25}^{+0.27}$ \\[5pt]
 $\alpha_{R1}$                                 & $-1.35_{-0.36}^{+0.35}$ & $-1.48_{-0.29}^{+0.29}$ & $-1.4$ & $-1.34_{-0.21}^{+0.21}$ & $-1.46_{-0.16}^{+0.16}$ & $-1.29_{-0.19}^{+0.19}$ & $-1.45_{-0.17}^{+0.18}$ & $-1.33_{-0.22}^{+0.24}$ & $-1.43_{-0.19}^{+0.21}$ \\[5pt]
 $\alpha_{R2}$                                 & $-4.69_{-0.67}^{+0.86}$ & $-4.92_{-0.56}^{+0.62}$ & $-5.2$ & $-5.24_{-0.43}^{+0.66}$ & $-5.43_{-0.33}^{+0.42}$ & $-5.31_{-0.37}^{+0.48}$ & $-5.41_{-0.34}^{+0.47}$ & $-5.28_{-0.39}^{+0.64}$ & $-5.28_{-0.40}^{+0.70}$ \\[5pt]
 $\sigma_{e,1}$*                               & $0.022_{-0.008}^{+0.009}$ & $0.016_{-0.008}^{+0.008}$ & 0.25 & $0.27_{-0.14}^{+0.13}$ & $0.30_{-0.15}^{+0.12}$ & $0.20_{-0.11}^{+0.16}$ & $0.25_{-0.14}^{+0.14}$ & $0.23_{-0.11}^{+0.13}$ & $0.30_{-0.16}^{+0.11}$ \\[5pt]
 $\sigma_{i,\rm high}$ ($^\circ$)      & $46_{-18}^{+18}$ & $48_{-18}^{+17}$ & - & - & - & - & - & - & - \\[5pt]
 $\sigma_{i,\rm low}$ ($^\circ$)       & $1.14_{-0.32}^{+0.33}$ & $1.24_{-0.33}^{+0.37}$ & - & - & - & - & - & - & - \\[5pt]
 $\sigma_R$                                    & $0.33_{-0.06}^{+0.06}$ & $0.32_{-0.08}^{+0.07}$ & 0.3 & $0.28_{-0.08}^{+0.08}$ & $0.31_{-0.07}^{+0.07}$ & $0.29_{-0.07}^{+0.07}$ & $0.33_{-0.07}^{+0.06}$ & $0.31_{-0.06}^{+0.05}$ & $0.34_{-0.06}^{+0.05}$ \\[5pt]
 $\sigma_P$                                    & $0.20_{-0.03}^{+0.03}$ & $0.18_{-0.04}^{+0.04}$ & 0.25 & $0.28_{-0.09}^{+0.12}$ & $0.22_{-0.05}^{+0.06}$ & $0.26_{-0.08}^{+0.12}$ & $0.21_{-0.05}^{+0.07}$ & $0.24_{-0.07}^{+0.11}$ & $0.17_{-0.04}^{+0.05}$ \\
\enddata
\tablecomments{While we trained the emulator on the transformed parameters $\ln(\lambda_c \lambda_p)$ and $\ln(\lambda_p/\lambda_c)$, we transform back to $\ln(\lambda_c)$ and $\ln(\lambda_p)$ for reporting the credible intervals. Unlogged rates $\lambda_c$ and $\lambda_p$ are shown for interpretability, and are equivalent to the rows with log-values.}
\tablenotetext{*}{In the two--Rayleigh model, this is the eccentricity (Rayleigh) scale for all planets.}
\label{tab:param_fits}
\end{deluxetable*}

Table \ref{tab:param_fits} shows the 68.3\% credible regions for the best-fitting values of each free parameter in our new model, derived from the ABC posterior distributions using each of the distance functions defined in \S\ref{Distance_functions}. 
We show the same credible regions as a ``corner plot'' \citep{Fm2016} in Figure \ref{fig:d3_KS_corner} for our analysis using $\mathcal{D}_{W,3}$ (KS terms), which takes into account all the marginal distributions of the \Kepler{} observables, as well as the new metrics from \citet{GF2020}. The ABC posteriors from the other distance functions (KS and AD terms) are shown in Figures \ref{fig:d1_KS_corner}-\ref{fig:d3_AD_corner}.

\subsubsection{Fraction of stars with planets \\
($f_{\rm swpa,med}$, $df_{\rm swpa}/d(b_p-r_p-E^*)$)} \label{secFracStarsWPlanets}

The fraction of solar-type (G2V) dwarfs hosting at least one planet between 3 and 300 days is $f_{\rm swpa}(0.823) = 0.86_{-0.06}^{+0.08}$ in the maximum AMD model, even higher than in our two--Rayleigh model. 
The overall increase in $f_{\rm swpa}$ is likely due to a change in the intrinsic multiplicity distribution, which we discuss in \S\ref{sec:lambda_candp}.

We find a trend of increasing $f_{\rm swpa}$ toward later type dwarfs (higher $b_p-r_p-E^*$). The maximum AMD model suggests that the fraction of stars with planets for the hottest stars in our sample (mid-F dwarfs) is $f_{\rm swpa}(0.5) = 0.59_{-0.15}^{+0.14}$ and increases sharply toward $\sim 1$ by early-K dwarfs, sooner than in the two--Rayleigh model.
This trend is consistent across all our distance functions. Using the same distance function $\mathcal{D}_{W,1}$ (KS), we find that $f_{\rm swpa,med} = 0.87_{-0.08}^{+0.07}$ and $df_{\rm swpa}/d(b_p-r_p-E^*) = 1.02_{-0.47}^{+0.45}$. Similar values for $f_{\rm swpa,med}$ are found for the other distance functions. The positive slope is similar in both the maximum AMD and the two--Rayleigh models.\footnote{Interestingly, the slope is poorly constrained in two of the distance functions involving AD terms.  This can be explained by the fact that most weights for the AD terms are significantly larger than those for the KS terms (Table \ref{tab:weights}), as the AD distance is more sensitive to deviations from a perfect model. Since the dependence of occurrence rates on color is only constrained by the difference in observed multiplicities for the bluer and redder stars and the relevant distance terms ($D_f$ and $D_{\rm mult}$) do not involve a KS or AD distance, the higher weights for the AD terms effectively result in less influence for $D_f$ and $D_{\rm mult}$.}

The overall increase in the fraction of stars with planets toward later stellar types is generally in agreement with previous studies \citep{H2012, DC2013, Mu2015, YXZ2020}. However, these studies also find that planet occurrence around M dwarfs is higher than that of K dwarfs, whereas $f_{\rm swpa}$ already rises to 100\% by early-K dwarfs in our maximum AMD model (and we do not include M dwarfs in this study). There are several explanations for this difference. First, while we allow for the fraction of stars with planets ($f_{\rm swpa}$) to vary as a function of color, we do not directly test for possible differences in the mean number of planets per system (i.e. $\lambda_c$ and $\lambda_p$), which could also be higher for M dwarf systems. Many of the aforementioned studies only considered the planet occurrence rate as a function of spectral type, but do not distinguish between planet occurrence and the planetary system occurrence (the exception is \citealt{YXZ2020}). These studies also used larger samples of \Kepler{} target stars in their analyses, whereas we selected a cleaner sample of FGK dwarfs filtering out stellar binaries; the inclusion of such stars may drive down the inferred planet occurrence rate. Finally, our parameter $f_{\rm swpa}$ is limited to a linear function of $b_p-r_p-E^*$ (bounded between 0 and 1; equation \ref{eq_fswp_bprp}); given the large slope and early plateau at unity, a more flexible model is necessary to describe any differences in occurrence at later (e.g. K and M) spectral types.

\begin{figure}
\centering
 \includegraphics[scale=0.425,trim={0 0.4cm 0 0.2cm},clip]{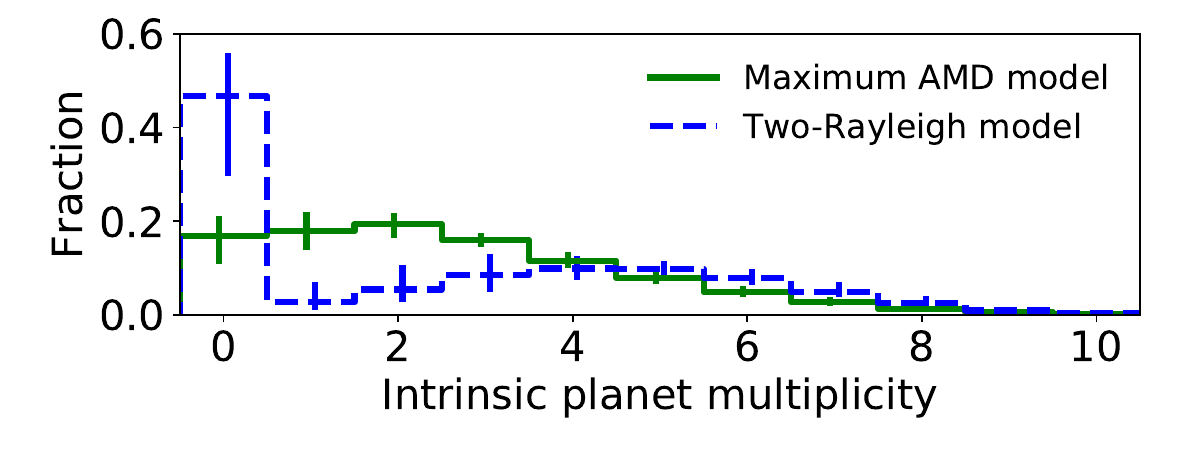}
 \includegraphics[scale=0.425,trim={0 0.4cm 0 0.2cm},clip]{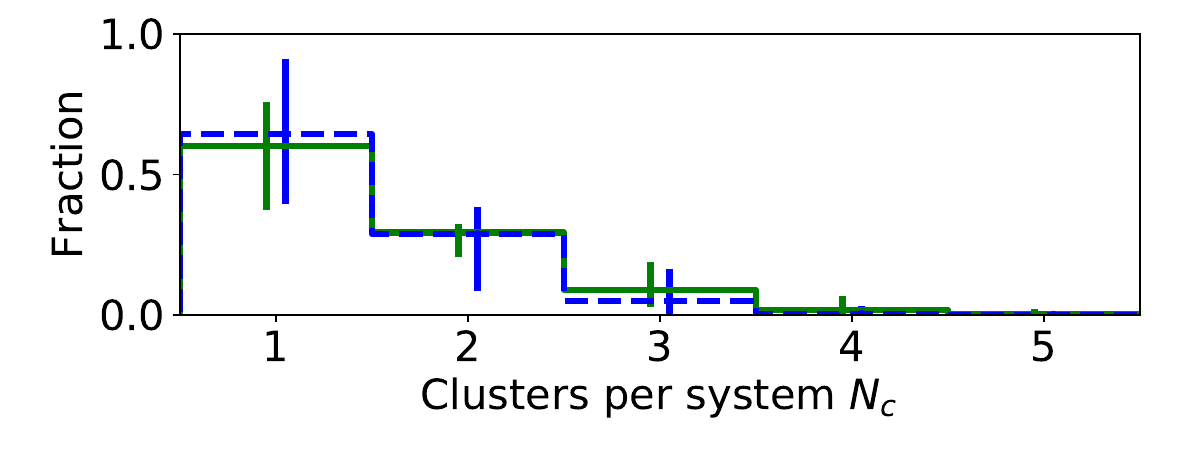}
 \includegraphics[scale=0.425,trim={0 0.4cm 0 0.2cm},clip]{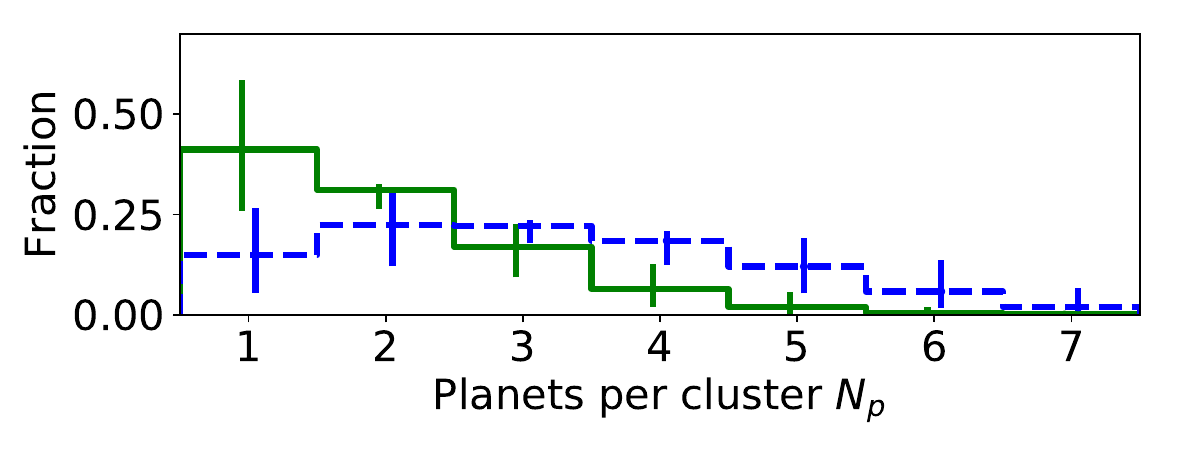}
\caption{Distributions of intrinsic planet multiplicity (\textbf{top panel}), cluster multiplicity $N_c$ (\textbf{middle panel}), and planets per cluster $N_p$ (\textbf{bottom panel}) for our models between 3 and 300 days. In each panel, the dashed blue histogram denotes the two--Rayleigh model, while the solid green histogram denotes our new maximum AMD model. In both cases, the error bars denote the 68.3\% credible region in each bin computed from 100 catalogs passing the (KS) distance threshold for each model.}
\label{fig:intrinsic_mults}
\end{figure}

\subsubsection{Numbers of clusters and planets per cluster ($\lambda_c$, $\lambda_p$)}
\label{sec:lambda_candp}

We find that $\lambda_c = 0.74_{-0.36}^{+0.79}$ and $\lambda_p = 1.62_{-0.68}^{+0.86}$ using $\mathcal{D}_{W,1}$ (KS); similar values are found for the other distance functions, although the uncertainties are large in all cases. 
These are lower in our maximum AMD model compared to those in the two--Rayleigh model, despite their identical parameterizations. 
While these parameters represent the mean numbers of attempted clusters per star and attempted planets per cluster, respectively, the rejection-sampling means that the true mean values for the number of clusters for star and planets per cluster could differ from $\lambda_c$ and $\lambda_p$.
In Figure \ref{fig:intrinsic_mults}, we plot the posterior predictive distributions of intrinsic planet multiplicity, cluster multiplicity $N_c$, and planets per cluster $N_p$ (all between 3 and 300 days), for our two-Rayleigh (blue) and maximum AMD (green) models.
For the maximum AMD model, the mean number of planets (in this period range with $R_p > 0.5 R_\oplus$) per star is $2.61_{-0.32}^{+0.32}$, and the mean number of such planets per planetary system (i.e. star with at least one planet) is $3.12_{-0.28}^{+0.36}$.
While the distribution for the number of clusters per system is very similar between the two models, the fraction of clusters with a single planet increases significantly compared to the previous model.

As a result, the overall intrinsic multiplicity distribution is very different. The numbers of true single, double, and triple planet systems are significantly higher in this model than in our two-Rayleigh model. The occurrence of higher multiplicity ($n \geq 5$) systems declines even more quickly. 
This result can be understood by considering our results for the overall fraction of stars with planets as previously discussed in \S\ref{secFracStarsWPlanets}. In order to produce the same overall number of observed planets, the higher $f_{\rm swpa}$ implies that each planetary system should have slightly fewer total planets. However, this is complicated by the detection biases that also depend on other architectural properties of the systems, especially the mutual inclinations of the planets. In particular, the mutual inclination distribution provides an additional constraint on the intrinsic multiplicity distribution in this model, since it is derived from the critical AMD of each system which is a function of the number of planets, as we will show in \S\ref{Results_ecc_incl}.

\begin{figure*}
\centering
\begin{tabular}{cc}
 \includegraphics[scale=0.425,trim={0 0.4cm 0 0.2cm},clip]{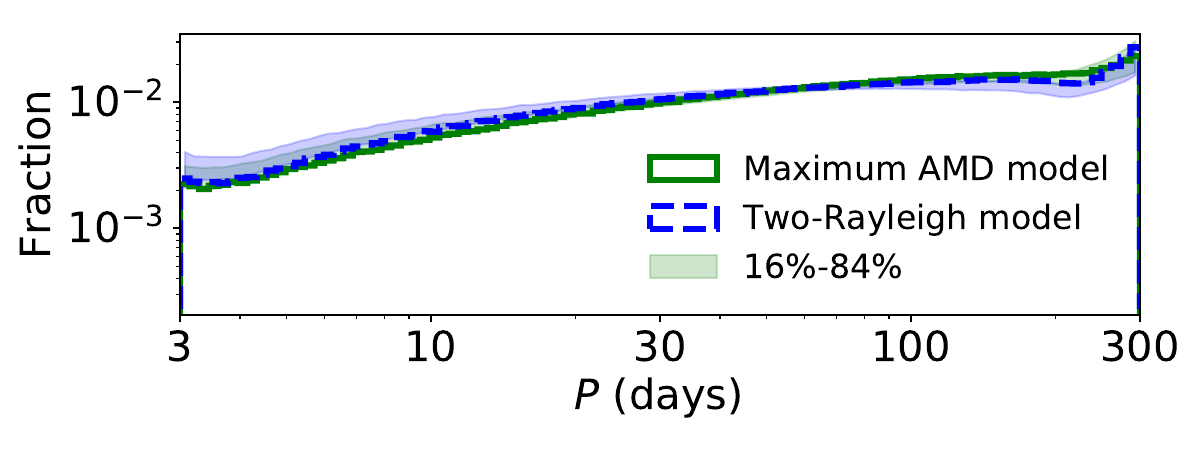} &
 \includegraphics[scale=0.425,trim={0 0.4cm 0 0.2cm},clip]{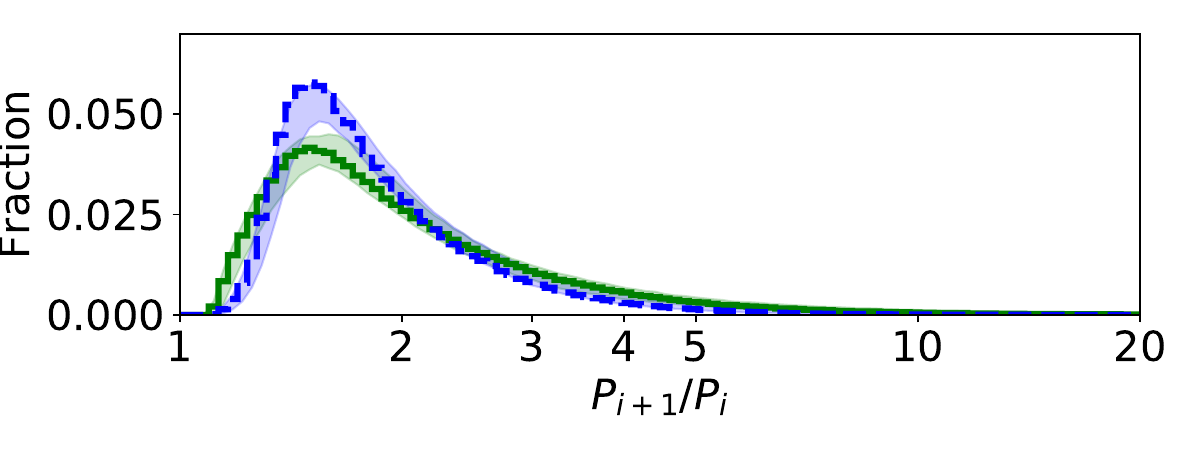} \\
 \includegraphics[scale=0.425,trim={0 0.4cm 0 0.2cm},clip]{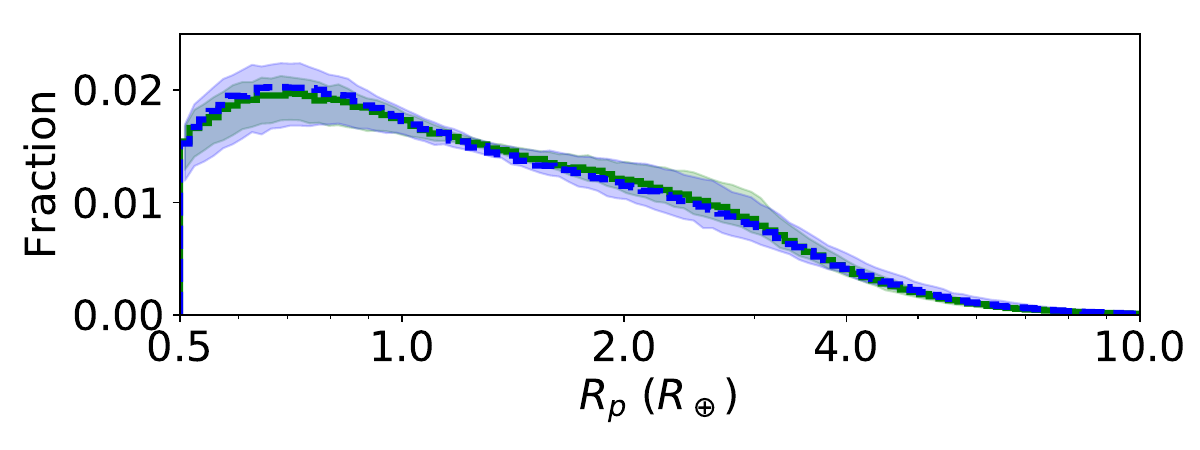} &
 \includegraphics[scale=0.425,trim={0 0.4cm 0 0.2cm},clip]{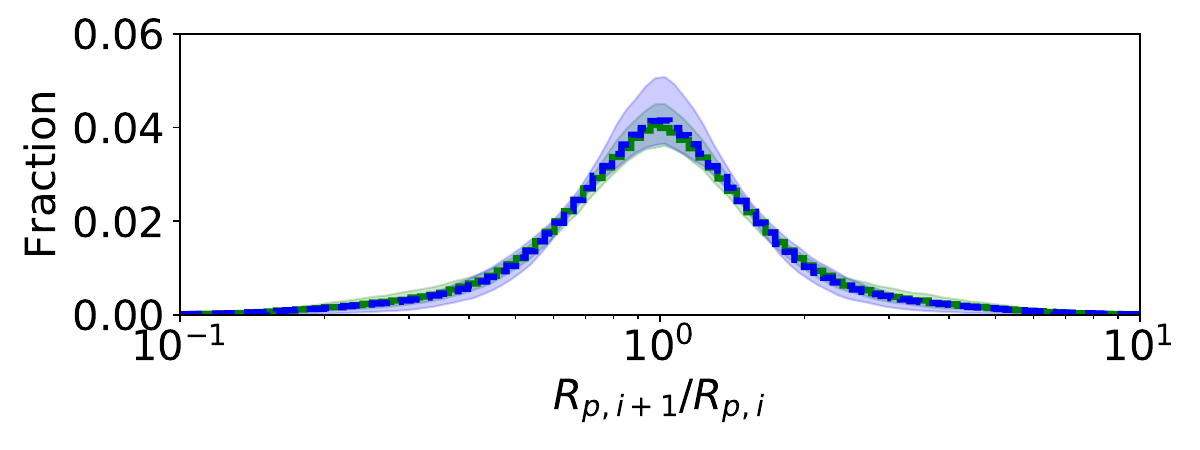} \\
 \includegraphics[scale=0.425,trim={0 0.4cm 0 0.2cm},clip]{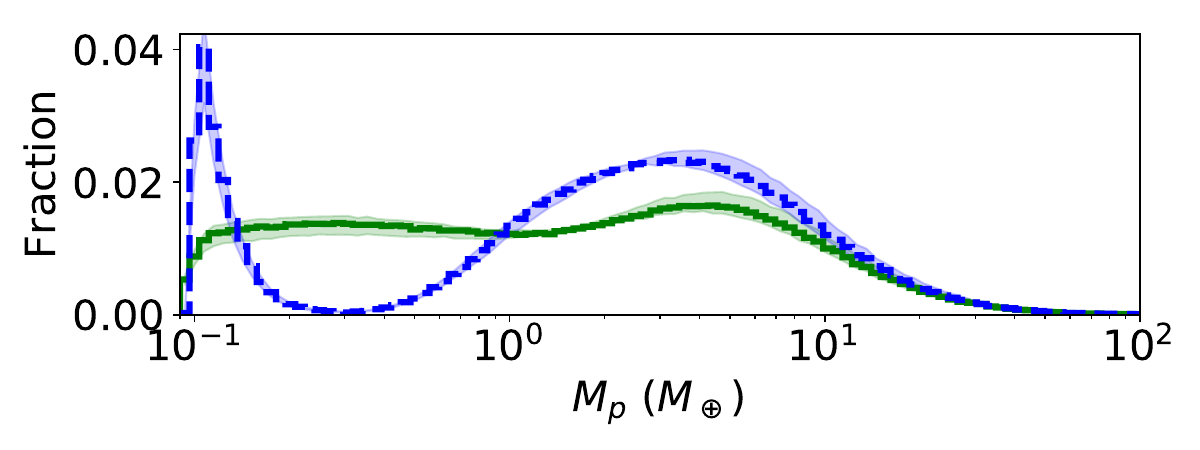} &
 \includegraphics[scale=0.425,trim={0 0.4cm 0 0.2cm},clip]{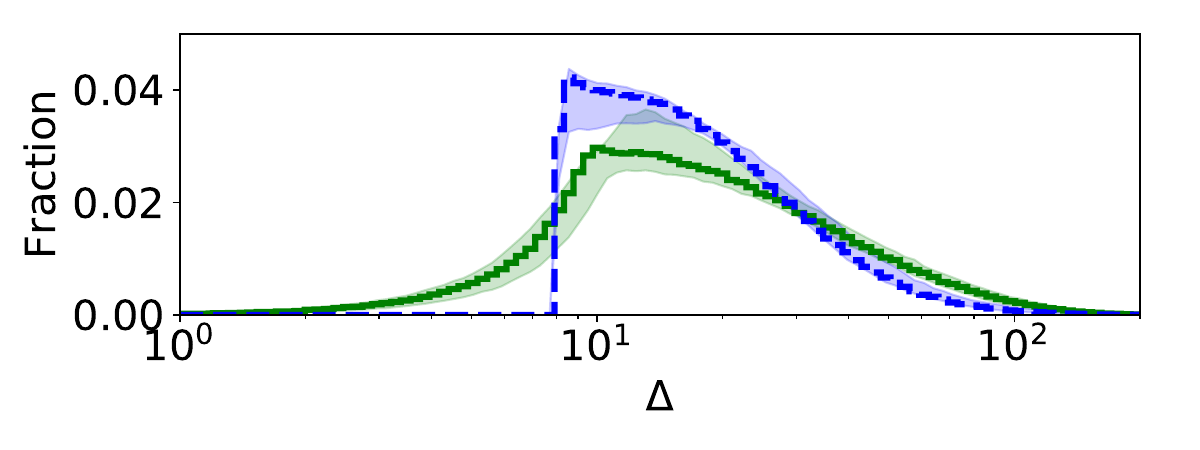} \\
 \includegraphics[scale=0.425,trim={0 0.4cm 0 0.2cm},clip]{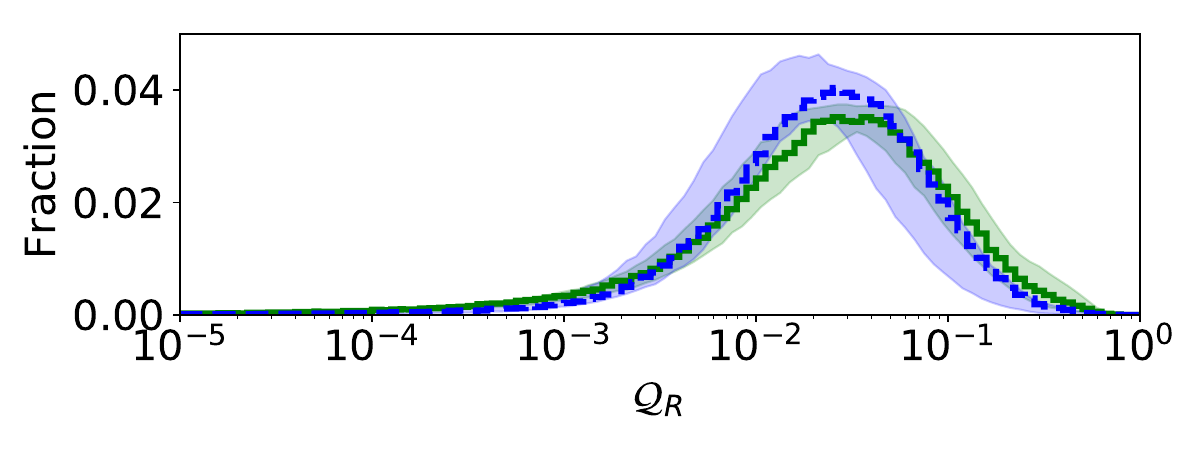} &
 \includegraphics[scale=0.425,trim={0 0.4cm 0 0.2cm},clip]{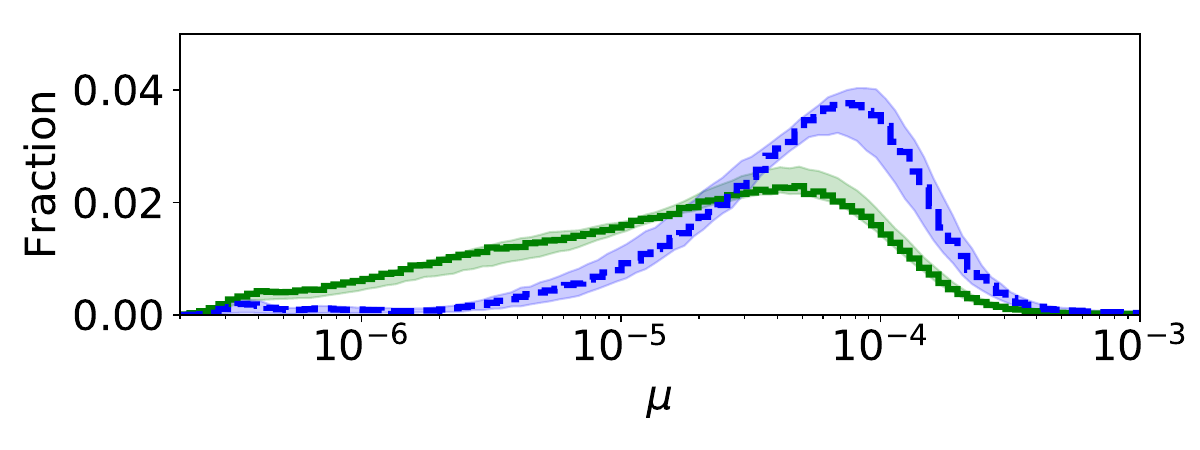} \\
 \includegraphics[scale=0.425,trim={0 0.4cm 0 0.2cm},clip]{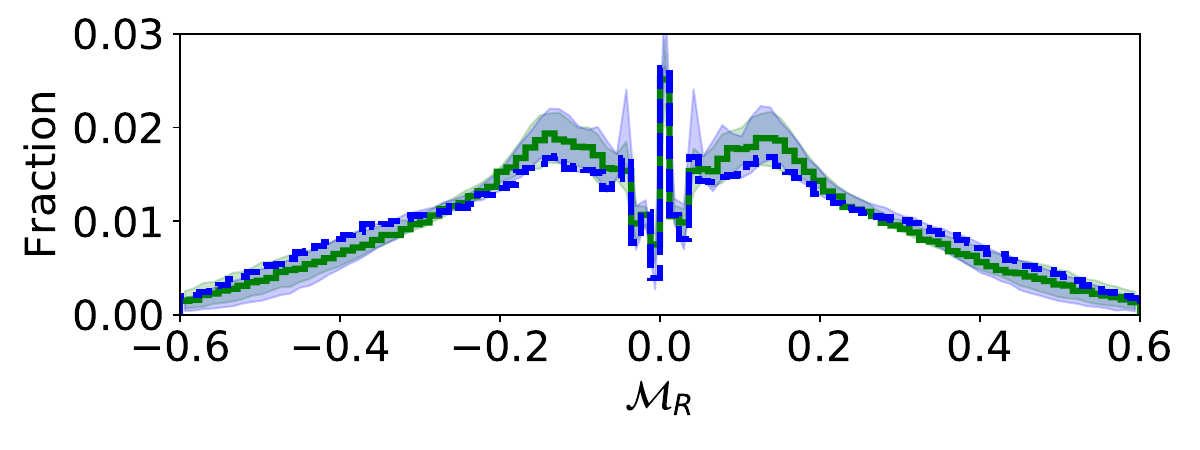} &
 \includegraphics[scale=0.425,trim={0 0.4cm 0 0.2cm},clip]{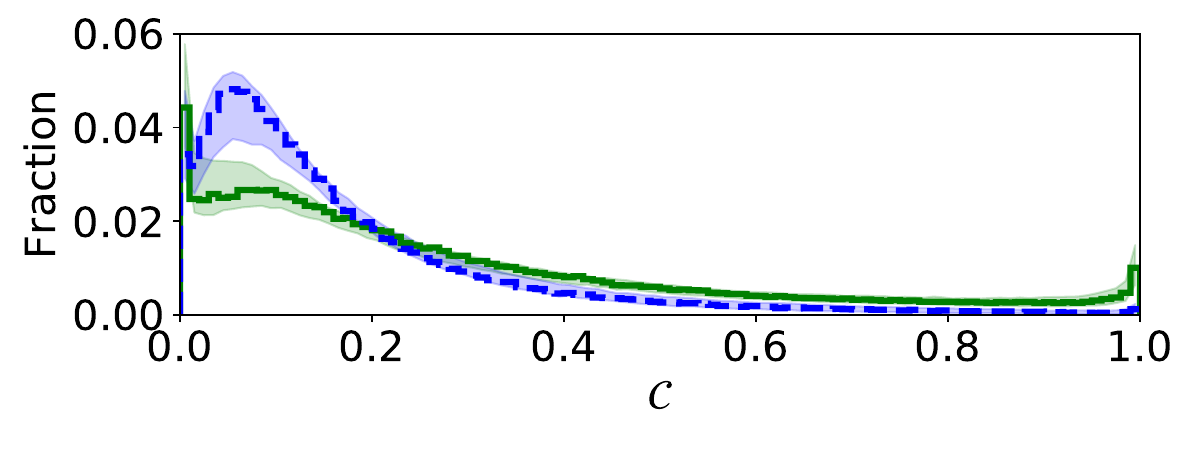} \\
\end{tabular}
\caption{Marginal distributions of intrinsic properties for the \textit{physical catalogs} drawn from our models. In each panel, the dashed blue histogram shows one simulated catalog from the two--Rayleigh model while the solid green histogram shows a simulated catalog from our maximum AMD model (with parameter values listed in Table \ref{tab:param_fits}); shaded regions denote the 68.3\% credible interval around the median in each bin, computed from 100 catalogs passing our (KS) distance threshold for each model. Note that our new mass-radius relationship for small planets creates a much more plausible mass distribution for small planets than the NWG18 relation used in the two--Rayleigh model as discussed in Section \ref{secMR}. The four bottom-most panels are additional system-level metrics inspired by \citet{GF2020}: dynamical mass ($\mu = \sum_k{M_{p,k}/M_\star}$), radii partitioning ($\mathcal{Q}_R$), radii monotonicity ($\mathcal{M}_R$), and gap complexity ($\mathcal{C}$). Note that these panels show the intrinsic distributions of these metrics; the observed distributions are shown in Figure \ref{fig:model_obs_GF2020} (for the maximum AMD model only).}
\label{fig:models_underlying}
\end{figure*}

\subsubsection{Minimum spacing ($\Delta_c$)}
\label{secDeltaC}

The $\Delta_c$ parameter denotes the minimum spacing in mutual Hill radii for any pair of planets. 
We find that $\Delta_c \simeq 9$ for distance functions $\mathcal{D}_{W,1}$ and $\mathcal{D}_{W,2}$ (with both KS and AD terms). This is very similar to the value we set in \citetalias{HFR2019}, $\Delta_c = 8$. The minimum spacing parameter is somewhat higher for the distance function involving the new terms from \citet{GF2020}, $\mathcal{D}_{W,3}$: $\Delta_c = 10.79_{-1.92}^{+1.84}$ and $\Delta_c = 11.27_{-1.35}^{+1.34}$ using KS and AD analyses, respectively. 
This is likely caused by the \textit{gap complexity} term, as discussed in more detail in \S\ref{Peas_in_a_pod}.

We note that there is a subtle difference in the interpretation of $\Delta_c$ between our new model and the previous models.
Previously, the stability criterion for each planet pair was based on the ratio of the periastron distance of the outer planet over the apastron distance of the inner planet. In the new model, the stability criterion is based only on the ratio of semi-major axes.\footnote{In the two--Rayleigh model, we test the mutual Hill stability criteria (equation \ref{eq_Nhill}) and sample the periods of the planets after their eccentricities have been drawn. In our maximum AMD model, the order is reversed, since the eccentricities (and mutual inclinations) are set by the AMD budget resulting from the critical AMD, which can only be computed after the semi-major axes are set. Thus, we first set the periods of the planets by requiring all adjacent planet pairs to be separated by a minimum $\Delta_c$ for circular orbits (i.e. equation \ref{eq_Nhill} with $e_{\rm in} = e_{\rm out} = 0$), before distributing the AMD amongst their orbits.}
Therefore, we expect the new model to prefer a slightly larger $\Delta_c$ than the two--Rayleigh model.

In Figure \ref{fig:models_underlying}, we show the distributions of a number of physical properties and system metrics, including $\Delta$, for both our two-Rayleigh model (blue) and our maximum AMD model (green). The solid green and dashed blue lines show one simulated catalog (with parameter values listed in Table \ref{tab:param_fits} for the maximum AMD model), while the shaded regions denote the 68.3\% credible regions over many models drawn from the ABC posteriors. The distribution of $\Delta$ for the two-Rayleigh model exhibits a sharp cut-off at $\Delta_c = 8$ by construction. For the maximum AMD model, the distribution exhibits a tail toward smaller separations due to the eccentricities being drawn after the periods have been set. While planets with very small separations (e.g. $\Delta \lesssim 3.46$; \citealt{G1993}) are almost certainly unstable, the eccentricity--induced tail falls rapidly at this point and only affects a small fraction of the planets.

\subsubsection{Period distribution ($\alpha_P$)}

We find that $\alpha_P$ is consistent with zero for all distance functions considered (a flat distribution in log-period corresponds to a power-law index of $-1$). While this is a slightly shallower slope than what we found for the two-Rayleigh model, the period distribution (top left panel Figure \ref{fig:models_underlying}) is very similar.
 
\subsubsection{Radius distribution ($\alpha_{R1}$, $\alpha_{R2}$)}

As in \citetalias{HFR2019}, we assume a broken power-law with clustered radii for the radius distribution, where the break radius is fixed at $R_{p,\rm break} = 3 R_\oplus$. We find similar results with our previous clustered models for both the power-law indices below and above the break: $\alpha_{R1} = -1.34_{-0.21}^{+0.21}$ and $\alpha_{R2} = -5.24_{-0.43}^{+0.66}$, respectively, using $\mathcal{D}_{W,1}$ (KS). These results are consistent across all our distance functions.

\subsubsection{Mass distribution (M-R relation)}

We adopt a new M-R relation for the maximum AMD model as described in \S\ref{secMR}, consisting of the NWG18 relation and a lognormal distribution around the Earth--like rocky model from \citet{Zeng2019}, above and below $R_p = 1.472 R_\oplus$, respectively. While the intrinsic planet radius distribution remains the same, the resulting planet mass distribution is very different, as shown in the middle--left panel of Figure \ref{fig:models_underlying}. Instead of the strong bimodal distribution of (resulting from solely using the NWG18 M-R relation), the new distribution is smooth and relatively flat below $\sim 2 M_\oplus$.

\subsubsection{Period and radius clustering ($\sigma_P$, $\sigma_R$)}

We quantify the degree of period clustering with $\sigma_P$ (the width in log-period of each cluster, per planet in the cluster; equation \ref{eq_P_unscaled}) and the degree of planet radius clustering with $\sigma_R$ (the width in log-radius for each cluster, regardless of the number of planets; equation \ref{eq_Rp}). 
Smaller values indicate more significant intra--cluster correlations in periods and in planet sizes, respectively. 
The value of $\sigma_R$ is consistently around $\sim 0.3$ across both models and all distance functions considered.
In our maximum AMD model, we find some variation in $\sigma_P$ across different distance functions; the value of $\sigma_P = 0.28_{-0.09}^{+0.12}$ using $\mathcal{D}_{W,1}$ is somewhat greater than in our two-Rayleigh model (although the uncertainties are also larger), while other distance functions give somewhat lower values. There is an (anti) correlation between $\sigma_P$ and $\Delta_c$ (Figure \ref{fig:d3_KS_corner}): we interpret this inverse correlation as a balance to match the observed period ratio distribution, as both of these parameters most directly affect the underlying period ratio distribution. 

\subsubsection{System--level metrics from \citet{GF2020}} \label{secIntrinsicGF2020}

We compute and plot the distributions of the system--level statistics inspired by \citet{GF2020} for our \textit{physical catalogs} in Figure \ref{fig:models_underlying} (bottom four panels). 
The \textit{radius partitioning} ($\mathcal{Q}_R$), \textit{radius monotonicity} ($\mathcal{M}_R$), and \textit{gap complexity} ($\mathcal{C}$) are defined in \S\ref{SummaryStats} (modified such that all planets in the system are included, instead of just the observed planets). We also include the \textit{dynamical mass} ($\mu$) from \citet{GF2020} (equation 6 therein), which is simply the sum of the planet masses $M_p$ divided by the stellar mass $M_\star$:
\begin{equation}
 \mu \equiv \sum_{k=1}^{n}{M_p/M_\star}. \label{eq_dynamical_mass}
\end{equation}

The $\mathcal{Q}_R$ distribution is similar in both models and peaks around $\sim 0.03$, highlighting the similarity in planet sizes within each system arising from the clustered radii (identically sized planets would yield $\mathcal{Q}_R = 0$). 
The $\mu$ distribution is broader and shifted to lower values for the maximum AMD model; this difference is due to a combination of the shift in the intrinsic multiplicity distribution toward smaller counts and the revised M-R relation compared to the two--Rayleigh model.
The distribution of $\mathcal{M}_R$ is symmetric because we have not introduced any correlation between planet size and period in either model, but it exhibits a peculiar shape. The sharp peak at zero monotonicity and dips on each side are due to the behaviour of the Spearman correlation coefficient at small multiplicities: while three planet systems can never result in $\mathcal{M}_R = 0$ for any ordering, four and five planet systems result in $\mathcal{M}_R = 0$ especially often from random ordering alone. 
Finally, the $\mathcal{C}$ distribution is highly weighted toward low complexity (i.e. near uniform spacings) in both models, although the behaviour near zero is different and the maximum AMD model generates slightly more systems with larger $\mathcal{C}$. This result is likely due to the slightly broader distributions of period ratios (and $\Delta$) in the new model, which would lead to more variations in the spacings between planets.

\begin{figure*}
\centering
 \includegraphics[scale=0.46,trim={0.7cm 1.3cm 0.4cm 0},clip]{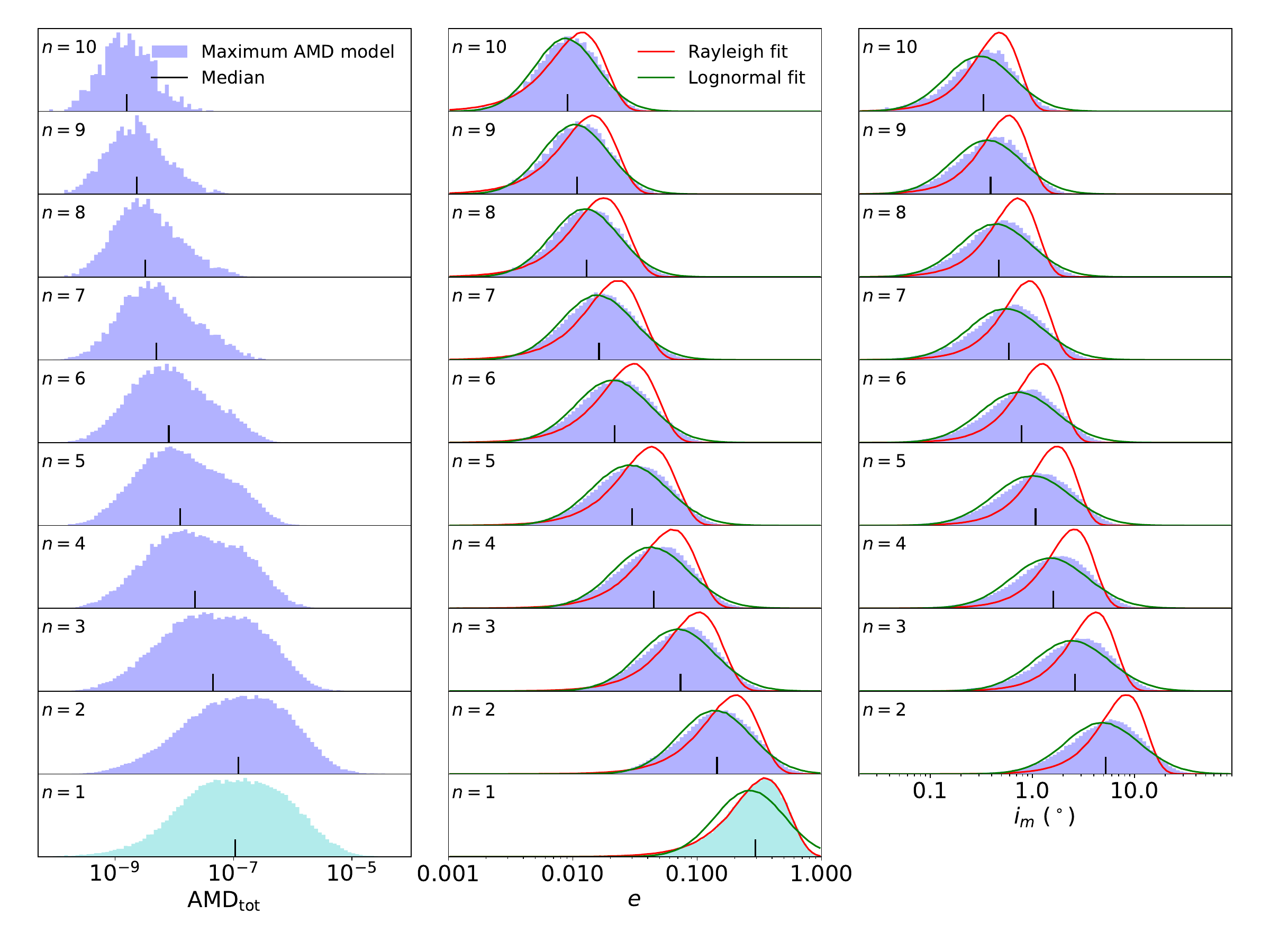}
\caption{Distributions of total system AMD$_{\rm tot}$ (\textbf{left-hand panels}), eccentricities $e$ (\textbf{middle panels}) and mutual inclinations $i_m$ (\textbf{right-hand panels}), as a function of the intrinsic planet multiplicity $n$. One simulated physical catalog drawn from our maximum AMD model (with parameters listed in Table \ref{tab:param_fits}) is shown here, as the colored shaded histograms. The eccentricity (and AMD) distribution for $n = 1$ is plotted in a different color (cyan) to remind the reader that we draw $e$ for the intrinsic singles separately, directly from a Rayleigh distribution. In each individual panel, the vertical black tick denotes the median value. For each $e$ and $i_m$ panel, the red and green curves show the best-fits for a Rayleigh distribution and a lognormal distribution, respectively. Note that we fit these distributions to the (unlogged) $e$ and $i_m$ distributions themselves, but plot them as histograms with log-uniform bins. We only plot panels up to $n = 10$ for clarity, but higher multiplicity orders can and do exist in the simulated physical catalogs. The total (i.e. critical) system AMD (and thus also the distributions of $e$ and $i_m$) decreases as the total planet multiplicity $n$ increases.}
\label{fig:mult_vs_amd_ecc_incl_dists}
\end{figure*}

\begin{figure}
\centering
 \includegraphics[scale=0.42,trim={0cm 0.5cm 0cm 0cm},clip]{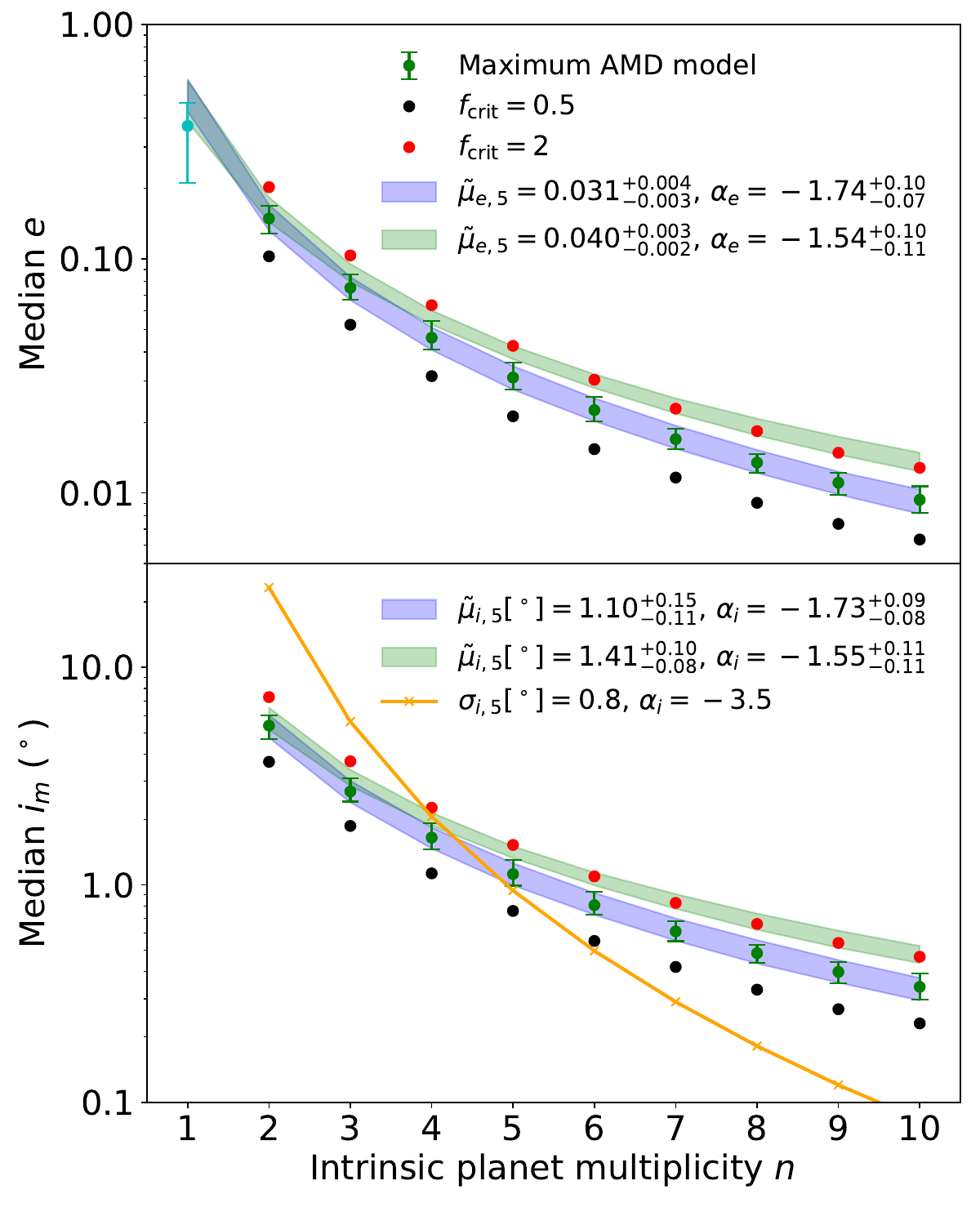}
\caption{Median eccentricities (\textbf{top panel}) and mutual inclinations (\textbf{bottom panel}) as a function of intrinsic planet multiplicity. The maximum AMD model is shown as green points where the error bars denote the 68.3\% credible regions from multiple simulated catalogs passing our $\mathcal{D}_{W,3}(\rm KS)$ distance threshold. Black and red points denote the median values from models with half ($f_{\rm crit} = 0.5$) and double ($f_{\rm crit}$ = 2) the critical AMD, respectively (the other model parameters are held fixed to the values listed in Table \ref{tab:param_fits}). The purple shaded regions denote the central 68.3\% from fitting the power--laws (equations \ref{eq_power_law_incl} and \ref{eq_power_law_ecc}) to $n \geq 2$ systems for each simulated catalog.
Likewise, the green shaded regions denote power-law fits for simulated catalogs from an alternative model in which the AMD of each system is distributed equally per planet, as opposed to equally per unit mass (see \S\ref{Distribute_AMD}).
The median eccentricity for intrinsic singles ($\simeq \sigma_{e,1}\sqrt{2\ln2}$) is plotted separately as the cyan point. For comparison, we also plot the median power--law relation from \citet{Z2018} as the orange curve.}
\label{fig:ecc_incl_mult_fits}
\end{figure}

\begin{deluxetable}{lcccccc}
\centering
\tablecaption{Eccentricity and mutual inclination distributions as a function of intrinsic planet multiplicity ($n$), of one simulated catalog (with parameters listed in Table \ref{tab:param_fits}) from the maximum AMD model.}
\tablehead{
 & \multicolumn3c{Eccentricity} & \multicolumn3c{Mutual inclination ($^\circ$)} \\
 & \colhead{Model} & \multicolumn2c{Lognormal fit} & \colhead{Model} & \multicolumn2c{Lognormal fit} \\
 \colhead{$n$} & \colhead{68.3\%} & \colhead{$e^\mu$} & \colhead{$\sigma$} & \colhead{68.3\%} & \colhead{$e^\mu$} & \colhead{$\sigma$}
}
\decimalcolnumbers
\startdata
 10 & $0.009_{-0.004}^{+0.007}$ & 0.009 & 0.587 & $0.33_{-0.18}^{+0.30}$ & 0.30 & 0.77 \\[5pt]
 9   & $0.011_{-0.005}^{+0.008}$ & 0.010 & 0.612 & $0.39_{-0.22}^{+0.39}$ & 0.36 & 0.79 \\[5pt]
 8   & $0.013_{-0.006}^{+0.011}$ & 0.013 & 0.632 & $0.47_{-0.27}^{+0.48}$ & 0.44 & 0.81 \\[5pt]
 7   & $0.016_{-0.008}^{+0.014}$ & 0.016 & 0.666 & $0.59_{-0.34}^{+0.64}$ & 0.55 & 0.84 \\[5pt]
 6   & $0.022_{-0.011}^{+0.019}$ & 0.021 & 0.685 & $0.78_{-0.46}^{+0.87}$ & 0.73 & 0.85 \\[5pt]
 5   & $0.030_{-0.016}^{+0.028}$ & 0.029 & 0.704 & $1.08_{-0.64}^{+1.23}$ & 0.99 & 0.86 \\[5pt]
 4   & $0.045_{-0.024}^{+0.040}$ & 0.043 & 0.701 & $1.61_{-0.96}^{+1.81}$ & 1.48 & 0.86 \\[5pt]
 3   & $0.073_{-0.038}^{+0.063}$ & 0.069 & 0.689 & $2.63_{-1.56}^{+2.89}$ & 2.42 & 0.85 \\[5pt]
 2   & $0.144_{-0.074}^{+0.127}$ & 0.138 & 0.670 & $5.22_{-3.08}^{+5.77}$ & 4.84 & 0.84 \\[5pt]
 1   & $0.294_{-0.145}^{+0.185}$ & 0.265* & 0.641* & - & - & - \\[5pt]
\enddata
\tablecomments{The parameters $\mu$ and $\sigma$ refer to the mean and standard deviation of the normal distribution for the log quantities; we report the values of $e^\mu$ (the median of the unlogged quantities) for interpretability.}
\tablenotetext{*}{While we also fit a lognormal distribution here for comparison, the eccentricities of intrinsic singles are drawn from a true Rayleigh distribution with $\sigma_{e,1} = 0.25$ for this catalog.}
\label{tab:ecc_incl}
\end{deluxetable}

\begin{figure}
\centering
 \includegraphics[scale=0.45,trim={0.8cm 0.5cm 0.8cm 0.5cm},clip]{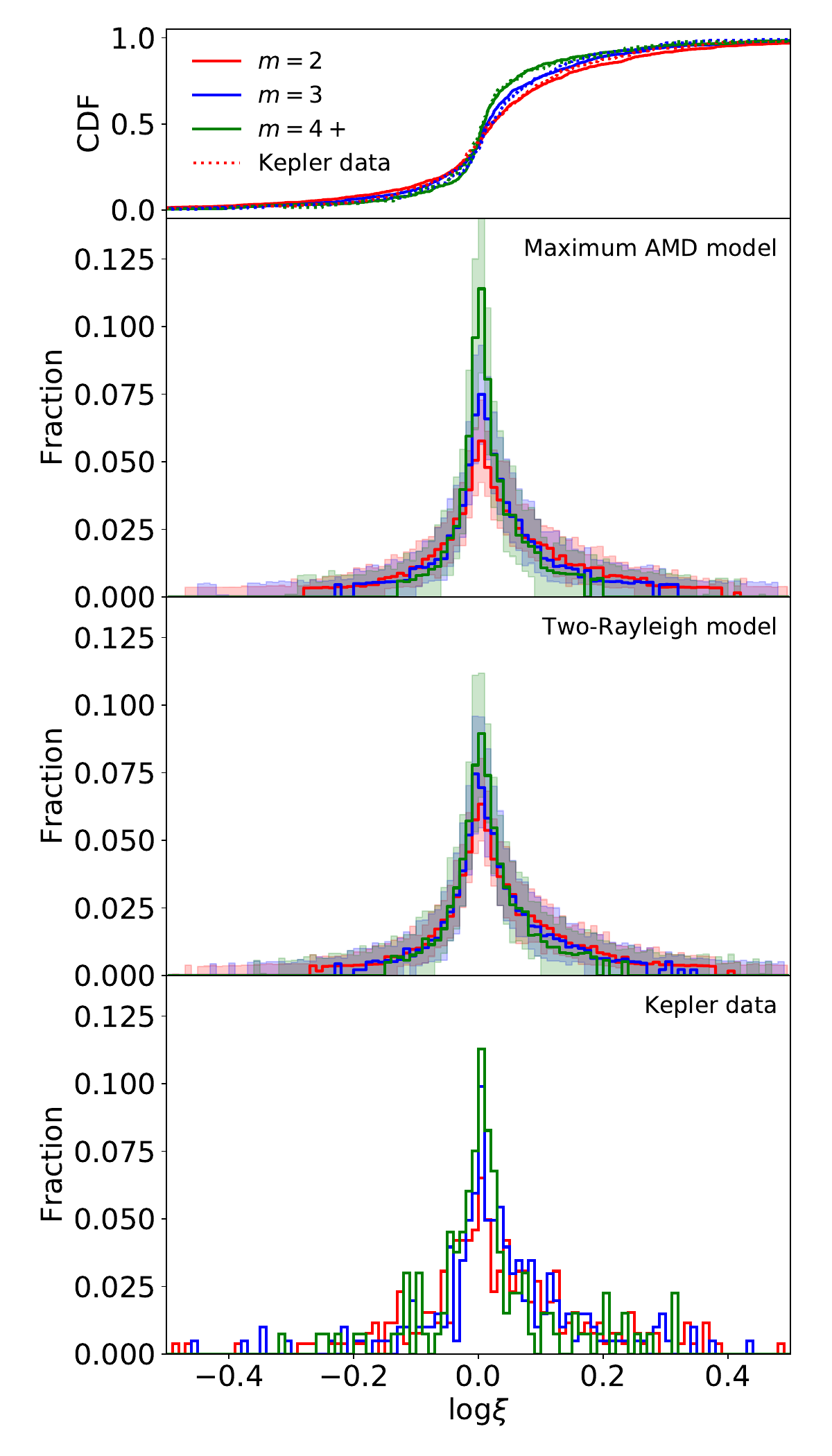}
\caption{Distributions of the (log) period--normalized transit duration ratios, $\xi = (t_{\rm dur,in}/t_{\rm dur,out})(P_{\rm out}/P_{\rm in})^{1/3}$, as a function of observed planet multiplicity $m$.
For the models (middle two panels), the bold line histograms show the medians in each bin while the shaded regions denote the central 68.3\% credible regions. Both models and the \Kepler{} data exhibit narrower and slightly more positively skewed distributions around $\log{\xi} = 0$ for higher observed multiplicities. The trend appears stronger in the maximum AMD model. There is an underlying anti-correlation between eccentricity (and mutual inclination) and intrinsic multiplicity in the maximum AMD model but not in the two-Rayleigh model. Thus, while some of the trend between $\xi$ and multiplicity is attributed to observational biases, the effect is amplified by the underlying anti-correlations (in the maximum AMD model). The top panel shows the cumulative distributions for a simulated catalog from the maximum AMD model (solid lines) and the \Kepler{} data (dotted lines), showing remarkable agreement.}
\label{fig:xi_per_mult}
\end{figure}
% Comparing p-values using 2-sample KS test (Max AMD vs. Two-Rayleigh):
% m=2: p=0.49_{-0.32}^{+0.38} vs. 0.53_{-0.34}^{+0.32}
% m=3: p=0.66_{-0.37}^{+0.22} vs. 0.51_{-0.37}^{+0.33}
% m=4+: p = 0.54_{-0.29}^{+0.36} vs. 0.66_{-0.41}^{+0.26}
% Thus, it is unclear if one model is a better fit than the other, for any m, and both models are consistent with the data for each m (cannot reject null hypothesis that the model is from the same distribution as the data, at p=0.05 significance).

% FYI: there are some statistically significant differences for each m in the data, again using 2-sample KS test:
% m=2 vs. m=3: p=0.384
% m=2 vs. m=4+: p=0.010
% m=3 vs. m=4+: p=0.033
% Thus, no statistically significant difference between m=2 and m=3, but statistically significant differences between m=2 and m=4+, and m=3 and m=4+ (at p=0.05 significance).

\subsection{The eccentricity and mutual inclination distributions} \label{Results_ecc_incl}

Our maximum AMD model results in very different distributions for the eccentricities and mutual inclinations of planets as compared to the two--Rayleigh model. As described in \S\ref{Max_AMD_model}, this new model provides a natural description for the orbital excitations (i.e. eccentricities and mutual inclinations) that does not require any free parameters, by assuming that all planetary systems are at the critical AMD.

\subsubsection{A multiplicity--dependent distribution} \label{secMult_vs_ecc_incl}

In Figure \ref{fig:mult_vs_amd_ecc_incl_dists}, we plot the distributions of total system AMD, eccentricity ($e$), and mutual inclination ($i_m$), from our maximum AMD model, for each intrinsic multiplicity order ($n$). The filled-in color histograms show the distributions for one simulated catalog (same one shown in Figure \ref{fig:models_underlying}, with model parameters listed in Table \ref{tab:param_fits}), where we have denoted the medians with vertical black ticks.

We find that the distributions are strong functions of the intrinsic planet multiplicity. While there is a wide distribution for each $n$, the median total system AMD (i.e. the critical AMD, for $n \geq 2$) decreases as the number of planets in the system increases (left--hand panels in Figure \ref{fig:mult_vs_amd_ecc_incl_dists}). The distribution is also narrower for higher $n$.

The critical AMD trend translates into an even stronger function of planet multiplicity for the distributions of eccentricities $e$ (middle panels) and mutual inclinations $i_m$ (right-hand panels in Figure \ref{fig:mult_vs_amd_ecc_incl_dists}). The total system AMD must be shared amongst all the planets in the system; in addition to the lower AMD budget for higher $n$, this AMD budget is further divided between a greater number of planets (per unit mass).
The $e$ and $i_m$ distributions for each $n$ ($\geq 2$) appear to be nearly lognormally--distributed.
This is in contrast to the parameterizations used in many previous studies (including our two--Rayleigh model), where the Rayleigh distribution is typically assumed for the eccentricities\footnote{This result may be somewhat surprising for the eccentricity distribution, as the Rayleigh distribution can be motivated by a model where the excitation of $x = e\cos{\omega}$, $y = e\sin{\omega}$, and $z = \sin{i}$ arises from a series of random kicks to a planet's orbit. However, this picture of random kicks to a circular orbit is different from the picture of chaotic collisions between planets until the total system AMD is reduced below the critical value, which is a motivation for setting all systems at the critical AMD.} 
and mutual inclinations. To emphasize this, we also plot the best-fit lognormal and Rayleigh distributions, as green and red curves, respectively, for each multiplicity order in Figure \ref{fig:mult_vs_amd_ecc_incl_dists}. The parameters of the lognormal ($e^\mu$, $\sigma$) fits are also listed in Table \ref{tab:ecc_incl}.

The inverse trend with multiplicity is very similar to the results of \citet{Z2018}, who also found a multiplicity-dependent distribution of mutual inclinations. They used \Kepler{} transit data along with TTV multiplicities to constrain the dispersion in orbital inclinations, modelling the mutual inclinations per multiplicity as a Fisher distribution (a generalization of the Rayleigh distribution to a sphere). By assuming a power--law of the form $\sigma_{i,n} = \sigma_{i,5}(n/5)^{\alpha_i}$ (where $\sigma_{i,n}$ is the inclination dispersion parameter; closely related to the Rayleigh scale parameter), \citet{Z2018} found a steep inverse relation of $\alpha_i = -3.5$ and $\sigma_{i,5} = 0.8^\circ$ (note that they chose to normalize at $n = 5$, and we have replaced their notation of $k$ with $n$ for the planet multiplicity). We find that our mutual inclination distribution is also well modelled by a power-law function of the intrinsic multiplicity, although the power--law index is shallower than what was found in \citet{Z2018}.
We fit a power--law to the median mutual inclination ($\tilde{\mu}_{i,n}$) of each $n = 2,3,...,10$, for each simulated catalog:
\begin{equation}
 \tilde{\mu}_{i,n} = \tilde{\mu}_{i,5}\bigg(\frac{n}{5}\bigg)^{\alpha_i}, \label{eq_power_law_incl}
\end{equation}
and find that the central 68.3\% values are $\tilde{\mu}_{i,5} = 1.10_{-0.11}^{+0.15}$ deg and $\alpha = -1.73_{-0.08}^{+0.09}$. We note that our combination of $\tilde{\mu}_{i,5}$ and $\alpha_i$ is between the $2\sigma$ and $3\sigma$ log-likelihood contours of \citet{Z2018} (Figure 6 therein).

Similarly, we fit a power--law for the median eccentricity (which we represent with $\tilde{\mu}_{e,n}$, analogously) as a function of the intrinsic multiplicity:
\begin{equation}
 \tilde{\mu}_{e,n} = \tilde{\mu}_{e,5}\bigg(\frac{n}{5}\bigg)^{\alpha_e}. \label{eq_power_law_ecc}
\end{equation}
We find that $\tilde{\mu}_{e,5} = 0.031_{-0.003}^{+0.004}$ and $\alpha_e = -1.74_{-0.07}^{+0.11}$ from repeated simulated catalogs. Thus, both the eccentricity and mutual inclination scales seem to follow a very similar ($\alpha \simeq -1.74$) scaling with multiplicity.

In Figure \ref{fig:ecc_incl_mult_fits}, we plot the median mutual inclinations (bottom panel) and eccentricities (top panel) as a function of the intrinsic multiplicity, along with the power--law fits of this work and of \citet{Z2018} (the orange curve). While there is a broad distribution of eccentricities and inclinations for any given multiplicity, the median values near perfectly follow power--law distributions with $n$. We also show how the median values of the eccentricity and inclination distributions shift up (down) if the total AMD is increased (decreased) relative to the critical AMD by a factor of $f_{\rm crit} = {\rm AMD}_{\rm tot}/{\rm AMD}_{\rm crit} = 2$ (red points) or 0.5 (black points).
Thus, the power--law trend persists for any fixed values of $f_{\rm crit}$ both above and below the critical value. We discuss the distribution of $f_{\rm crit}$ further in \S\ref{AMD_parameter}.

Our model predictions for the inverse relation between eccentricities and multiplicity are qualitatively in agreement with the observed correlations found by \citet{LT2015} and \citet{ZT2017}, who used samples of mostly RV planets in multi--planet systems with measured eccentricities. \citet{LT2015} considered a broad dataset of 403 RV exoplanets with non--zero point estimates of eccentricities and found a power--law relation of $e(m) = 0.584m^{-1.20}$ (where we have used $m$ to denote the observed multiplicity count), also fitting to the median eccentricities. \citet{ZT2017} selected a more restricted, heterogeneous sample of 258 RV and transiting planets with estimated uncertainties on their eccentricities and found that the mean eccentricities weighted by their relative errors follow an even tighter power--law fit of $e(m) = 0.630m^{-1.02}$. The latter study also used a small subset of their dataset consisting of systems with known planet masses, semi-major axes, eccentricities, and mutual inclinations to compute the AMD of each system and found a tentative anti--correlation between AMD and multiplicity. Thus, the qualitatively similar trend for the critical AMD values of our simulated planetary systems hints at the physical nature of this correlation arising from the AMD stability criteria itself. We emphasize that while these previous studies focused on rather heterogenous datasets of \textit{observed} planets (and mostly from RV observations), we show that these correlations also arise in the underlying planetary systems with the \textit{intrinsic} numbers of planets per system and provide an excellent description of the \Kepler{} observed multi-planet systems after accounting for observational biases through our forward model.

%\bigskip
\subsubsection{Observational constraints on the trend with multiplicity} 
Our findings that the maximum AMD model naturally predicts 
a dependence of the eccentricity and mutual inclination distributions on the intrinsic multiplicity raise an interesting question: 
is there evidence for this correlation in the observed data of multi--transiting systems?
A key summary statistic for addressing this question is the period--normalized transit duration ratio, $\xi = (t_{\rm dur,in}/t_{\rm dur,out})(P_{\rm out}/P_{\rm in})^{1/3}$, as listed in \S\ref{SummaryStats}. The distribution of log $\xi$ encodes information about both the eccentricities and impact parameters (and indirectly inclinations and orbital spacing) of transiting planets \citep{S2010, Li2011a, FM2012, F2014, M2016phd, HFR2019}. Larger eccentricities lead to wider distributions, due to more disparate and randomized velocities during transit (extreme values of duration ratios become more common).
Higher mutual inclinations lead to more symmetric distributions, as the ordering of impact parameters becomes randomized.
On the other hand, coplanar orbits imply $\log{\xi} \geq 0$, so lower mutual inclinations cause the $\log{\xi}$ distribution to be skewed to positive values.

In Figure \ref{fig:xi_per_mult}, we plot distributions of $\log{\xi}$ for multi--planet systems from our maximum AMD model, the two-Rayleigh model, and the \Kepler{} data, for $m = 2, 3$, and $4+$ observed planet systems. In both the \Kepler{} data and the simulated catalogs from the maximum AMD model, there is evidence for a multiplicity correlation that is consistent with the expected trends arising from the eccentricities: the distributions for higher $m$ are more sharply peaked (i.e. narrower) around $\log{\xi} = 0$ due to their lower eccentricities, compared to lower $m$. This effect is robust in our simulated catalogs (upper middle panel); the shaded regions denote the 68.3\% credible intervals for 100 catalogs drawn from our maximum AMD model. There may also be a hint of the skewness arising from the mutual inclination--multiplicity trend. We plot cumulative distribution functions (CDFs) of both the simulated catalog (solid lines) and the \Kepler{} data (dotted lines) in the top panel. 
The excellent fit for each $m$ is unexpected since we did not include the individual distributions of $\log{\xi}$ for each $m$ in any of our distance functions, but only included the overall distribution (split by planets near MMRs and not-near-MMRs).

Interestingly, there is a similar but weaker trend between the distribution of $\log{\xi}$ and $m$ in the two-Rayleigh model (lower middle panel of Figure \ref{fig:xi_per_mult}). This is despite the lack of any real (anti-)correlation between the eccentricities or mutual inclinations of the planets and the intrinsic planet multiplicity in this model. These results indicate that the observed $\log{\xi}$ trend with $m$ is at least partially due to detection biases. An intuitive explanation is that systems with lower eccentricities and mutual inclinations favor being observed as higher multiplicity transiting systems. To compare the two models in terms of their fits to the $\log{\xi}$ distribution, we compute the KS distances and $p$-values between the distributions of 100 simulated catalogs from each model and the \Kepler{} data, for each $m$. We find no statistically significant differences: the $p$-values for the maximum AMD model are $p = 0.49_{-0.32}^{+0.38}, 0.66_{-0.37}^{+0.22}$, and $0.54_{-0.29}^{+0.36}$ for $m = 2$, 3, and 4+, respectively. The $p$-values for the two-Rayleigh model are very similar, at $p = 0.53_{-0.34}^{+0.32}, 0.51_{-0.37}^{+0.33}$, and $0.66_{-0.41}^{+0.26}$. Thus, while the KS tests do not favor one model over the other, both models are consistent with the data.

\begin{figure}
\centering
 \includegraphics[scale=0.425,trim={0 0.8cm 0 0.2cm},clip]{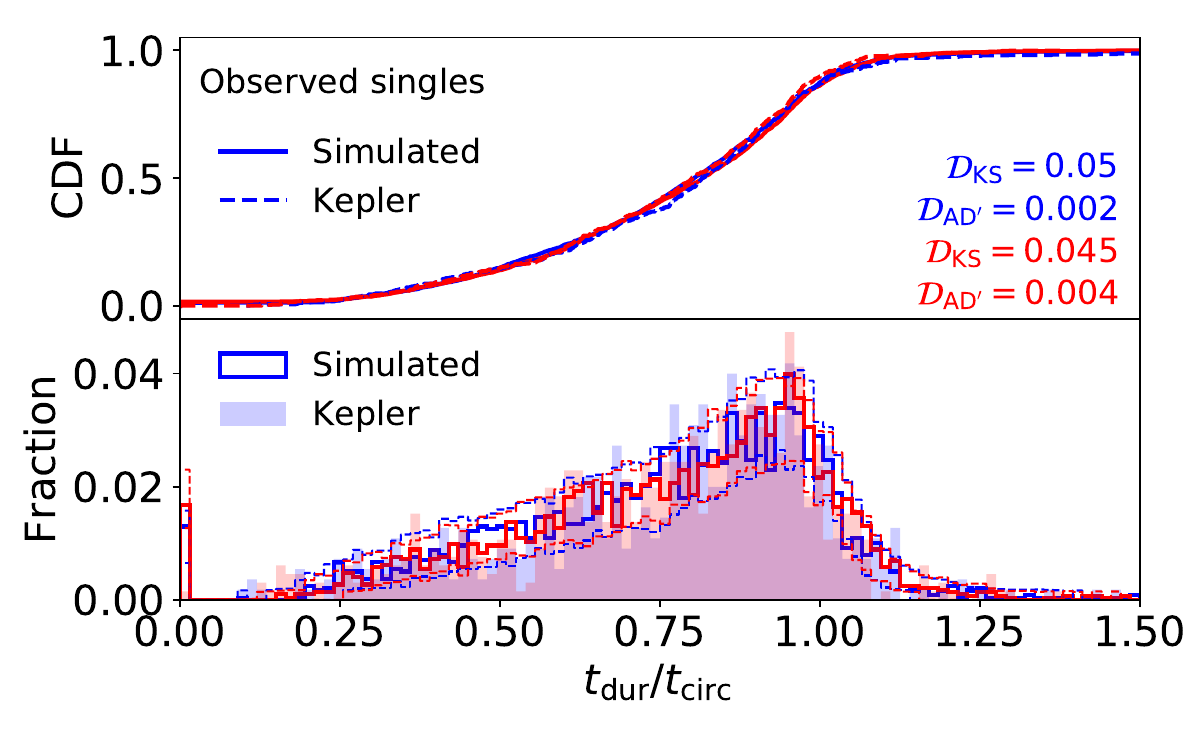}
 \includegraphics[scale=0.425,trim={0 0.8cm 0 0.2cm},clip]{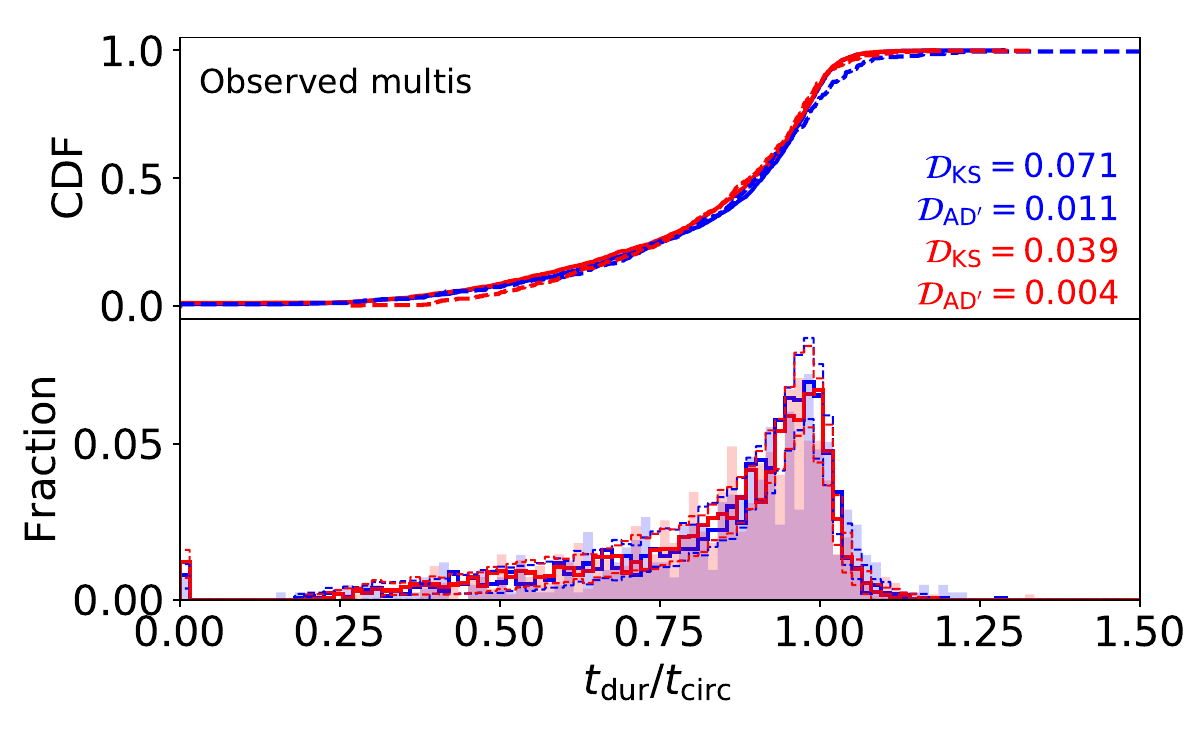}
\caption{Distributions of the circular--normalized transit duration, $t_{\rm dur}/t_{\rm circ}$, for observed singles (\textbf{top two panels}) and multis (\textbf{bottom two panels}). As in Figure \ref{fig:model_split_bprp}, the bold lines show one simulated catalog from the maximum AMD model while the dashed lines denote the 68.3\% credible regions for each bin. The shaded histograms denote the \Kepler{} data. The CDFs show the same distributions as the histograms. Blue and red colors refer to the bluer and redder samples, respectively. The distribution is wider for the observed singles (in both bluer and redder halves) than for observed multis, suggesting that they have larger eccentricities. These distributions are included in the $\mathcal{D}_{W,2}$ and $\mathcal{D}_{W,3}$ distance functions.}
\label{fig:tdur_tcirc}
\end{figure}

\subsubsection{The eccentricity distribution of single planet systems} 
\label{secEccSingles}

In Figures \ref{fig:mult_vs_amd_ecc_incl_dists} and \ref{fig:ecc_incl_mult_fits}, the results for $n = 1$ are colored differently to remind the reader that these systems are treated separately in this model, where the eccentricities are drawn from a Rayleigh$(\sigma_{e,1})$ distribution; in this case, the ``critical'' AMD is based on the eccentricity that leads to a collision with the host star.  
Therefore, intrinsic single planets are generally well below the critical AMD in this model. 
We find that $\sigma_{e,1}$ is around $\sim 0.25$, although it is poorly constrained with any of the distance functions (Table \ref{tab:param_fits}).

We plot distributions of the circular--normalized transit durations ($t_{\rm dur}/t_{\rm circ}$), for observed singles and multis, in Figure \ref{fig:tdur_tcirc}. The distribution is broader for observed singles than observed multis, suggesting that the eccentricities of single planet systems are larger than those in multi-planet systems. This result corroborates the study by \citet{M2011}, who also found a statistically significant difference in the distributions of observed singles and multis.
However, our inclusion of distance terms for fitting these distributions (i.e. $\{t_{\rm dur}/t_{\rm circ}\}_1$ and $\{t_{\rm dur}/t_{\rm circ}\}_{2+}$) in $\mathcal{D}_{W,2}$ and $\mathcal{D}_{W,3}$ evidently did not provide strong enough constraints on the eccentricity scale of intrinsic single planet systems.
The uncertainties on $\sigma_{e,1}$ using $\mathcal{D}_{W,2}$ and $\mathcal{D}_{W,3}$ are not any lower than those from using $\mathcal{D}_{W,1}$. We find that a significant fraction ($\gtrsim 90\%$) of \textit{apparently} single--planets are in multi--planet systems, which makes it difficult to distinguish the distinct eccentricity distribution of true single planets. Only $7.7_{-4.6}^{+5.6}\%$ of systems with a single observed--transiting planet are the sole planets in the period and size range explore. This small fraction is largely due to the fact that we consider planets with sizes down to $R_{p,\rm min} = 0.5 R_\oplus$, when assessing intrinsic multiplicity, and many of the small planets would not be detected around most \Kepler{} targets, even if they were transiting.  If we consider only planets larger than $1 R_\oplus$ when assessing multiplicity, then the fraction of apparent singles that are true singles rises to $16_{-7}^{+7}\%$. Likewise, the fraction rises to $35_{-9}^{+8}\%$ considering only planets larger than $2 R_\oplus$. These results indicate that the properties of intrinsic single planets (within 300 days) are generally difficult to constrain with transit surveys alone, as it is difficult to disentangle these planets from those in true multi-planet systems.

Nevertheless, we find evidence that the eccentricity scale for intrinsic singles is greater than the typical eccentricity of planets in the multi-planet systems, and significantly greater than the eccentricity scale found in the two--Rayleigh model (which is for all planets in that model). 
A value of $\sigma_{e,1} \simeq 0.25$ (corresponding to a median eccentricity of $\tilde{\mu}_{e,1} = 0.25\sqrt{2\log{2}} \simeq 0.29$) produces a distribution of eccentricities and AMD$_{\rm tot}$ that qualitatively follows the trend for higher multiplicities (Figures \ref{fig:mult_vs_amd_ecc_incl_dists} and \ref{fig:ecc_incl_mult_fits}), although the power--law fits to the higher multiplicities appear to over--predict the eccentricity scale of singles (it extrapolates to $\tilde{\mu}_{e,1} = 0.50 \pm 0.08$). Our results for the eccentricity distribution of observed single planet systems are consistent with the findings of \citet{X2016} (who found a mean eccentricity of $\bar{e} \approx 0.3$), \citet{vE2019} (who reported $\sigma_e = 0.32 \pm 0.06$), and \citet{M2019} (who concluded that about a third of singles are drawn from $\sigma_{e,\rm high} > 0.3$).

\begin{figure*}
\centering
 \includegraphics[scale=0.45,trim={0.5cm 0.5cm 0.2cm 0},clip]{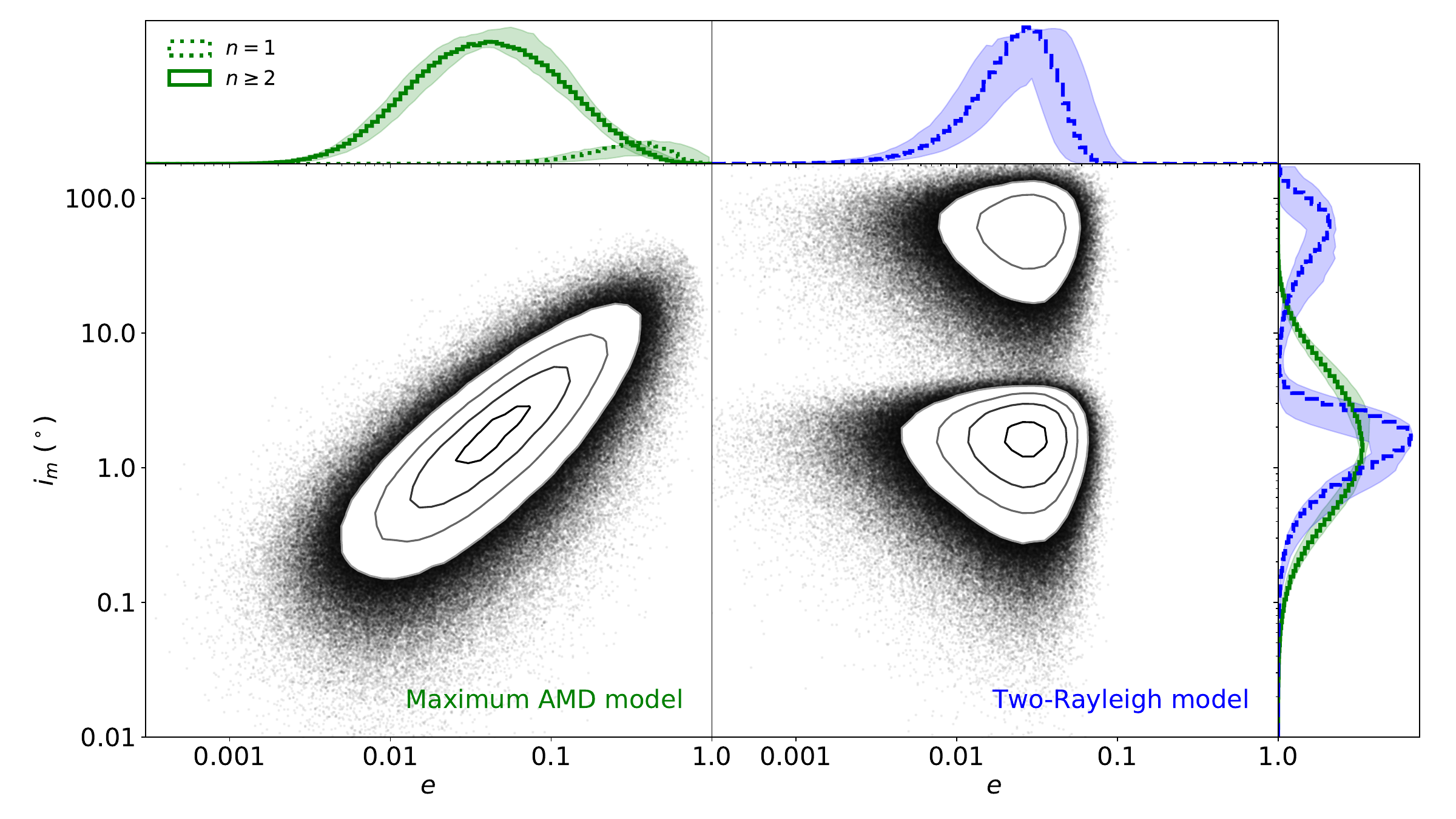}
\caption{Scatter plots of eccentricity ($e$) vs. mutual inclination ($i_m$), for intrinsic multi-planet ($n \geq 2$) systems in our maximum AMD model (green) and in our two--Rayleigh model (blue) as labeled. Four contour levels enclosing 11.8\%, 39.3\%, 67.5\%, and 86.5\% of the points (from innermost to outermost levels) are shown. On the top and right-hand panels, we plot the marginal distributions of the eccentricities and mutual inclinations, respectively, from both models. We also show the eccentricity distribution of singles ($n = 1$) as a separate histogram for the maximum AMD model. In this model, since we distribute the AMD of each planet randomly amongst the $x = e\sin{\omega}$, $y = e\cos{\omega}$, and $z = \sin{i_m}$ components, there is a natural correlation between $e$ and $i_m$. In the two-population model, there is a clear bimodality of mutual inclinations. Unlike in our maximum AMD model, there is no correlation between $e$ and $i_m$ in this model.}
\label{fig:ecc_vs_incl}
\end{figure*}

\subsubsection{Correlated eccentricities and mutual inclinations}
\label{secCorEccIncl}

The joint distribution for eccentricities and mutual inclinations of a planet shows a strong correlation for the maximum AMD model. 
We show scatter plots of mutual inclinations vs. eccentricities for multi-planet ($n \geq 2$) systems in Figure \ref{fig:ecc_vs_incl}, from the maximum AMD model (left side) and from the two--Rayleigh model (right side). For the maximum AMD model, there is a locus of points exhibiting a positive correlation.
While the AMD of any single planet must be shared amongst its eccentricity and inclination components (producing an anti--correlation for fixed AMD values), the wide range of AMD across all the planets implies that the strong correlation is primarily a function of the total AMD budget (and thus also the total multiplicity, as previously discussed). Indeed, this central locus of points (left-hand panel) shifts as a function of $n$ (not shown). These properties are very different from the orbital distributions in the two--Rayleigh model, where the eccentricities and mutual inclinations are independent (middle--right panel), and also not dependent on the intrinsic multiplicity. The two modes clearly show the two populations of mutual inclinations (hence, ``two--Rayleigh''), where the higher mutual inclination population extends across arbitrarily high values even including many retrograde orbits (for the catalog shown in this plot, the Rayleigh scales are $\sigma_{i,\rm low} = 1.25^\circ$ and $\sigma_{i,\rm high} = 45^\circ$).

The overall distributions of eccentricity and mutual inclination, marginalizing over all multiplicity orders, are also shown in Figure \ref{fig:ecc_vs_incl} (top and side panels, respectively). The eccentricity distribution is significantly broader in the maximum AMD model than in the two--Rayleigh model.

\begin{figure*}
\centering
 \includegraphics[scale=0.45,trim={1.2cm 0.2cm 0.4cm 0},clip]{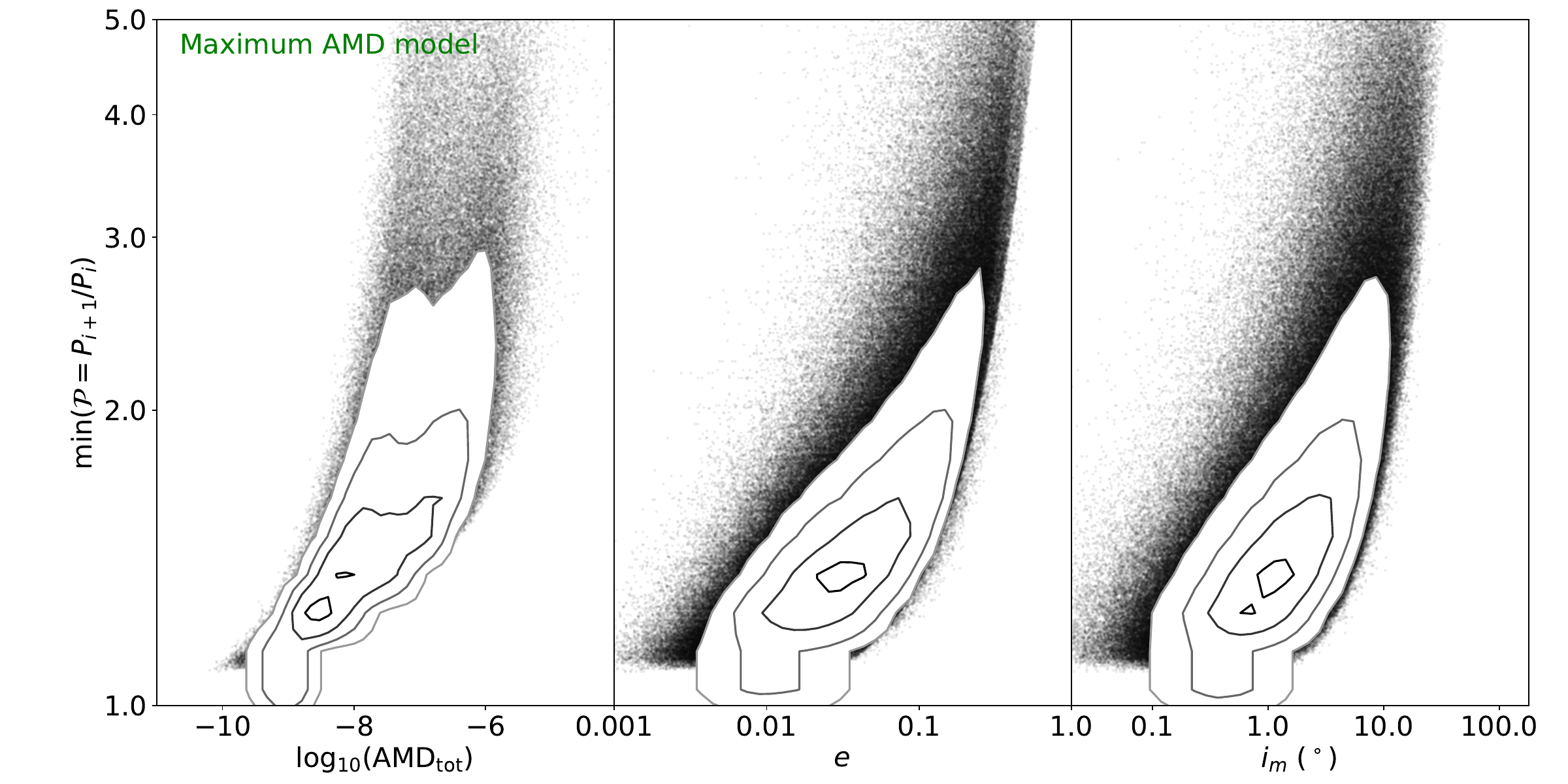}
 \includegraphics[scale=0.45,trim={1.2cm 0.2cm 0.4cm 0},clip]{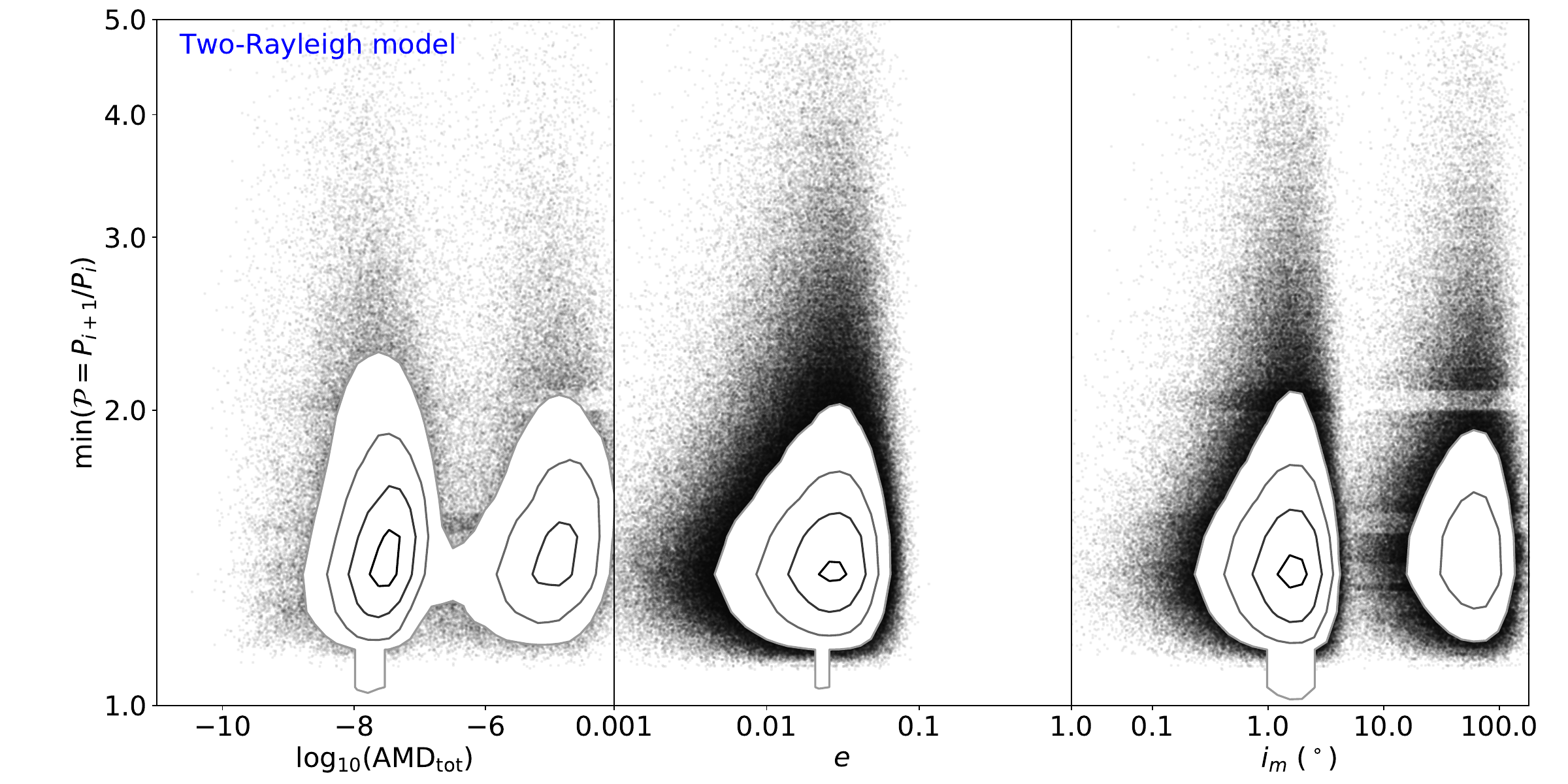}
\caption{Scatter plots of total system AMD (\textbf{left-hand panels}), eccentricity $e$ (\textbf{middle panels}), and mutual inclination $i_m$ (\textbf{right-hand panels}), vs. minimum period ratio $\mathcal{P}$ in the system. The top panels show the results for the maximum AMD model, while the bottom panels show the results for the two-Rayleigh model. All planets in intrinsic multi-planet systems ($n \geq 2$) are plotted, where four contour levels enclosing 11.8\%, 39.3\%, 67.5\%, and 86.5\% of the points (from innermost to outermost levels) are shown. The contours appearing to intersect $\mathcal{P} = 1$ (the $x$-axis) are an artifact of the plotting function; we do not actually have any period rates equal to or less than one. In contrast to the maximum AMD model, there is no clear correlation between the minimum system period ratio and the total system AMD in the two-Rayleigh model, since in this model the AMD stability criteria is not assumed. The two populations of mutual inclinations $i_m$ exhibit bands of over and under densities for the low and high $i_m$ modes, respectively, due to our prescription for reproducing the observed peaks of period ratios near the first-order MMRs: planet pairs in the high $i_m$ population that are near an MMR have their mutual inclinations drawn from the low $i_m$ distribution instead. This is clearest for the 2:1 MMR. These gaps are not completely cleared out because we are plotting the minimum period ratio, not all the period ratios, in each system.}
\label{fig:pratio_vs_amd_ecc_incl}
\end{figure*}

\subsubsection{Correlations with minimum period ratio}
\label{secCorMinPeriodRatio}

The multiplicity dependence of the total (critical) AMD can be explained by a simple dynamical consideration: it arises from the degree to which planets in a given system are tightly spaced. In Figure \ref{fig:pratio_vs_amd_ecc_incl}, we show scatter plots of the minimum period ratio in each system vs. AMD$_{\rm tot}$ (left-hand panels), eccentricities (middle panels), and mutual inclinations (right-hand panels). 
In our maximum AMD model (top panel), we compute and distribute the critical AMD after drawing the periods of each planet in a given system. 
The AMD-stability criterion requires that no pair of planets have crossing orbits (or resonance overlap), even if the total system AMD were ``dumped'' into those two planets. Thus, we expect a strong correlation between the minimum period ratio and the critical AMD. Systems with intrinsically more planets (with periods within 3--300 days) tend to be more dynamically compact, thereby inducing a correlation between the total planet multiplicity and the total AMD (and thus planet eccentricities and mutual inclinations), as previously discussed (Figure \ref{fig:mult_vs_amd_ecc_incl_dists}). 
In Figure \ref{fig:pratio_vs_amd_ecc_incl}, the absence of points in the bottom right corner of the top--left panel denotes the region of AMD-instability, which intersects the mutual Hill stability boundary toward the smallest allowed period ratios.

In contrast, the two--Rayleigh model (bottom panel) involves no such correlations with (minimum) period ratios or multiplicities. 
Here, the bimodal distribution of AMD$_{\rm tot}$ (left panel) primarily arises from the two Rayleigh distributions of mutual inclinations ($\sigma_{i,\rm low}$ and $\sigma_{i,\rm high}$; right panel). 
In this model, only the mutual Hill stability criteria ($\Delta \geq \Delta_c = 8$) is enforced. 
Indeed, a large fraction of the systems in this model are AMD--unstable; comparison of the top and bottom left panels reveals that almost the entire mode of points corresponding to the larger peak (i.e., $\sigma_{i,\rm high}$) fall outside of the AMD--stable boundary. 
This highlights a key limitation of the two--Rayleigh model, which we have fully addressed in the maximum AMD model. 
Two other features of the two--Rayleigh model are evident from Figure \ref{fig:pratio_vs_amd_ecc_incl}: 
(1) there is a weak correlation between eccentricity and minimum period ratio (middle panel) due to mutual Hill stability (since the periods are drawn \textit{after} the eccentricities in this model), and 
(2) there are horizontal bands of lower density just wide of the first order MMRs for the high mutual inclination population (right--hand panel).
The latter features arises because of our treatment of the planets near resonance (i.e., planets near resonance that were initial assigned to the high inclination population get reassigned to the low inclination population, so as to create the increased number of observed planets near period ratios of 2/1, 3/2, and 4/3.
Since we have not applied a similar procedure in our new model, these artifacts are not present in our maximum AMD model; we discuss this further in \S\ref{MMRs}.

\section{Discussion} \label{Discussion}

\subsection{Is there a \Kepler{} dichotomy?  What does a \Kepler{} dichotomy mean?} \label{Kepler_dichotomy}

Analysis of early \Kepler{} data by \citet{Li2011b} reported an excess of systems with a single transiting planet in the observed multiplicity distribution.  
More precisely, a population of planetary systems that reproduces the observed ratios of systems with multiple transiting planets predicts significantly fewer stars with a single transiting planet detected than observed by \Kepler{}.  
This finding is commonly called the ``\Kepler{} dichotomy''.
Perhaps the simplest interpretation would be that a single population of planetary systems does not simultaneously explain the observed properties of the systems with single and multiple detected planets.  
Several subsequent studies have supported this claim (e.g., \citealt{J2012}), showing that planet formation simulations also struggle to produce the excess of singles \citep{HM2013}, and that this over-abundance extends to planets around M-dwarfs as well \citep{BJ2016}. 
The most common astrophysical explanation has been to invoke at least two populations of planetary systems, with either different intrinsic multiplicity or mutual inclination distributions (e.g., \citealt{FM2012, Mu2018, Z2018, HFR2019}) to explain the two populations.   
Subsequently, \citet{ZCH2019} argued that an unmodeled reduction in the transit detection efficiency for multi-planet systems might contribute to the \Kepler{} dichotomy, and proposed that the excess of planetary systems with a single transiting planet might not even be physical.  
Alternatively, authors have proposed more creative distributions of inclination and/or intrinsic multiplicity that could nearly reproduce the observed multiplicity distribution (e.g., \citealt{Li2011b, TD2012, BL2017, SKC2019}).
These studies had not accounted for the overall number of planets detected by \Kepler{} (due to concerns about the reliability of planet candidates early in the \Kepler{} mission) and/or the transit duration ratio distribution (due to the need for a sophisticated modeling procedure).

\begin{deluxetable*}{lccccccccc}
%\centering
\tablecaption{A comparison of the observed multiplicity distribution between the \Kepler{} data and our models.}
\tablewidth{0pt}
\tablehead{
 \colhead{Observed multiplicity $m$} & \multicolumn3c{\Kepler{} data} & \multicolumn3c{Two-Rayleigh model*} & \multicolumn3c{Maximum AMD model} \\
 & \colhead{All} & \colhead{Bluer} & \colhead{Redder} & \colhead{All} & \colhead{Bluer} & \colhead{Redder} & \colhead{All} & \colhead{Bluer} & \colhead{Redder}
}
\decimalcolnumbers
\startdata
 1 (singles) & 1205 & 550 & 655 & $1252_{-109}^{+110}$ & $525_{-69}^{+68}$ & $726_{-78}^{+80}$ & $1158_{-86}^{+107}$ & $525_{-54}^{+57}$ & $633_{-56}^{+69}$ \\[5pt]
 2 (doubles) & 252 & 115 & 137 & $269_{-29}^{+29}$ & $116_{-18}^{+18}$ & $152_{-18}^{+21}$ & $284_{-28}^{+26}$ & $132_{-16}^{+16}$ & $151_{-19}^{+17}$ \\[5pt]
 3 (triples) & 97 & 37 & 60 & $93_{-14}^{+14}$ & $39_{-8}^{+9}$ & $53_{-8}^{+10}$ & $87_{-12}^{+15}$ & $41_{-8}^{+8}$ & $46_{-8}^{+9}$ \\[5pt]
 4 (quadruples) & 29 & 12 & 17 & $29_{-6}^{+9}$ & $13_{-4}^{+4}$ & $17_{-5}^{+5}$ & $25_{-6}^{+7}$ & $12_{-4}^{+4}$ & $13_{-4}^{+5}$ \\[5pt]
 5 (quintuples) & 7 & 4 & 3 & $8_{-4}^{+3}$ & $3_{-1}^{+3}$ & $4_{-2}^{+3}$ & $7_{-3}^{+3}$ & $3_{-2}^{+2}$ & $4_{-2}^{+2}$ \\[5pt]
 6 (sextuples) & 3 & 1 & 2 & $1_{-1}^{+2}$ & $1_{-1}^{+1}$ & $1_{-1}^{+1}$ & $2_{-2}^{+1}$ & $1_{-1}^{+1}$ & $1_{-1}^{+1}$ \\[5pt]
 7 (septuples) & 0 & 0 & 0 & $0_{-0}^{+1}$ & $0_{-0}^{+0}$ & $0_{-0}^{+0}$ & $0_{-0}^{+1}$ & $0_{-0}^{+1}$ & $0_{-0}^{+1}$ \\[5pt]
 8 (octuples) & 0 & 0 & 0 & $0_{-0}^{+0}$ & $0_{-0}^{+0}$ & $0_{-0}^{+0}$ & $0_{-0}^{+0}$ & $0_{-0}^{+0}$ & $0_{-0}^{+0}$ \\[5pt]
 \hline
 Total planets: $\sum{m N(m)}$ & 2169 & 965 & 1204 & $2241_{-170}^{+166}$ & $951_{-112}^{+114}$ & $1284_{-116}^{+124}$ & $2142_{-155}^{+149}$ & $977_{-78}^{+100}$ & $1162_{-111}^{+96}$ \\[5pt]
\enddata
\tablecomments{The ``Bluer'' and ``Redder'' columns add up to the ``All'' columns. For each model, the 68.3\% credible intervals are computed from 1000 simulated catalogs passing the (KS) distance threshold.}
\tablenotetext{*}{The results of our two--Rayleigh model were fit using a slightly (2.5\%) larger stellar catalog and \Kepler{} sample of exoplanet candidates, since some targets were lost due to cross--matching with the \citet{B2020} catalog in this study.}
\label{tab:mult}
\end{deluxetable*}

\citetalias{HFR2019} presented our two--Rayleigh model where planetary systems were assigned mutual inclinations drawn from a mixture of two Rayleigh distributions.  Using two different mutual inclination scale parameters could produce simulated catalogs that closely match the observed multiplicity distribution for planets around FGK dwarfs, with a very small fraction of intrinsic single--planet systems. \citetalias{HFR2019} not only found the need for a significant fraction ($\sim 30\%$) of planetary systems belonging to the high inclination population in order to produce enough single transiting systems, but also argued that an alternative theory involving a population of intrinsic single--planet systems is unlikely to explain the \Kepler{} dichotomy, due to constraints from the total number of available stars and other multi-planet distributions (e.g. period ratios).

Here, we find that the mutual inclination distribution does not have to be dichotomous, but can be characterized by a broad and multiplicity--dependent distribution. As discussed in \S\ref{Results_ecc_incl}, this is most similar to the results of \citet{Z2018}, who assumed a similar distribution of mutual inclination dispersions as a function of multiplicity. In Table \ref{tab:mult}, we show the observed multiplicity counts of the \Kepler{} catalog and of our two--Rayleigh and maximum AMD models. While the two component model fits the multiplicity distribution slightly better (marginally smaller distances for $\rho_{\rm CRPD}$; Figure \ref{fig:dists1_KS}), both models are consistent with the overall multiplicity distribution given the $1\sigma$ credible regions. While there are modest differences in the intrinsic multiplicity distributions (Figure \ref{fig:intrinsic_mults}) between these two models, neither model produces an excess of intrinsically single--planet systems in the period range considered.   
We consider our maximum AMD model to be preferred model, because:
\begin{enumerate}[leftmargin=*]
 \item the resulting distributions of mutual inclination and AMD (see Figure \ref{fig:pratio_vs_amd_ecc_incl}) are more physically plausible based on long--term stability considerations; 
 \item the joint eccentricity--mutual inclination distribution (see Figure \ref{fig:ecc_vs_incl}) is more physically plausible based on planet formation theory; and
 \item the observed joint $\log \xi$--multiplicity distribution (see Figure \ref{fig:xi_per_mult}) is naturally predicted by the maximum AMD model.
\end{enumerate}
Additionally, the maximum AMD model requires fewer model parameters than the two--Rayleigh model, yet both models provide a similar goodness of fit (Figures \ref{fig:dists1_KS} and \ref{fig:dists1_AD}).  
Therefore, we conclude that our maximum AMD model is the best available explanation for the so--called \Kepler{} dichotomy, yet it does not require a second population of planetary systems with high mutual inclinations or a significant number of stars hosting a single planet (with radius larger than 0.5 $R_\oplus$ and orbital period less than 300 days).

\begin{figure*}
\centering
\begin{tabular}{cc}
 \includegraphics[scale=0.425,trim={0 0.4cm 0 0.2cm},clip]{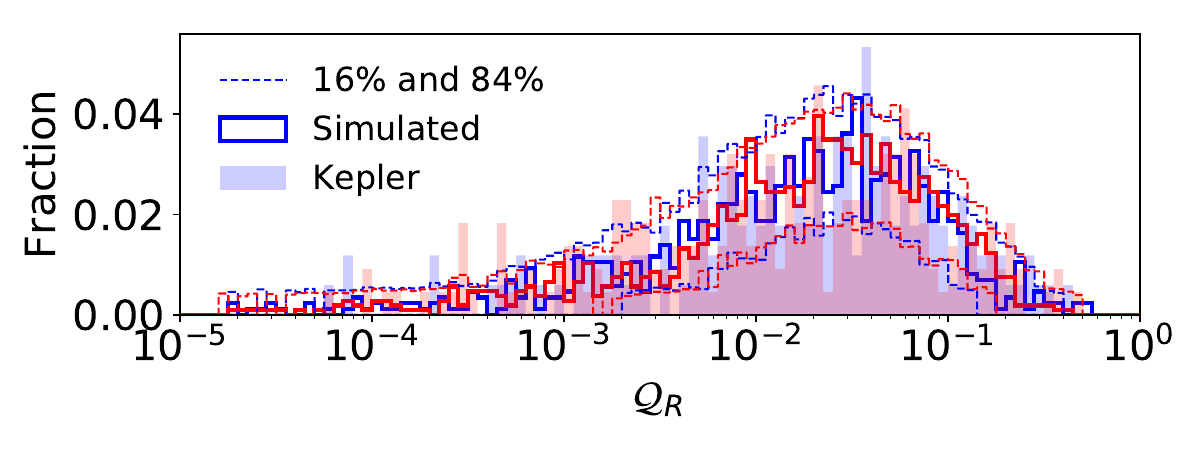} &
 \includegraphics[scale=0.425,trim={0 0.4cm 0 0.2cm},clip]{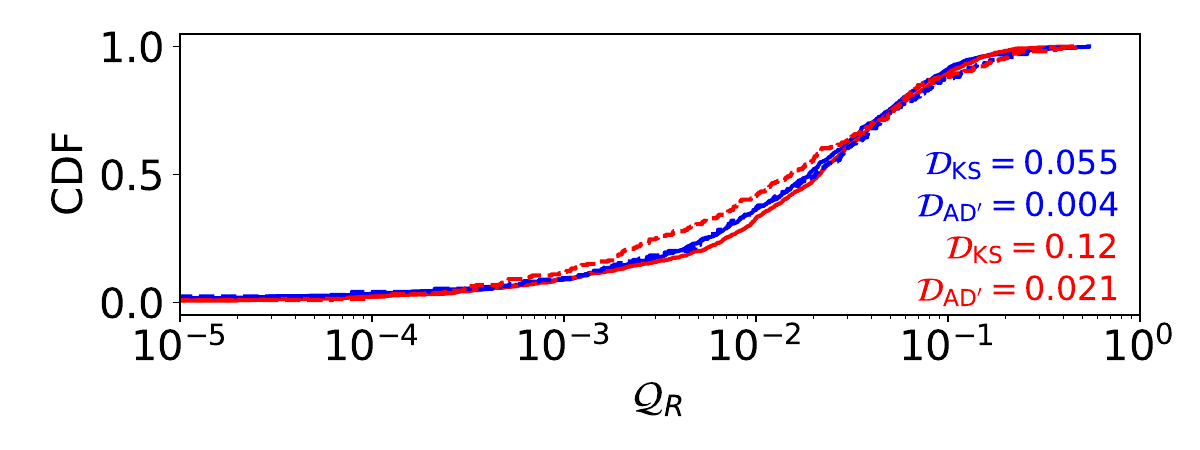} \\
 \includegraphics[scale=0.425,trim={0 0.4cm 0 0.2cm},clip]{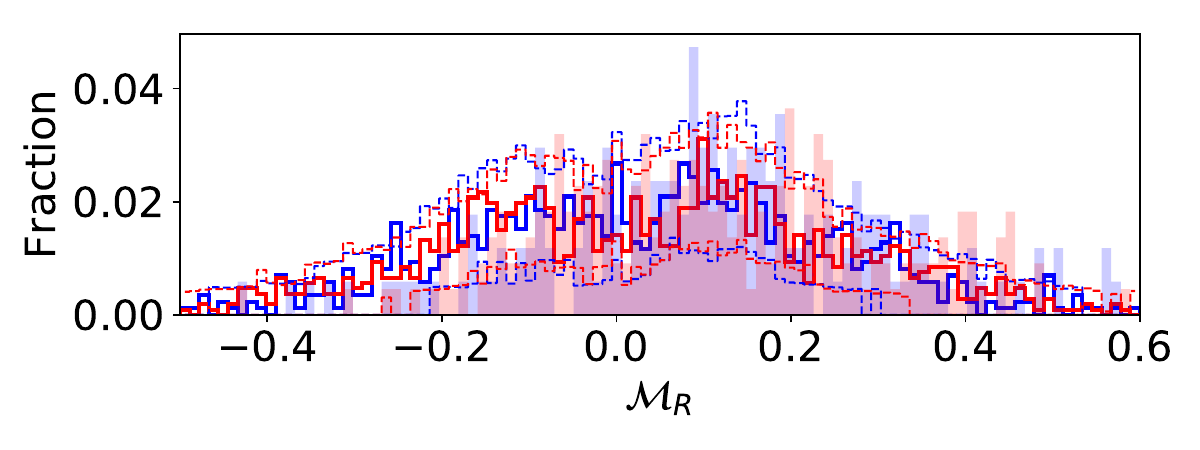} &
 \includegraphics[scale=0.425,trim={0 0.4cm 0 0.2cm},clip]{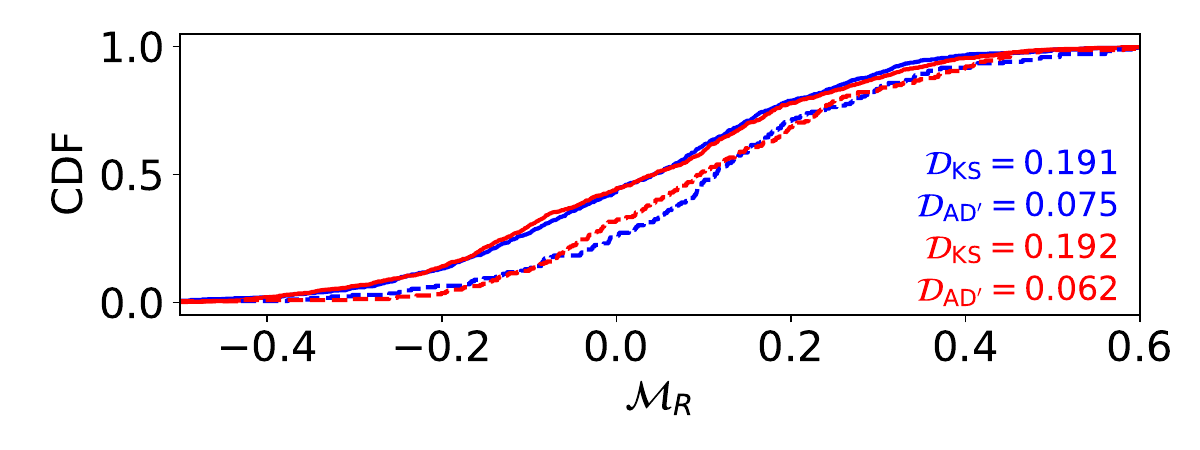} \\
 \includegraphics[scale=0.425,trim={0 0.4cm 0 0.2cm},clip]{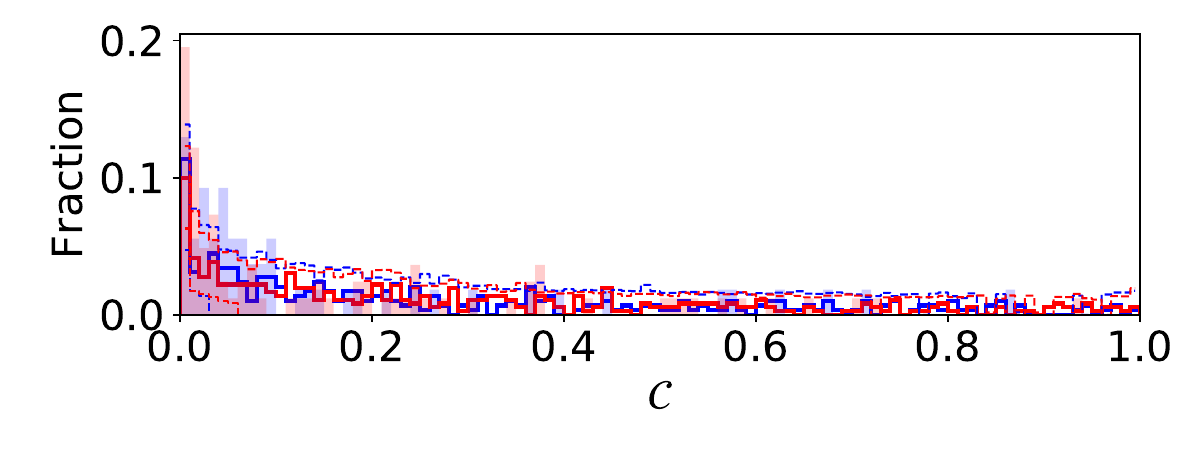} &
 \includegraphics[scale=0.425,trim={0 0.4cm 0 0.2cm},clip]{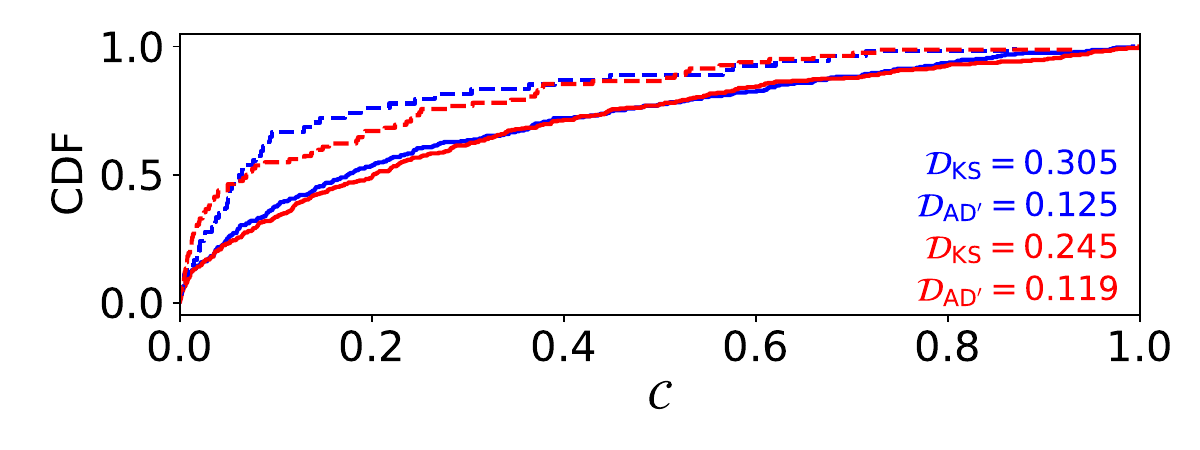} \\
\end{tabular}
\caption{Marginal distributions of additional metrics for our maximum AMD model as compared to the \Kepler{} data, split into bluer and redder halves as likewise colored. As in Figure \ref{fig:model_split_bprp}, the left--hand panels show a simulated catalog (solid bold lines) with 16th and 84th percentiles from 100 catalogs (dashed lines) as compared to the \Kepler{} data (shaded histograms), while the right--hand panels show the same simulated catalog (solid bold lines) and \Kepler{} data (dashed lines) as CDFs. The system-level metrics inspired by \citet{GF2020} are, from top to bottom: radii partitioning ($\mathcal{Q}_R$), radii monotonicity ($\mathcal{M}_R$), and gap complexity ($\mathcal{C}$). \textbf{Top panels:} the observed distribution of $\mathcal{Q}_R$ is well matched by our models (which have clustered radii), suggesting that planets in the same system tend to be similarly sized. \textbf{Middle panels:} the observed distribution of $\mathcal{M}_R$ for the \Kepler{} data is more skewed toward positive monotonicity than our models, implying that planetary systems have a preferred size ordering (our models do not produce any true ordering, as shown in Figure \ref{fig:models_underlying}). \textbf{Bottom panels:} the observed distribution of $\mathcal{C}$ is significantly more weighted toward low values than what our models produce, suggesting that planetary systems are very uniformly spaced.}
\label{fig:model_obs_GF2020}
\end{figure*}

\subsection{On the observed uniformity of planet sizes, monotonicity, and spacings in multi-planet systems} \label{Peas_in_a_pod}

Recently, there has been some debate surrounding the so--called ``peas in a pod'' pattern of multi-planet systems \citep{MWL2017, W2018a, Z2019, WP2019, MT2020, GF2020}. This term was invoked by \citet{W2018a} to describe the observations that planets within a given system tend to be fairly uniform in both size and orbital spacing.
Subsequent studies on this subject disagree about whether the observed trends are primarily due to the intrinsic distribution of planetary systems \citep{WP2019, GF2020} or primarily due to detection biases affecting the observed \Kepler{} multi-planet systems \citep{Z2019, MT2020}.

Our forward modeling procedure allows us to directly address these concerns. In Figure \ref{fig:model_obs_GF2020}, we show the observed distributions of system \textit{radius partitioning} ($\mathcal{Q}_R$), \textit{radius monotonicity} ($\mathcal{M}_R$), and \textit{gap complexity} ($\mathcal{C}$). As in Figure \ref{fig:model_split_bprp}, the maximum AMD model is shown as the bold and dashed line (for 68.3\% credible region) histograms, the \Kepler{} distributions are over-plotted as shaded histograms, and both are divided into bluer and redder stellar halves.

\subsubsection{Size uniformity}
We find that the observed $\mathcal{Q}_R$ distribution is well matched with our model, for both bluer and redder samples. This is an encouraging result of our clustering in planet radii, which we have retained from the previous clustered models of \citetalias{HFR2019}. There is no difference between the bluer and redder samples. In comparison to the intrinsic $\mathcal{Q}_R$ distribution (Figure \ref{fig:models_underlying}), we see that the observed distribution peaks at the same small value of $\mathcal{Q}_R \simeq 0.03$, suggesting that the \Kepler{} transit observations do not strongly bias this distribution. We remind the reader that while our partitioning metric $\mathcal{Q}_R$ is inspired by the mass partitioning statistic in \citet{GF2020}, we use the planet radii instead of planet masses. While planet mass is arguably a more fundamental property from a planet formation perspective, we prefer radius as a more readily measurable property of transiting planets that does not directly rely on the assumed M-R relationships. Taken together, the results of \citet{MWL2017}, \citet{W2018a}, \citet{GF2020}, and this paper are complementary in that there is a high level of intra--system similarity using either planet radii or masses for observed multi-planet systems.

\subsubsection{Size monotonicity}
Over 70\% of the \Kepler{} multi-planet systems in our sample exhibit a positive monotonicity trend ($\mathcal{M}_R > 0$), consistent with the findings of \citet{C2013}, \citet{W2018a}, and \citet{GF2020} that larger planets tend to be exterior to smaller planets within a given system. This trend persists in both our bluer and redder samples. Focusing on our models, we find that our simulated observed catalogs also tend to include slightly more systems with positive monotonicity compared to negative monotonicity; the fraction of observed systems with $\mathcal{M}_R > 0$ from our maximum AMD model is $58 \pm 3 \%$. This is despite the lack of any preferential size ordering in the underlying planetary systems (the intrinsic distribution of $\mathcal{M}_R$ is symmetric around zero; Figure \ref{fig:models_underlying}).\footnote{The observed distribution is also fairly smooth and unimodal, unlike the intrinsic distribution. As discussed in \S\ref{secIntrinsicGF2020}, the sharp spike at $\mathcal{M}_R = 0$ in the intrinsic distribution is due to the behaviour of the Spearman correlation coefficient for four and five planet systems. This feature is not seen in the observed distribution because these multiplicities are relatively rare for the observed systems (there are very few $m \geq 4$ systems compared to $m = 2$ and 3).} This finding indicates that detection biases do tend to contribute to the observed positive monotonicity trend. However, the observed preference for $\mathcal{M}_R > 0$ in our models (which arise purely from observational effects) is significantly weaker than that of the \Kepler{} data. Of 100 simulated catalogs, the maximum fraction of systems with $\mathcal{M}_R > 0$ is 65\%, still not as extreme as that of the \Kepler{} distribution. These results imply that a true monotonicity trend in the intrinsic planetary systems is necessary to match the magnitude of the observed trend.

\subsubsection{Uniform spacing}
\citet{GF2020} defined the \textit{gap complexity} ($\mathcal{C}$) term to capture global patterns in the distribution of orbital periods within a given system, and observed that the \Kepler{} multi-planet ($3+$) systems are significantly more uniformly--spaced ($\mathcal{C} \to 0$) than those in our simulated catalogs from \citetalias{HFR2019}. In this paper, we adopted the same definition of $\mathcal{C}$ and attempt to fit to the observed distribution. Despite including it in a distance function ($\mathcal{D}_{W,3}$), we find that our models provide a poor match to the observed distribution. Over half (60\%) of the \Kepler{} $3+$ planet systems have $\mathcal{C} < 0.1$; for comparison, only $35_{-5}^{+6}\%$ of systems have such low values in our simulated catalogs (Figure \ref{fig:model_obs_GF2020}). A similar result was found in \citetalias{HFR2019}, by comparing the distribution of ratios of period ratios. We echo the conclusion of \citet{GF2020} that planetary systems are very evenly spaced. A more detailed model is needed to further study and explain these features.

\subsection{Planets near resonances} \label{MMRs}

Studies focusing on the period ratio distribution of adjacent planet pairs in \Kepler{} multi-planet systems have found statistically significant peaks near first-order MMRs, especially just wide of the 2:1 and 3:2 resonances (e.g., \citealt{Li2011b, PMT2013, F2014, SH2015}). While planet migration theories predict an abundance of planet pairs trapped at low-order resonances, the exact details depend on the details of the disk and other planets in the same system.  
Furthermore, the true fraction of planet pairs near an MMR may be shrouded by observational biases. 
In our previous models from \citetalias{HFR2019}, we explored whether the observed period ratio distribution could be explained entirely by the mutual inclination distribution.
Our models produced similar peaks in the observed period ratio distribution by setting planet pairs near MMRs to have lower mutual inclinations than planets at any other arbitrary period ratios.\footnote{Since systems with multiple low mutual inclination planets are more likely to manifest as multi-transiting systems due to their orientations, this has the effect of producing apparent spikes near MMRs even without an underlying excess of planets in or near resonance.} For example, in the two--Rayleigh model we set the planets near an MMR with another planet (as defined in \S\ref{secOldModel}) to have mutual inclinations drawn from the low inclination Rayleigh scale ($\sigma_{i,\rm low}$) regardless of whether the systems they were originally assigned to the $\sigma_{i,\rm high}$ or $\sigma_{i,\rm low}$ population.

In this study, we have opted to present a model in which the mutual inclination (and eccentricity) distributions of multi--planet systems are purely described by randomly distributing the critical AMD of each system (the maximum AMD model). Thus, in this model there is no special treatment for planet pairs near resonances and (by construction) no statistically significant spikes near the MMRs in the observed period ratio distribution (Figure \ref{fig:model_split_bprp}). 
In principle, we could again reset the orbital inclination of near--MMR planets to be coplanar.  We find that when applying this to the physical catalogs computed in \S\ref{secNewProc}, we are able to recover spikes in the period ratio distribution similar to that observed.  However, we have not refit models with this effect, because we find that the distance functions used in this paper are not sufficiently sensitive to the resulting changes in the period ratio distribution in order to distinguish between such models. 
In particular, while we still include distance terms for the $\xi$ distributions of planets near and not-near MMRs ($\{\xi\}_{\rm res}$ and $\{\xi\}_{\rm non-res}$, respectively), our best--fit distances for these terms are very similar between the models described in this paper (Figures \ref{fig:dists1_KS} and \ref{fig:dists1_AD}). Given the dynamical importance of resonances and their observable effects, especially for TTVs, further work is needed to model these MMR features and should adopt a distance function that is more sensitive to these such features in the period ratio distribution.

\subsection{How is the AMD distributed in each planetary system?} \label{Distribute_AMD}

In the primary model described in this paper, we chose to distribute the critical AMD of each planetary system amongst the planets in proportion to their masses (Equation \ref{eq_amd_per_pl}). Our motivation for sharing the AMD this way is to provide the same level of dynamical ``excitation'' for each planet in a given system, since the AMD of a planetary orbit is proportional to its mass. In other words, the eccentricities and mutual inclinations of the planets do not depend on the planet masses in this model. A reasonable alternative is to distribute the AMD of a system equally per planet, such that less massive planets have much more excited orbits compared to more massive planets. This can also be motivated by the dynamical interactions of planets over time, where larger planets may impart more angular momentum to smaller planets than the other way around during interactions.

To test this model, we re-run the full analysis using the same distance functions defined in \S\ref{Distance_functions}. We find that while the distributions of eccentricities and mutual inclinations for each intrinsic multiplicity $n$ are still close to lognormal, the widths of the distributions are increased, due to the less massive planets being more excited and the more massive planets being less excited. The power-law slopes with $n$ are also slightly shallower: $\alpha_e = -1.54_{-0.11}^{+0.10}$ and $\alpha_i = -1.55_{-0.11}^{+0.11}$. The green shaded regions in Figure \ref{fig:ecc_incl_mult_fits} show the central 68.3\% for power-law fits to many simulated catalogs from this alternative model also passing the $\mathcal{D}_{W,3}$ distance threshold. There is also an overall shift to larger $e$ and $i_m$ at higher $n$.  This can be understood by considering that systems at high multiplicities tend to have more smaller planets, which are assigned larger eccentricities and inclinations in this model compared to if the AMD were distributed per unit mass. However, despite these minor differences in the planets' dynamical excitations, we find no differences in the best-fitting model parameters or distances, for any of our distance functions. While we cannot distinguish between these two models using the data and our methods, this analysis shows that our key results (e.g. the underlying anti-correlations between eccentricity/mutual inclination and multiplicity) are relatively insensitive to such model choices.

\subsection{Are all planetary systems at the critical AMD?} \label{AMD_parameter}

\begin{figure}
\centering
 \includegraphics[scale=0.42,trim={0.3cm 0.2cm 0.3cm 0.2cm},clip]{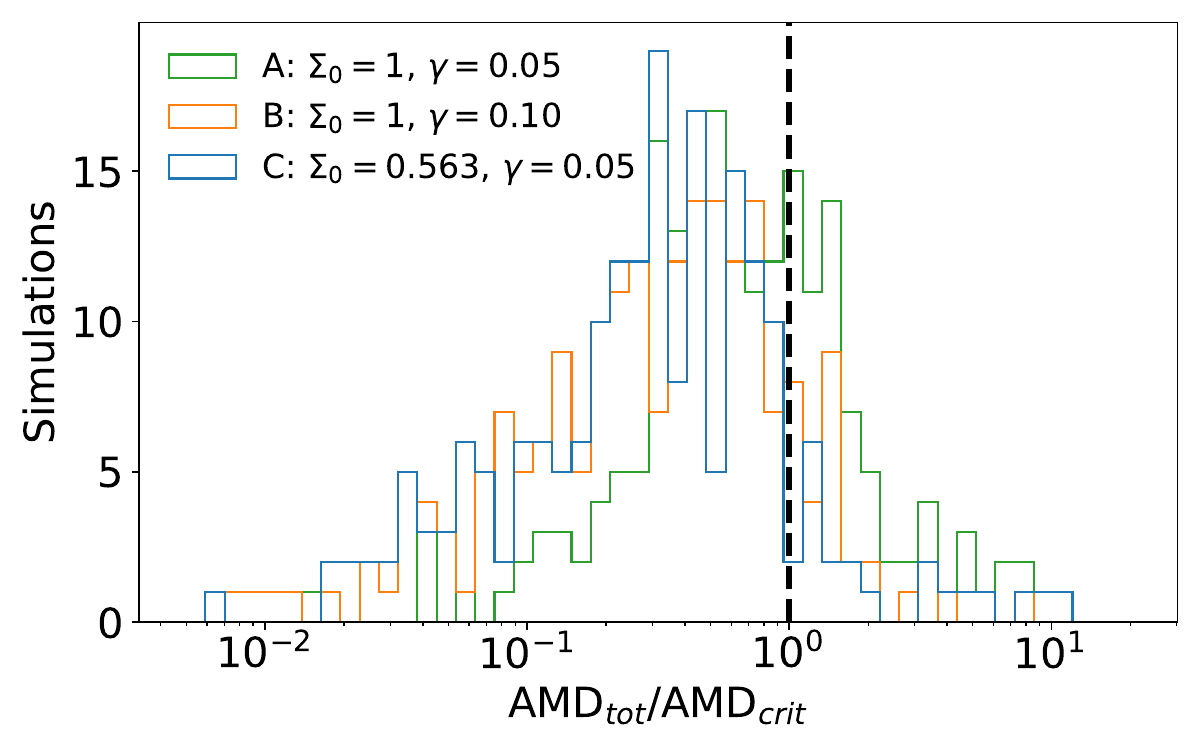}
\caption{Distribution of the total AMD relative to the critical AMD for each system, AMD$_{\rm tot}$/AMD$_{\rm crit}$, from planet formation simulations using Mercury \citep{C2018}. Three sets of simulations are shown here, each with 200 runs and assuming a solid surface density power--law of $\Sigma = \Sigma_0 r^{-\gamma}$ where $r$ is the distance from the star ($1 M_\odot$ in all runs). Each simulation starts with 200 planet embryos (totaling 43.8, 24.1, and 24.8 $M_\oplus$ for sets A, B, and C, respectively) and is evolved to include collisions which merge planets. The total and critical AMD are computed from the surviving planets at the end, including only planets within 0.877 AU (corresponding to a 300 day period around a solar mass star, i.e. similar to the period range considered in this study). While the majority of these planetary systems are AMD-stable, there is a tail of AMD-unstable systems (to the right of the vertical dashed line) in all three sets of simulations with some as large as AMD$_{\rm tot}$/AMD$_{\rm crit} \sim 10$.}
\label{fig:amd_amd_crit}
\end{figure}

\begin{figure}
\centering
 \includegraphics[scale=0.42,trim={0.3cm 0 0.3cm 0},clip]{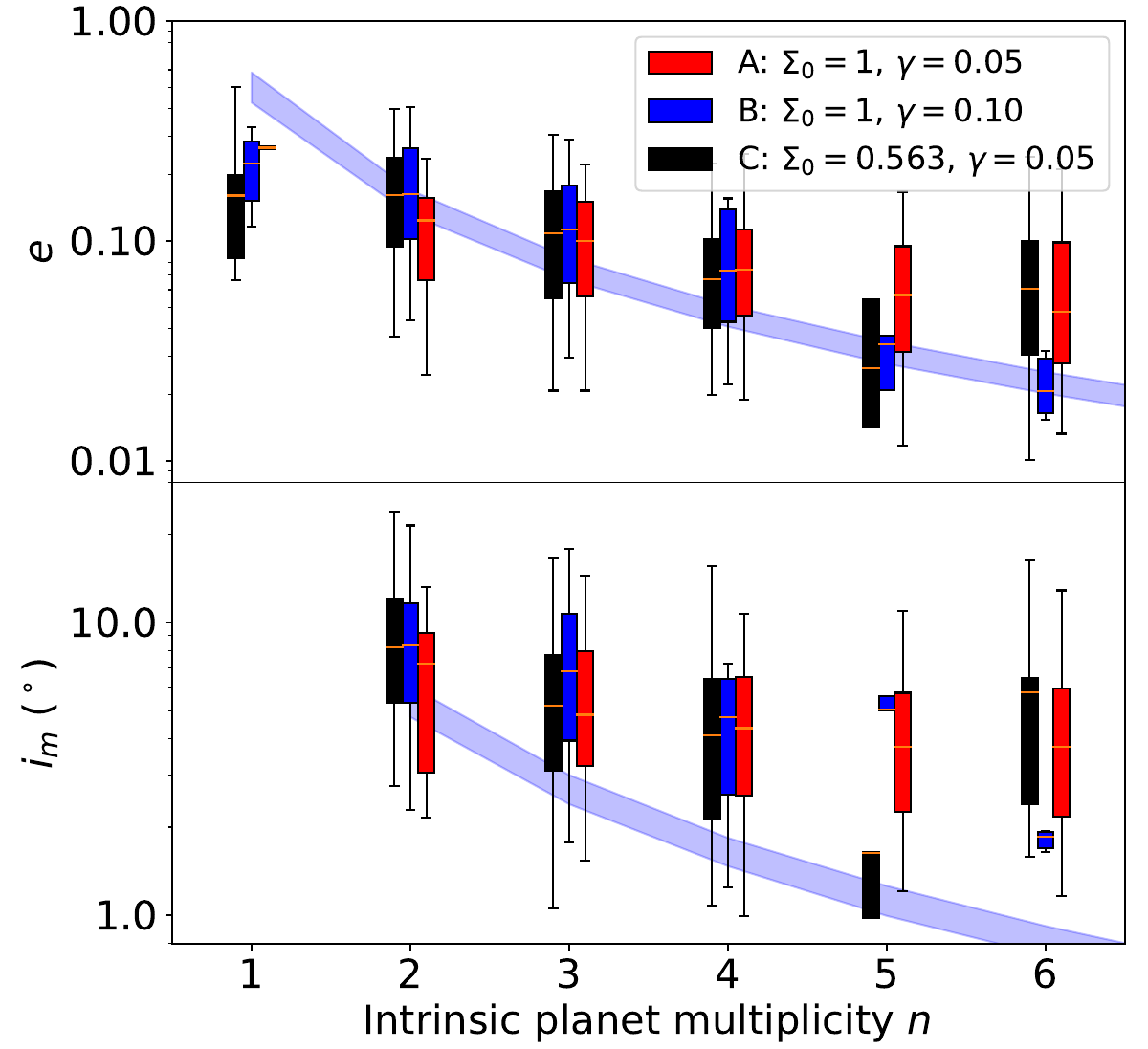}
\caption{Eccentricity (\textbf{top panel}) and mutual inclination (\textbf{bottom panel}) distributions from the planet formation simulations in Figure \ref{fig:amd_amd_crit}, as a function of intrinsic planet multiplicity $n$. The box and whiskers show the 25th-75th and 5th-95th percentiles, respectively, for each $n$ and each set of simulations. For comparison, the power-law fits to the simulated catalogs from our maximum AMD model are also shown as the purple shaded regions (same as in Figure \ref{fig:ecc_incl_mult_fits}).}
\label{fig:ecc_incl_mult_formation}
\end{figure}

The new model presented in this paper assumes that all multi-planet systems are at the AMD stability limit. A reasonable question to ask is whether all systems indeed have close to the critical (i.e. maximum) AMD, or if there is a significant fraction of systems that are considerably below the critical AMD.
Another point of interest is to consider the fraction of systems \textit{above} the critical AMD value. Indeed, a significant fraction of known planetary systems in the Extrasolar Planet Encyclopaedia database appear to be AMD-unstable based on best-fit orbital configurations \citep{LP2017, PLB2017}. Even the solar system as a whole is formally AMD--unstable, as the outer (Jovian) planets have enough AMD to cause overlapping orbits of the innermost planets, if all of the AMD were transferred to those planets \citep{L1997, LP2017}. However, this transfer of AMD between the inner and outer Solar System is slow and the system is long--lived as a result. Intriguingly, the inner four (terrestrial) planets and the outer four (giant) planets are AMD--stable when treated separately, which \citet{LP2017} further classify as ``hierarchically AMD--stable systems'' (a few exoplanetary systems also fit this definition). 
Because our current model focuses on inner planetary systems (orbital periods less than 300 days), it is not affected by such concerns about the potential impact of much more distant giant planets.
To compare the orbital excitations of planetary systems with very different orbital architectures, one should also consider the ``normalized'' angular momentum deficit (NAMD), as described in \citet{TZB2020}.

In order to gain some theoretical insight into the distribution of total system AMD relative to critical AMD, we consider a number of planet formation simulations, using the same planet formation model as in \citet{C2018}.
In Figure \ref{fig:amd_amd_crit}, we plot the final distributions of AMD$_{\rm tot}$/AMD$_{\rm crit}$ from these simulations. 
While the final states of these simulations result in many planets at a wide range of semi-major axes (out to $100$ AU), we restrict our calculation to planets within 0.877 AU ($\simeq 300$ days around a $1 M_\odot$ star). 
There is a broad distribution of AMD relative to the critical AMD, peaking at around $\sim 0.5$. While most of these systems are AMD--stable, there is a tail of AMD--unstable systems in all three sets of simulations, out to AMD$_{\rm tot}$/AMD$_{\rm crit} \sim 10$ for the range considered.
Despite the wide range of initial conditions, the simulations result in configurations with AMD$_{\rm tot}$ that are typically within a factor of 2 of AMD$_{\rm crit}$.
We also find that these simulations tend to produce an anti-correlation between the final eccentricities (and mutual inclinations) and the number of planets (within 0.877 AU). In Figure \ref{fig:ecc_incl_mult_formation}, we plot the median and quantiles of the eccentricity and mutual inclination distributions from these simulations as box and whisker plots, for multiplicities $n \leq 6$ (there are too few systems with higher multiplicities). The trend with multiplicity appears for each set of simulations, although it is weaker than that of our maximum AMD model (plotted again as the purple shaded regions for comparison). Given that our model was fit to match many of the observed properties of the \Kepler{} systems, our catalogs can in theory be used to constrain the initial conditions of such simulations. These results also provide additional support for the physical nature of the correlations seen in our maximum AMD model as arising from planet formation processes.

To test our assumption of the maximum AMD model, we briefly explore two additional models where an additional parameter is introduced: 
(1) all multi-planet systems have AMD$_{\rm tot} = f_{\rm crit} \times {\rm AMD}_{\rm crit}$ for a fixed value of $f_{\rm crit}$ (thus, our maximum AMD model can be considered a special case where $f_{\rm crit} = 1$), and (2) each multi-planet system has AMD$_{\rm tot} = f \times {\rm AMD}_{\rm crit}$ where $f \sim {\rm Unif}(f_{\rm crit,min},2)$.
For each model, we repeat the optimization process, exploring all the free parameters, along with $f_{\rm crit}$ (or $f_{\rm crit,min}$) in the range $[0,2]$. In model (1), we find a slight preference for $f_{\rm crit} \gtrsim 1$ (peaking around $f_{\rm crit} \simeq 1.25$) in the optimization stage, although $f_{\rm crit}$ is not well constrained and values between 0.4 and 2 are all acceptable. In model (2), we find even more mixed results and do not constrain $f_{\rm crit,min}$ in the range explored. Even values close to $f_{\rm crit,min} \sim 0$ cannot be ruled out. This could have been anticipated, as even $f_{\rm crit,min} = 0$ results in a mean value of $f = 1$. We do not observe any clear correlation between $f_{\rm crit}$ or $f_{\rm crit,min}$ with any of the other model parameters. To show the effect of decreasing or increasing the total AMD relative to the critical AMD on the underlying eccentricity and mutual inclination distributions, we simulate two catalogs, one with $f_{\rm crit} = 0.5$ and one with $f_{\rm crit} = 2$, and plot their median values as a function of intrinsic multiplicity in Figure \ref{fig:ecc_incl_mult_fits}. The inverse relation with multiplicity remains and the power--law slopes appear unchanged. In any case, these models do not significantly improve the best--fitting distances found over our maximum AMD model, even with the extra parameter. We conclude that while we cannot easily constrain the true distribution of AMD$_{\rm tot}$/AMD$_{\rm crit}$ from \Kepler{} data, our maximum AMD model's approach of setting all multi-planet systems to be at the AMD-stability limit is both a physically plausible assumption and a good match to most \Kepler{} observations.

\subsection{Implications for radial velocity (RV) surveys} \label{RVs}

\begin{figure}
\centering
 \includegraphics[scale=0.42,trim={0.3cm 0.5cm 0.3cm 0.3cm},clip]{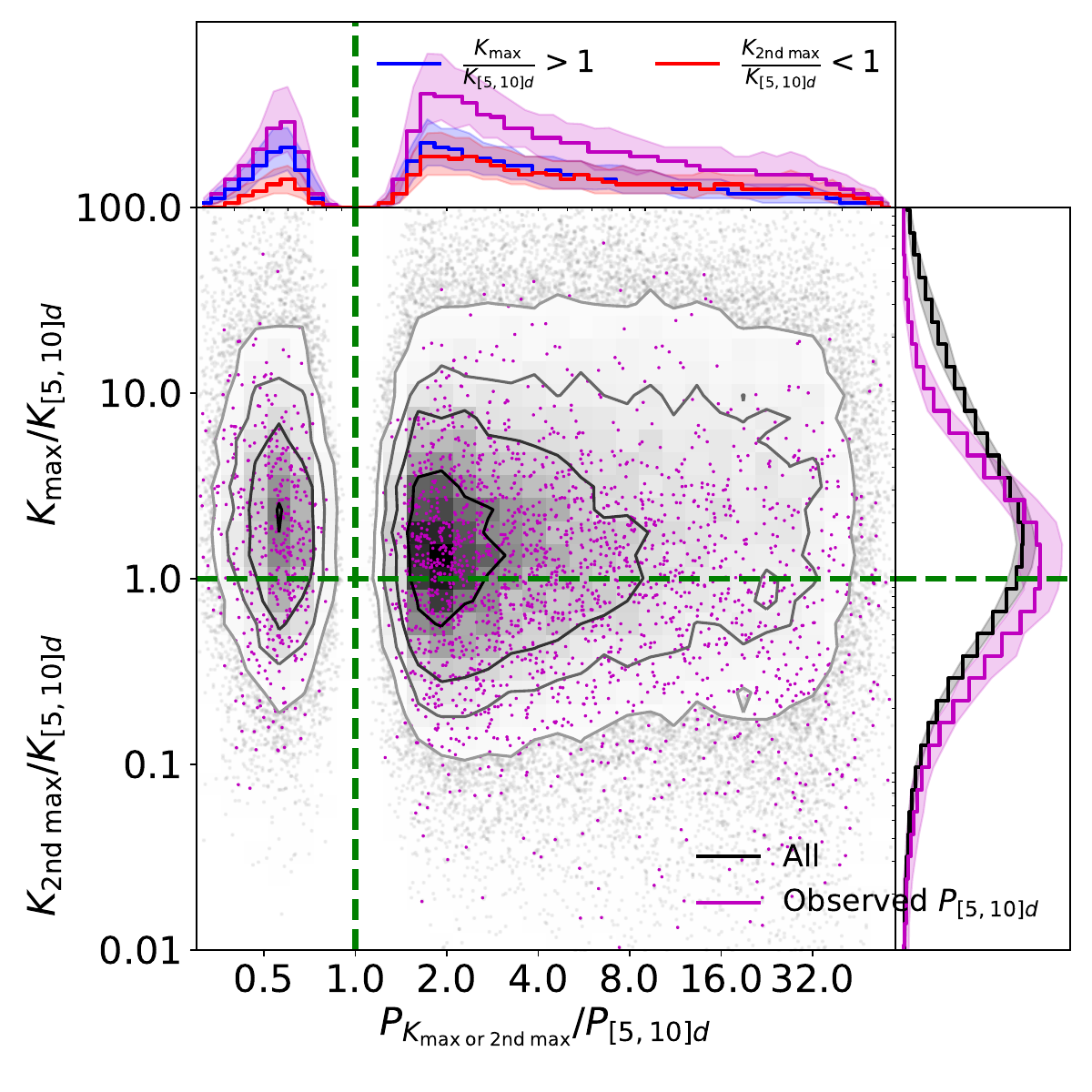}
\caption{Scatter plot of RV semi-amplitude ($K$) ratio vs. period ratio, for simulated physical catalogs drawn from our maximum AMD model. For each planet between $P=5-10$d, we plot either the ratio of the maximum $K$ in the system to the $K$ of the $P=5-10$d planet, $K_{\rm max}/K_{[5,10]d}$, if the $P=5-10$d planet is not the dominant RV signal (upper half; $K_{\rm max}/K_{[5,10]d} > 1$), or the ratio of the second maximum $K$ in the system to the $K$ of the $P=5-10$d planet, $K_{\rm 2nd\:max}/K_{[5,10]d}$, if the $P=5-10$d planet is the maximum $K$ (lower half; $K_{\rm 2nd\:max}/K_{[5,10]d} < 1$), on the $y$-axis. The $x$-axis denotes the period ratio of the planet with the maximum or second maximum $K$ relative to the $P=5-10$d planet (i.e, points to the left (right) of the vertical dashed line denote planets interior (exterior) to the $P=5-10$d planet). Grey points and contours show period and $K$ ratios for all planet pairs that include at least one planet with $P=5-10$d.  Systems containing two planets with period in $P=5-10$d appear as two dots. Magenta points show period and $K$ ratios only for planet pairs where the $P=5-10$d planet transits and would be detected by \Kepler{} in our simulations. Histograms to the top and right sides show the distributions of period ratios and $K$ ratios, respectively, where the shaded regions denote the 68.3\% credible regions from many simulated catalogs.}
\label{fig:RVratios}
\end{figure}

Radial velocity (RV) follow--up observations of stars with planets previously detected via the transit method can provide complementary information about the planet mass and bulk density.  
Additionally, RV observations may detect additional non--transiting or long--period planets.  
If observers knew that a star hosted exactly one planet, then they could choose observing times to measure the planet's mass efficiently (in terms of number of RV observations) by making use of the orbital period and phase measured by transit observations.  
However, this and other studies of the \Kepler{} population demonstrate that the typical planetary system contains multiple short--period planets.  
The planet with the largest RV amplitude ($K$) may not be the planet initially discovered via transits.  
Both the intrinsic architectures of planetary systems and the prevalence of multi--planet systems affect the yields of RV follow--up programs.
If multiple planets cause the host star to wobble with similar amplitudes and orbital periods, then the number of RV observations necessary to accurately characterize the mass of any single planet in the system may increase substantially.  
On the other hand, if the RV amplitudes are sufficiently disparate, then RV observations may measure the mass of the planet inducing the largest stellar wobble but not detect other planets in the system.  
When planning an RV observing program to follow--up transit discoveries, it would be very helpful to know how often the RV signal would be dominated by the transiting planet and how often a multiple planet system will induce a complex RV signal that necessitates many RV observations.

Predictions of our model can be used to inform the planning of RV follow--up campaigns, e.g., number of observations per star, number of stars to survey, and the selection of targets for RV follow--up based on properties of the transiting planet(s).
Additionally, the results of such RV follow--up can be compared to predictions of our model to test predictions about the distributions of orbital eccentricities, mutual inclinations, and orbital spacing of non--transiting planets.
Computing the predicted RV signatures of the full physical catalog could be useful for planning a ``blind'' RV survey, i.e., with targets chosen irrespective of any transiting planets.  
In order to make predictions for transit follow--up observations, one should instead consider the 
\textit{conditional} distribution of planetary architectures given the properties of the transiting planet.  
While the complex nature of our physical model and detection biases preclude an analytic solution, such conditional probabilities can be estimated by generating a large number of planetary systems with our forward model and selecting a subset of systems that include a planet closely matching the characteristics being conditioned on.

Unfortunately, RV follow--up of planets identified by \Kepler{} is often challenging due to the faintness (or other properties) of the host star.  This motivated NASA's TESS mission to perform an all--sky transit survey, so as to find many more transiting planets that would be amenable to follow--up observations.  
Recently, TESS began finding many transiting planet candidates with periods less than $\sim 10$ days around relatively nearby, bright stars, some of which will be prime targets for RV follow--up observations. The physical catalogs generated by our model can be analyzed to make predictions about the frequency and properties of additional planets conditional on the properties of the planet candidate(s) identified by TESS. A recent study by \citet{DA2020} has begun to explore this direction, albeit with a somewhat different approach (they fit parametric models, including a variation of our two-Rayleigh model, to the observed planets to infer the existence of additional planets, on a system-by-system basis). A complete analysis is beyond the scope of this paper, due to the large number of possible properties to condition on (e.g., number of transiting planet candidates, sizes and orbital periods of putative planets, properties of host star).

Here we demonstrate the forecasting capabilities of our model by computing the distribution of orbital periods and RV amplitudes conditional on there being a planet with an orbital period between 5 and 10 days, common among TESS planet candidates.  
In Figure \ref{fig:RVratios}, we present a scatter plot of RV amplitude ($K$) ratio vs. period ratio for all planets in such systems as grey points and contours.  The planet with $P=5-10$d serves as the reference and is in the denominator for both ratios.  
We overlay a scatter plot with magenta points after conditioning on the presence of a $P=5-10$d planet that would have been detected by \Kepler{}.  
In the latter case, the distribution of $K$ ratios shifts lower, because the orbit of the reference planet must be very nearly edge--on for it to have been observed in transit and more massive planets tend to be larger and thus easier to detect via transit.
We also show histograms for the distribution of the $K$ ratios and period ratios.
We find that for each $P=5-10$d planet in our physical catalogs, there is another planet with a greater RV amplitude in the same system $66\% \pm 2 \%$ of the time, but that this fraction decreases to $53\% \pm 3 \%$ if the $P=5-10$d planet is also transiting and detectable by \Kepler{}. This is also dependent on whether the other planet is interior or exterior to the transiting planet (left or right of the vertical dashed line, respectively). The largest $K$ planet is not the $P=5-10$d planet in $69_{-7}^{+4}\%$ of such systems if it is interior to the transiting planet, and about $50\% \pm 3 \%$ of the time if it is exterior. Thus, our results show that conditioning on transiting $P=5-10$d planets detected by \Kepler{}, about half or more of the time there will be another planet in the same system (transiting or not) that induces a larger RV amplitude.

\section{Conclusions} \label{Conclusions}

This study presents a state-of-the-art model for the distribution of planetary architectures in which all multi--planet systems are at the AMD--stability limit (the ``maximum AMD model''). 
This model provides an excellent fit to most \Kepler{} observations, as it retains many features from our previous clustered model (the ``two-Rayleigh model''; \citetalias{HFR2019}).
We summarize how the new model distinguishes itself from the two-Rayleigh model below.
\begin{itemize}[leftmargin=*]
 \item \textbf{The critical AMD for each planetary system is distributed amongst all the planets' eccentricities and mutual inclinations.} The critical AMD can be computed from a set of planet radii, masses, and periods following the conditions against collisions and MMR overlap \citep{LP2017, PLB2017}, which we have summarized in \S\ref{secMaxAMDmodel}.
 \item \textbf{This model provides a dynamically motivated and parameter--free distribution of eccentricities and mutual inclinations.}
 In particular, it does not assume a bimodal distribution of mutual inclinations. Only the eccentricity distribution of true single--planet systems is assumed to follow a Rayleigh distribution and treated separately.
 It provides a broader distribution of eccentricities for multiple planet systems.
 \item \textbf{The distributions of eccentricities ($e$) and mutual inclinations ($i_m$) are implicitly functions of the intrinsic planet multiplicity ($n$).} There is a strong inverse relationship between both $e$ and $n$, and $i_m$ and $n$; systems with more planets have lower eccentricities and mutual inclinations on average. The median eccentricities ($\tilde{\mu}_{e,n}$) and mutual inclinations ($\tilde{\mu}_{i,n}$) are well modeled by power-law functions of $n$. For the eccentricity distribution, we find that $\tilde{\mu}_{e,n} = \tilde{\mu}_{e,5}(n/5)^{\alpha_e}$ where $\tilde{\mu}_{e,5} = 0.031_{-0.003}^{+0.004}$ and $\alpha_e = -1.74_{-0.07}^{+0.11}$. For the mutual inclination distribution, we fit $\tilde{\mu}_{i,n} = \tilde{\mu}_{i,5}(n/5)^{\alpha_i}$ where $\tilde{\mu}_{i,5} = 1.10_{-0.11}^{+0.15}$ deg and $\alpha_i = -1.73_{-0.08}^{+0.09}$. The fit for the mutual inclinations is similar to but a shallower function than the result of \citet{Z2018}, who also assumed a power-law and found $\sigma_{i,5} = 0.8^\circ$ and $\alpha = -3.5$ (here $\sigma_{i,n}$ is a Rayleigh scale). However, they assumed a Rayleigh distribution for each $n$, whereas we find that the distributions of $e$ and $i_m$ for each $n$ are close to lognormal.
 
 The trends with intrinsic multiplicity arise from the strong correlation of the critical AMD and the minimum period ratio in the system. In simple terms, the AMD stability criteria requires that the total system AMD is sufficiently low such that no pair of planets can have crossing orbits given the entire AMD budget. As such, we show that the critical AMD is a strong function of the minimum system period ratio. This provides a simple dynamical explanation for the multiplicity-dependence of $e$ and $i_m$; systems with more planets tend to be more tightly spaced, allowing for a lower total AMD in order to remain stable.
 \item \textbf{The eccentricities and mutual inclinations of planets are highly correlated with each other.} On one hand, within a single planetary system, the AMD budget must be divided amongst all the $e$ and $i_m$ components for all the planets.  However, a stronger effect is that the wide distribution of total AMD (either for all systems or systems of a given intrinsic multiplicity) tends to cause planets in systems with large AMD to have both high $e$ and $i_m$. This correlation is not present in the two--Rayleigh model, where $e$ and $i_m$ are drawn independently.
\end{itemize}

\bigskip
Having found a physically motivated model for the distribution of planetary architectures that is consistent with most \Kepler{} observations, we explore the predictions and implications of this model. Our key conclusions are listed as follows:
\begin{itemize}[leftmargin=*]
 \item \textbf{Our maximum AMD model demonstrates 
 that the apparent \Kepler{} dichotomy can be resolved with a single population.}  While both our dichotomous (two--Rayleigh) and single--population (maximum AMD) models can match the observed multiplicity distribution (and the numerous other marginal distributions we have adopted) well, the new model incorporates a much more detailed stability criteria, produces more physically plausible systems, and requires fewer parameters, making it the preferred model.  
 Further, we show that the anti--correlation between mutual inclination and intrinsic multiplicity can be interpreted as a natural outcome of the planet formation process and does not require an {\em ad hoc} assumption for the mutual inclination distribution. Our model also improves upon the similar finding from \citet{Z2018} by incorporating a much more detailed model for the \Kepler{} detection and vetting efficiency.
 \item \textbf{The observed transit duration ratio ($\xi$) distributions of our models as a function of observed multiplicity are consistent with \Kepler{} observations (Figure \ref{fig:xi_per_mult}).} Namely, the $\log{\xi}$ distribution appears narrower and slightly more asymmetric around zero for higher $m$, as expected of planets with lower $e$ and $i_m$.
 However, some of this observed trend is due to observational biases, as systems with lower eccentricities and mutual inclinations also tend to be observed as higher multiplicity transiting systems.
 \item \textbf{Intrinsic single planets likely have larger eccentricities than those in multi-planet systems.}
 With all our distance functions, we find that their eccentricity scale is $\sigma_{e,1} \simeq 0.25$. The circular--normalized transit duration ($t_{\rm dur}/t_{\rm circ}$) distribution for observed singles is well fit by this scale, and is broader than that of observed multis, consistent with their broader $e$ distribution. The fairly large uncertainty in $\sigma_{e,1}$ (over $\pm 0.1$ for the central 68.3\%) is primarily due to the difficulty in precisely characterizing the eccentricity distribution of intrinsically single--planet systems, because most systems with a single planet detected in transit actually contain multiple planets.
 \item \textbf{It is very difficult to characterize the population of intrinsically single systems via transit surveys, because most planetary systems with a single detectable transiting planet harbor additional undetected planets.} According to our maximum AMD model, the fraction of observed singles that are true singles (between 3 and 300 days) is just $7.7_{-4.6}^{+5.6}\%$.  The high rate of multiple planet systems is partially due to our models including planets down to $R_{p,\rm min} = 0.5 R_\oplus$, which are too small to have been detected around most \Kepler{} target stars even if they were transiting. Considering only planets larger than $1 R_\oplus$ ($2 R_\oplus$), the fraction rises to $16_{-7}^{+7}\%$ ($35_{-9}^{+8}\%$). Thus, the properties of the population of intrinsically single planets (over the period range 3--300 days) are especially difficult to probe as they cannot be easily disentangled from the multi-planet systems.  This affects their period--radius distribution, as well as their eccentricity distribution.
 \item \textbf{We find evidence supporting the ``peas in a pod'' trends} \citep{W2018a,WP2019}.  Adopting similar system--level metrics as defined in \citet{GF2020}, we show that our models provide an excellent fit to the observed \textit{radius partitioning} ($\mathcal{Q}_R$) distribution due to the clustered planet radii. Assuming no underlying monotonicity trend, we find that the simulated observed \textit{radius monotonicity} ($\mathcal{M}_R$) distribution produces only a slight preference for positive monotonicity ($58\% \pm 3 \%$ of systems). However, this effect is not nearly as strong as that of the \Kepler{} data (70\% of the systems exhibit positive monotonicity), suggesting that real planetary systems exhibit preferential size ordering. Finally, the distribution of \textit{gap complexity} ($\mathcal{C}$) observed for \Kepler{} systems is significantly more weighted toward low values than those of our simulated catalogs, implying that planets in \Kepler{}'s multiple planet systems are substantially more uniformly spaced than those in our model (which includes clustering of orbital periods, but does not enforce uniformity of spacing within a cluster).
 \item \textbf{Our results are insensitive to assumptions for the level of dynamical excitation, as parameterized by $f_{\rm crit} = {\rm AMD}_{\rm tot}/{\rm AMD}_{\rm crit}$.} Values of $f_{\rm crit}$ below 0.5 are disfavored based on \Kepler{} observations (e.g., transit durations and duration ratios). While $f_{\rm crit}$ larger than 2 may still provide an adequate fit to \Kepler{} observations, they are unlikely due to considerations of long-term dynamical stability.
 \item \textbf{About half of all transiting planets between 5 and 10d detected by \Kepler{} are in systems where another planet dominates the RV signal.} We use the physical catalogs drawn from our maximum AMD model to compute the RV amplitudes $K$ and conditional probabilities of other dominant RV planets given short--period transiting planets. For planets in $P=5-10$d detectable by \Kepler{}, we find that in $53\% \pm 3 \%$ of such systems, there is another planet with a larger $K$. Most of the time, this planet is exterior to the transiting planet.
\end{itemize}

The new catalogs generated from our models are available to the public, along with the core SysSim code (\url{https://github.com/ExoJulia/ExoplanetsSysSim.jl}), inputs collated from numerous data files (\url{https://github.com/ExoJulia/SysSimData}), and the code specific to the clustered models (\url{https://github.com/ExoJulia/SysSimExClusters}). We encourage other researchers to contribute model extensions via Github pull requests and/or additional public git repositories.

\acknowledgments

We thank the entire \Kepler{} team for years of work leading to a successful mission and data products critical to this study.  
We acknowledge many valuable contributions with members of the \Kepler{} Science Team's working groups on multiple body systems, transit timing variations, and completeness working groups.  
%We thank an anonymous referee for useful suggestions for improving the analysis and manuscript.  
We thank Keir Ashby, Danley Hsu, and Robert Morehead for contributions to the broader SysSim project.  
We thank Derek Bingham, Earl Lawrence, Ilya Mandell, Dan Fabrycky, Gregory Gilbert, Jack Lissauer, Gijs Mulders, Antoine Petit, Daniel Tamayo, and Sarah Millholland for useful discussions.

M.Y.H. acknowledges the support of the Natural Sciences and Engineering Research Council of Canada (NSERC), funding reference number PGSD3 - 516712 - 2018.
E.B.F. and D.R. acknowledge support from NASA Origins of Solar Systems grant \# NNX14AI76G and Exoplanet Research Program grant \# NNX15AE21G. 
E.B.F. acknowledges support from NASA Kepler Participating Scientist Program, grant \# NNX08AR04G, \# NNX12AF73G, and \# NNX14AN76G.
This work was supported by a grant from the Simons Foundation/SFARI (675601, E.B.F.).
E.B.F. acknowledges the support of the Ambrose Monell Foundation and the Institute for Advanced Study.
M.Y.H. and E.B.F. acknowledge support from the Penn State Eberly College of Science and Department of Astronomy \& Astrophysics, the Center for Exoplanets and Habitable Worlds, and the Center for Astrostatistics.  
E.B.F. acknowledges support and collaborative scholarly discussions during  residency at the Research Group on Big Data and Planets at the Israel Institute for Advanced Studies.  

The citations in this paper have made use of NASA's Astrophysics Data System Bibliographic Services.  
This research has made use of the NASA Exoplanet Archive, which is operated by the California Institute of Technology, under contract with the National Aeronautics and Space Administration under the Exoplanet Exploration Program.
This work made use of the stellar catalog from \citet{H2019} and thus indirectly the gaia-kepler.fun crossmatch database created by Megan Bedell.
Several figures in this manuscript were generated using the \texttt{corner.py} package \citep{Fm2016}.
We acknowledge the Institute for Computational and Data Sciences (\url{http://icds.psu.edu/}) at The Pennsylvania State University, including the CyberLAMP cluster supported by NSF grant MRI-1626251, for providing advanced computing resources and services that have contributed to the research results reported in this paper.
This study benefited from the 2013 SAMSI workshop on Modern Statistical and Computational Methods for Analysis of \Kepler{} Data, the 2016/2017 Program on Statistical, Mathematical and Computational Methods for Astronomy, and their associated working groups.
This material was based upon work partially supported by the National Science Foundation under grant DMS-1127914 to the Statistical and Applied Mathematical Sciences Institute (SAMSI). Any opinions, findings, and conclusions or recommendations expressed in this material are those of the author(s) and do not necessarily reflect the views of the National Science Foundation.

\software{ExoplanetsSysSim \citep{F2018b},
          SysSimData \citep{F2019},
          Numpy \citep{Numpy2011},
          Matplotlib \citep{Matplotlib2007},
          Corner.py \citep{Fm2016}
          }


\bibliography{sample63}{}
\bibliographystyle{aasjournal}



\appendix

% To prepend the letter 'A' for Appendix figures and tables:
\renewcommand{\thefigure}{A\arabic{figure}}
\renewcommand{\thetable}{A\arabic{table}}
\setcounter{figure}{0}
\setcounter{table}{0}

\begin{figure*}
\centering
\begin{tabular}{cc}
 \includegraphics[scale=0.28,trim={1.5cm 0.2cm 0.5cm 0.2cm},clip]{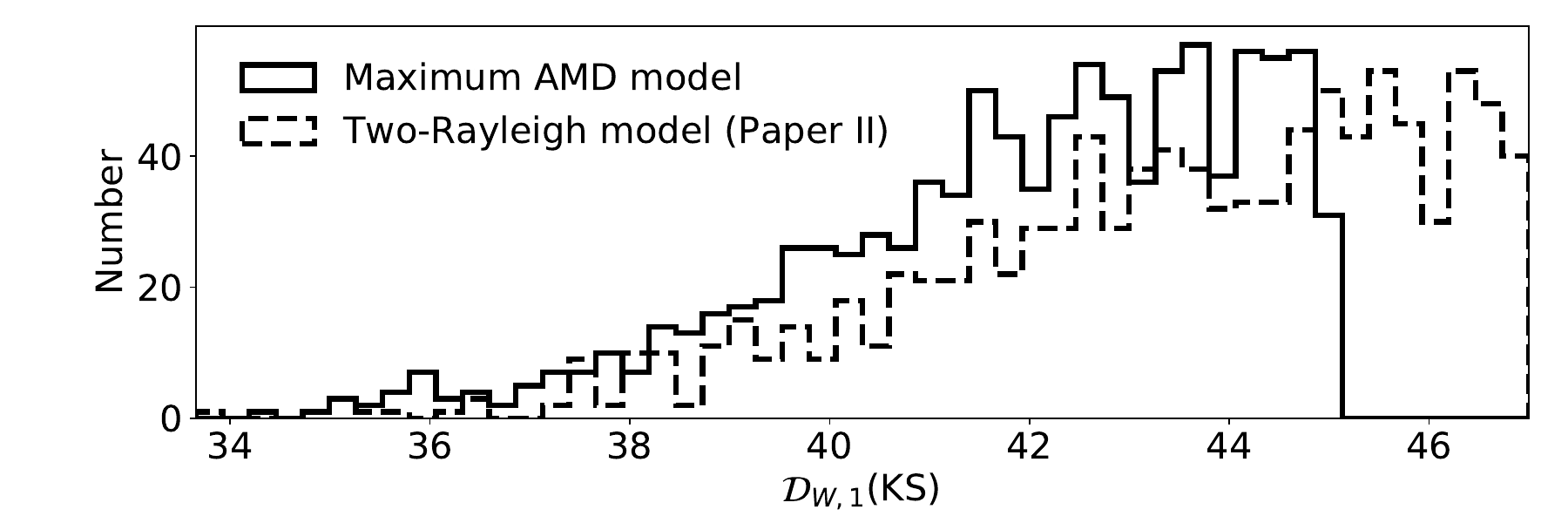} &
 \includegraphics[scale=0.28,trim={1.5cm 0.2cm 0.5cm 0.2cm},clip]{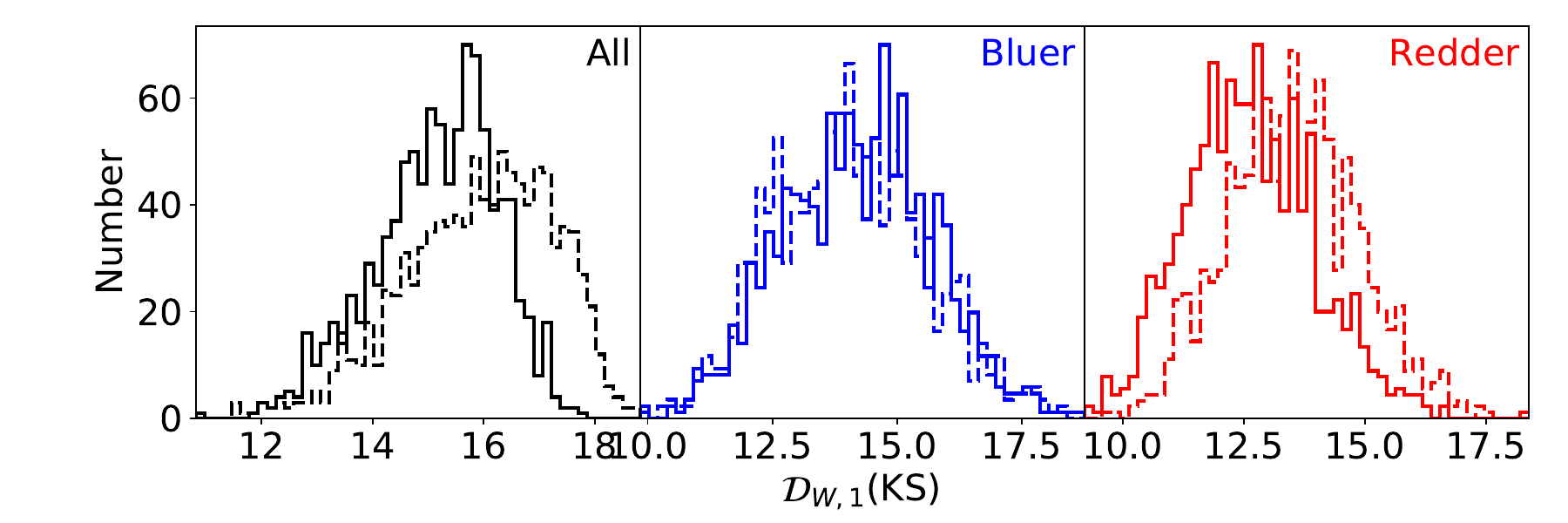} \\
\end{tabular}
\begin{tabular}{ccc}
 \includegraphics[scale=0.28,trim={0.2cm 0.2cm 0.4cm 0.2cm},clip]{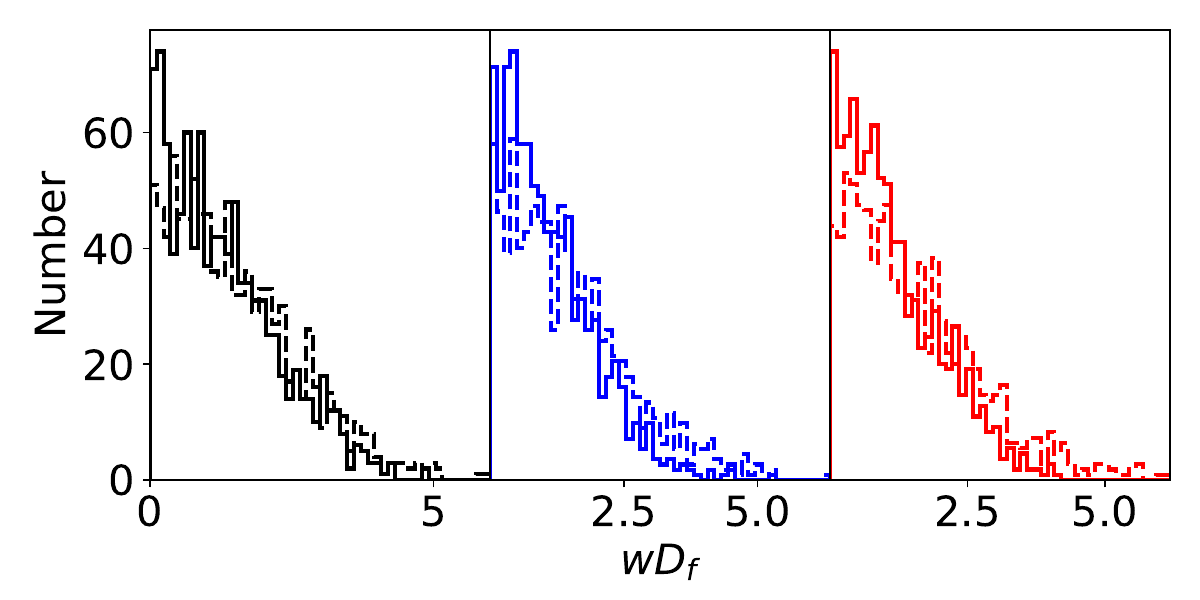} &
 \includegraphics[scale=0.28,trim={0.2cm 0.2cm 0.4cm 0.2cm},clip]{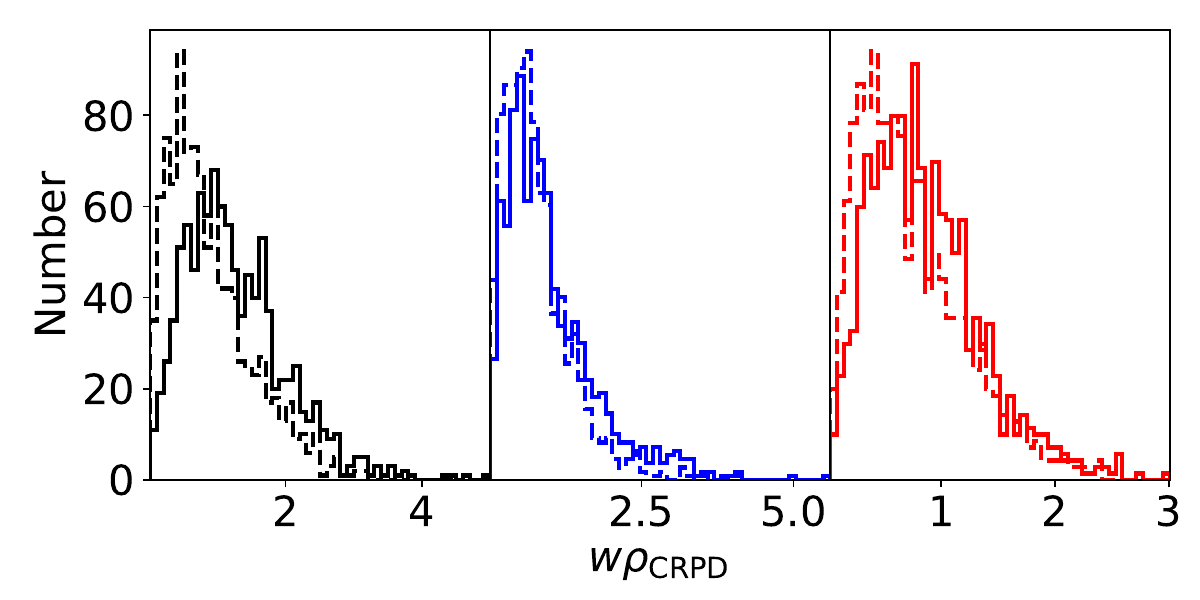} &
 \includegraphics[scale=0.28,trim={0.2cm 0.2cm 0.4cm 0.2cm},clip]{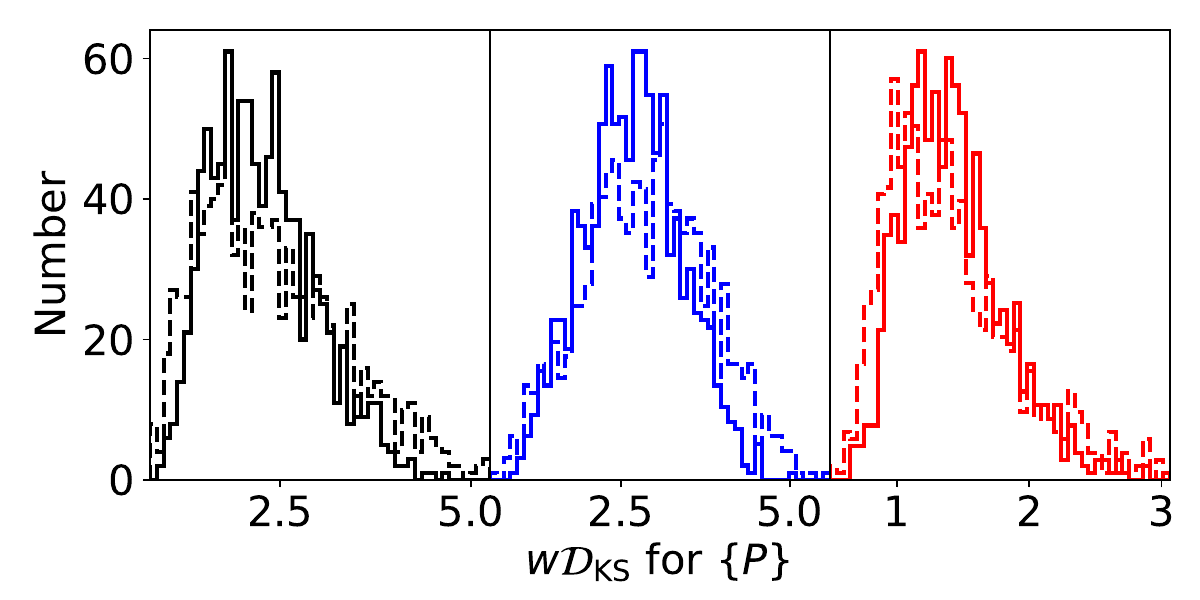} \\
 \includegraphics[scale=0.28,trim={0.2cm 0.2cm 0.4cm 0.2cm},clip]{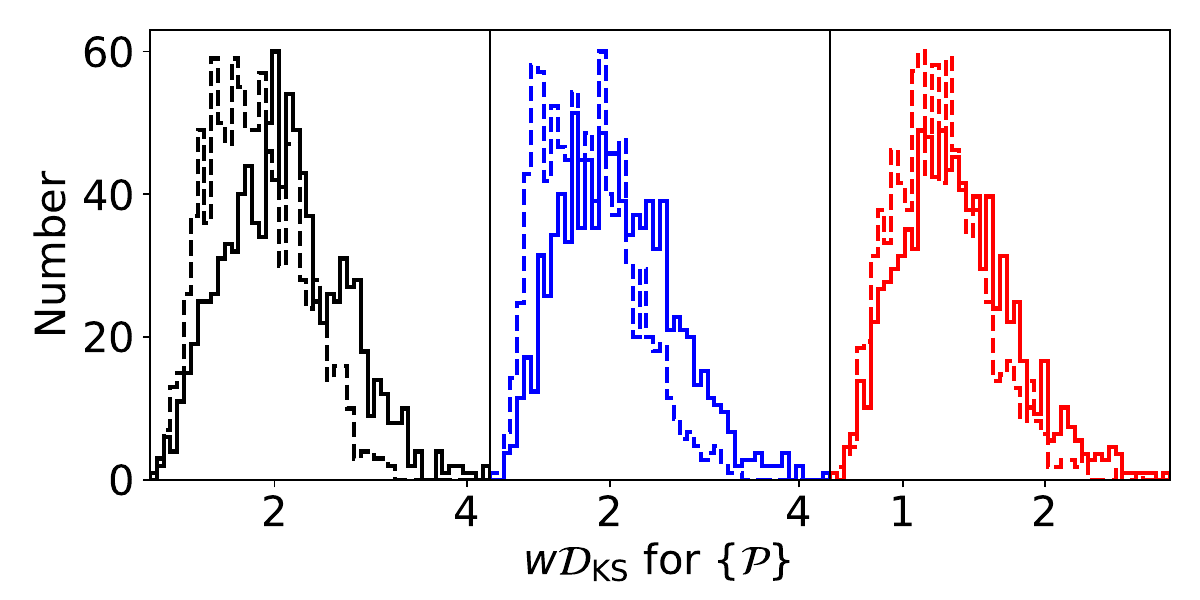} &
 \includegraphics[scale=0.28,trim={0.2cm 0.2cm 0.4cm 0.2cm},clip]{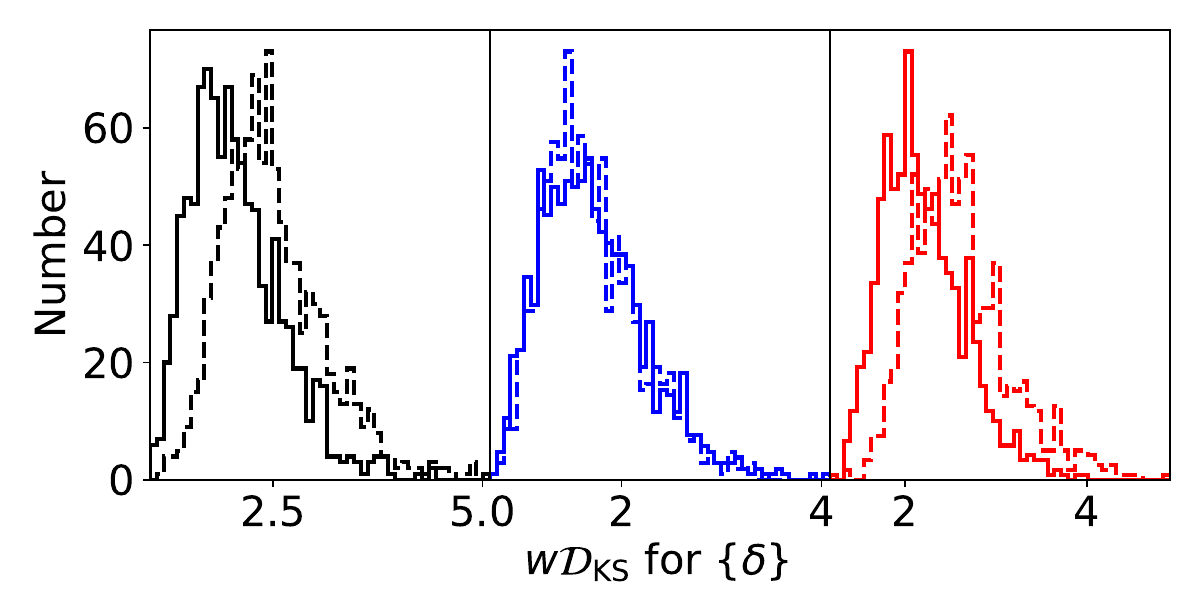} &
 \includegraphics[scale=0.28,trim={0.2cm 0.2cm 0.4cm 0.2cm},clip]{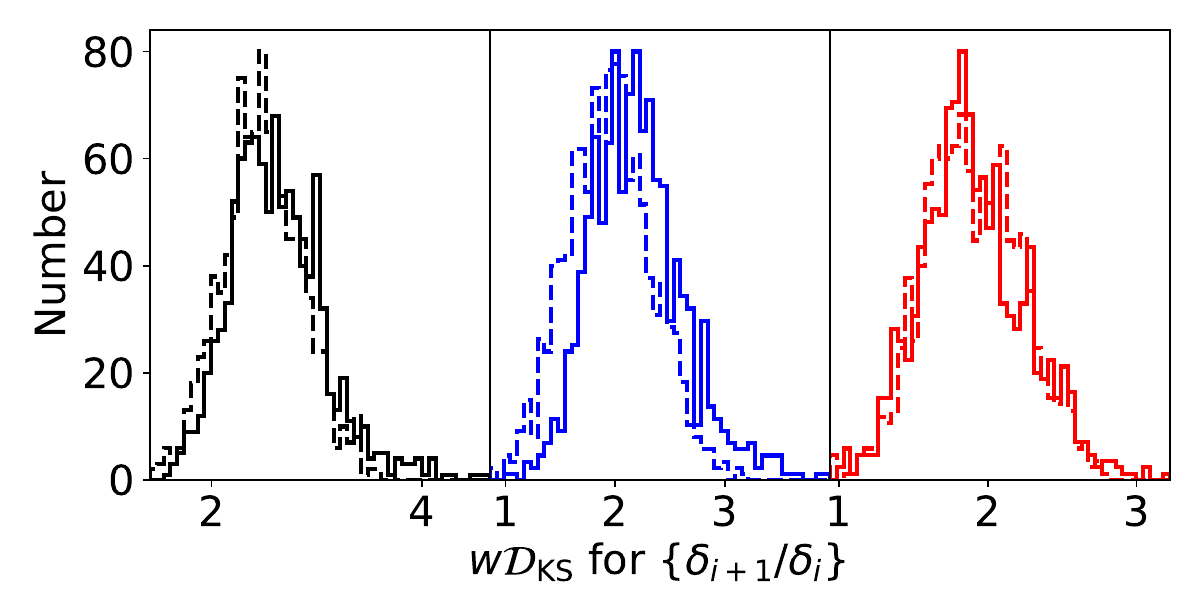} \\
 \includegraphics[scale=0.28,trim={0.2cm 0.2cm 0.4cm 0.2cm},clip]{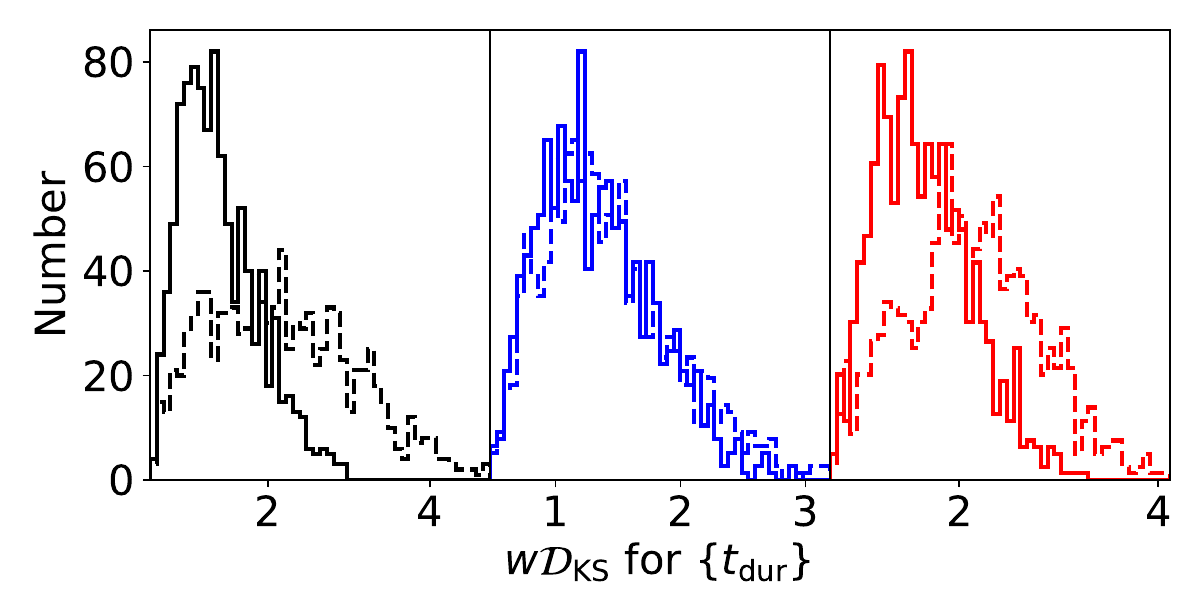} &
 \includegraphics[scale=0.28,trim={0.2cm 0.2cm 0.4cm 0.2cm},clip]{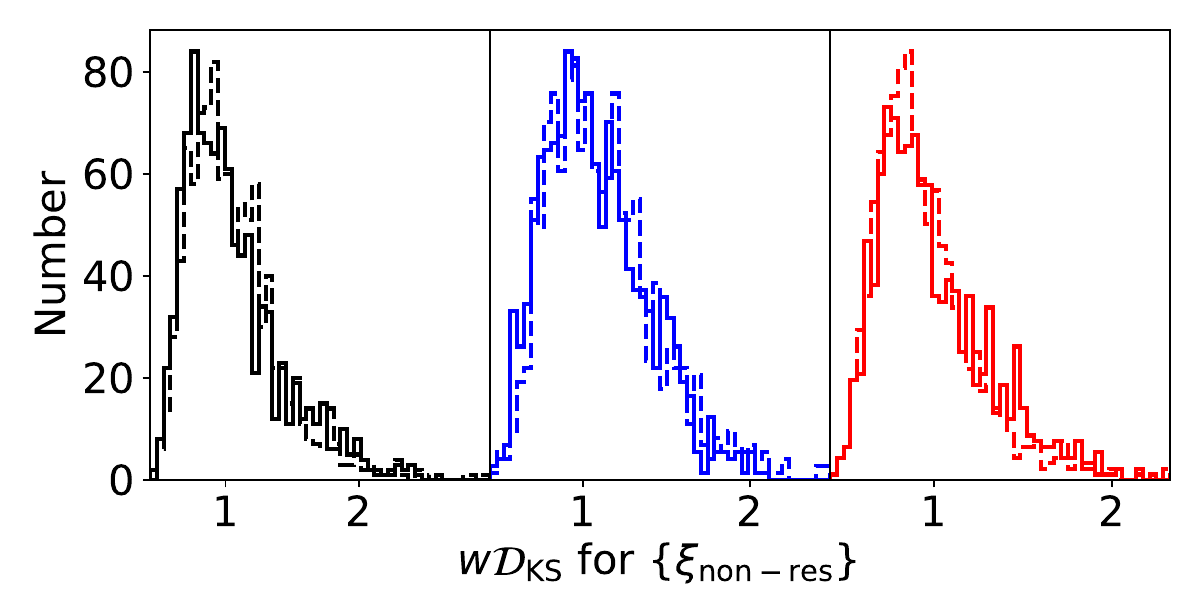} &
 \includegraphics[scale=0.28,trim={0.2cm 0.2cm 0.4cm 0.2cm},clip]{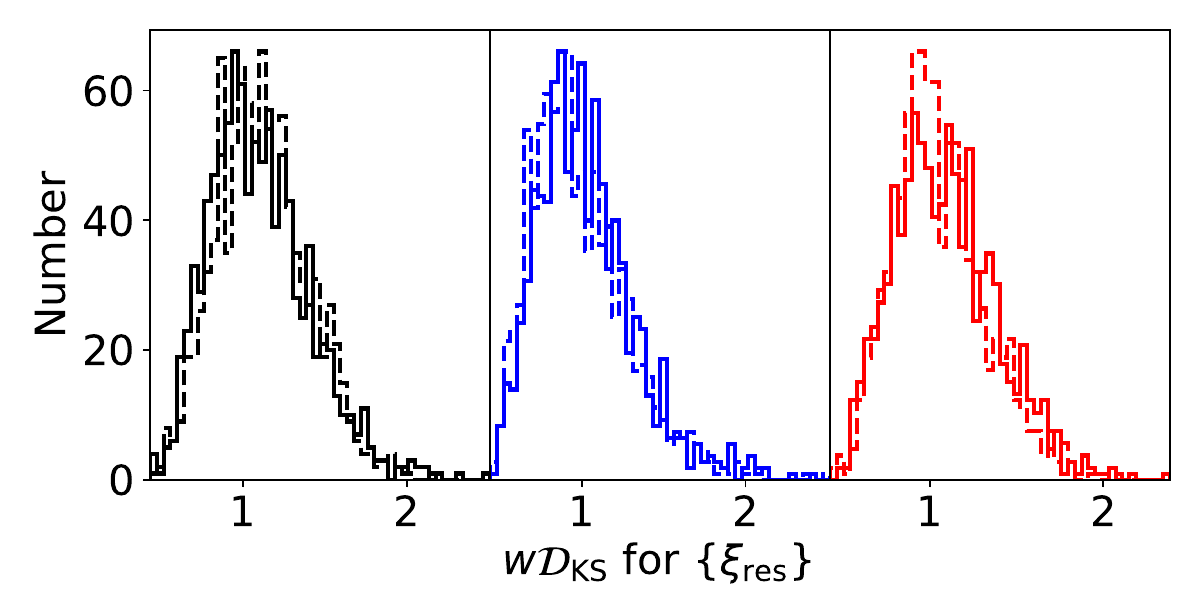} \\
\end{tabular}
\caption{Histograms of the weighted total distances using our $\mathcal{D}_{W,1}$ (KS) distance function (\textbf{top row}) and individual distances (\textbf{second row and below}), for the two-Rayleigh model (dashed lines) and the maximum AMD model (solid lines). 1000 simulated catalogs passing our distance thresholds ($\mathcal{D}_{W,1} = 47$ and 45) are included for each model.}
\label{fig:dists1_KS}
\end{figure*}

\begin{figure*}
\centering
\begin{tabular}{cc}
 \includegraphics[scale=0.28,trim={1.5cm 0.2cm 0.5cm 0.2cm},clip]{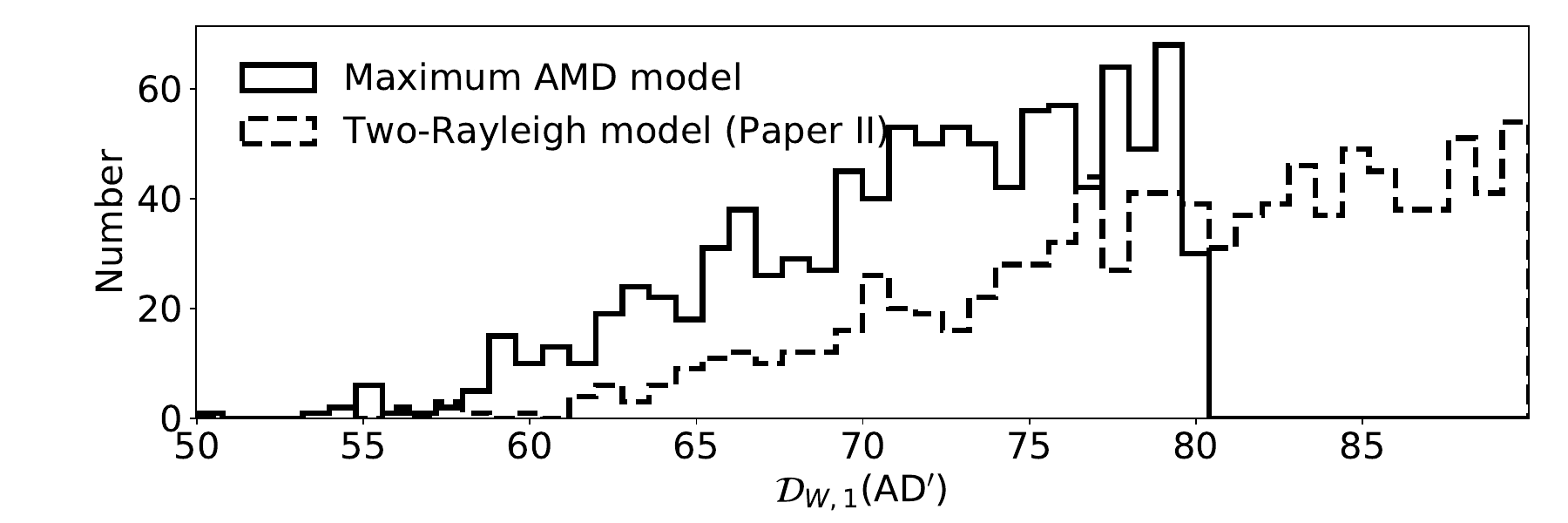} &
 \includegraphics[scale=0.28,trim={1.5cm 0.2cm 0.5cm 0.2cm},clip]{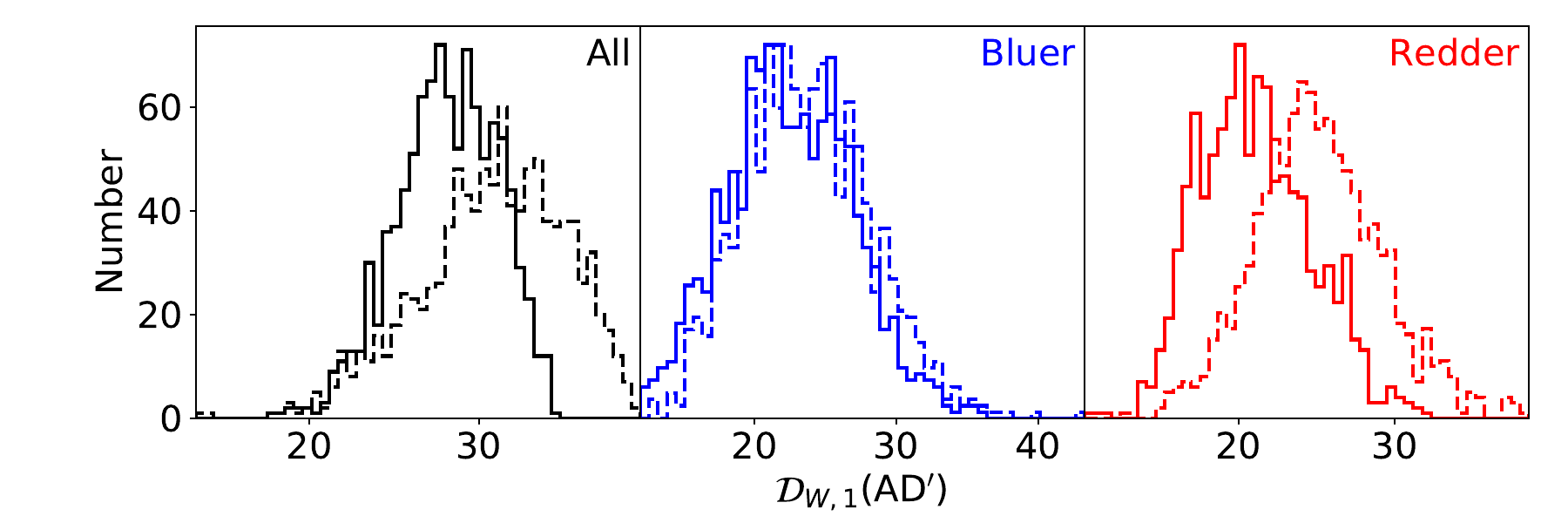} \\
\end{tabular}
\begin{tabular}{ccc}
 \includegraphics[scale=0.28,trim={0.2cm 0.2cm 0.4cm 0.2cm},clip]{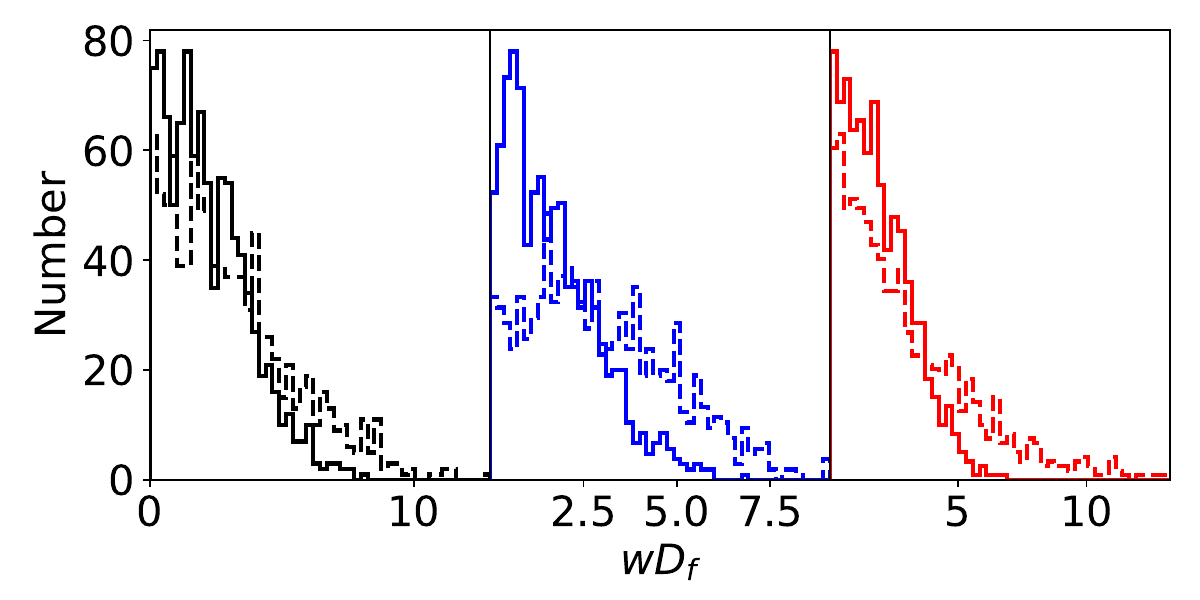} &
 \includegraphics[scale=0.28,trim={0.2cm 0.2cm 0.4cm 0.2cm},clip]{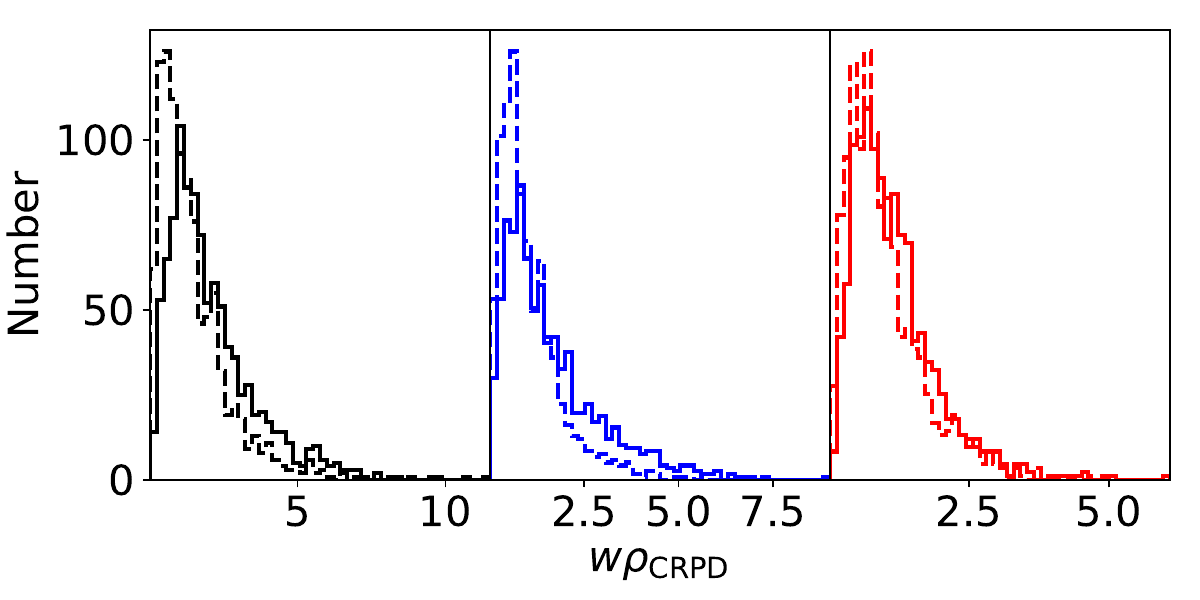} &
 \includegraphics[scale=0.28,trim={0.2cm 0.2cm 0.4cm 0.2cm},clip]{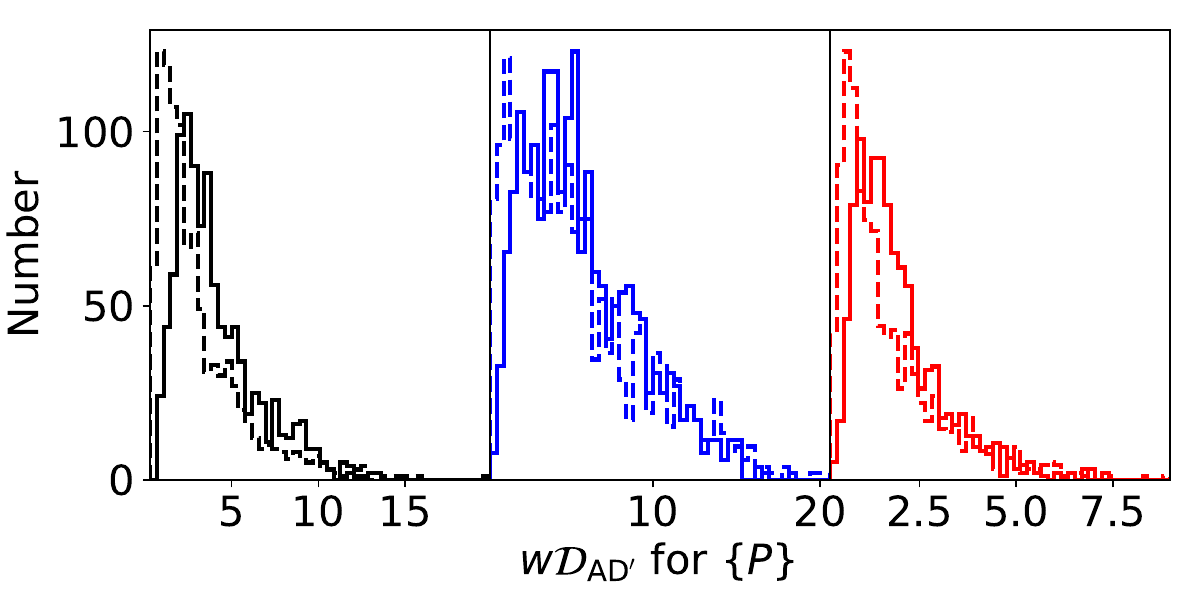} \\
 \includegraphics[scale=0.28,trim={0.2cm 0.2cm 0.4cm 0.2cm},clip]{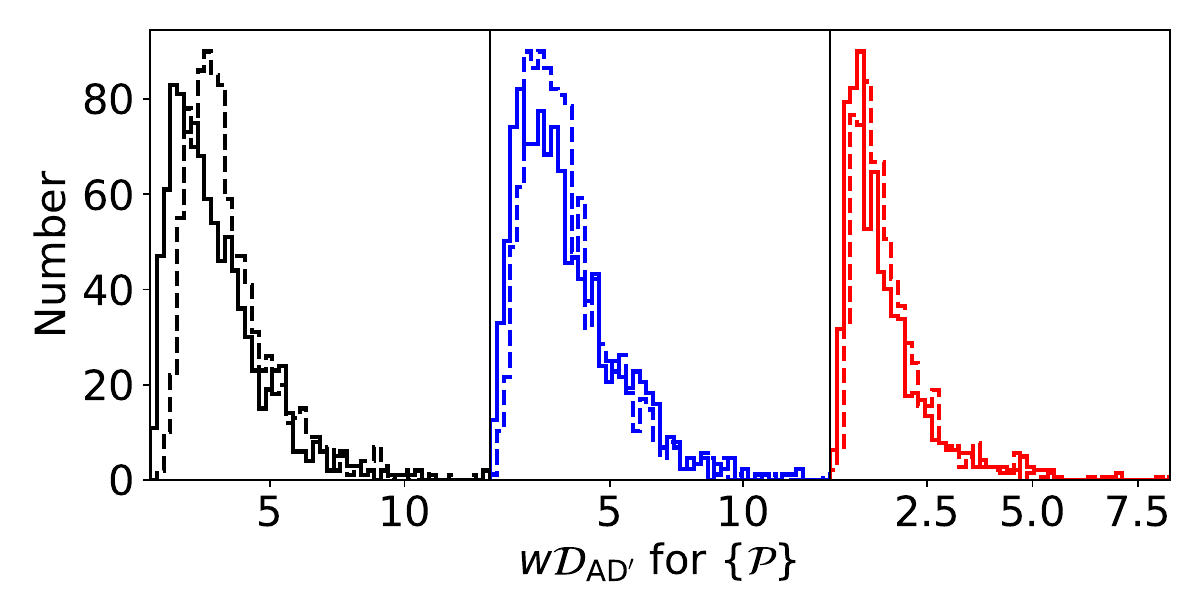} &
 \includegraphics[scale=0.28,trim={0.2cm 0.2cm 0.4cm 0.2cm},clip]{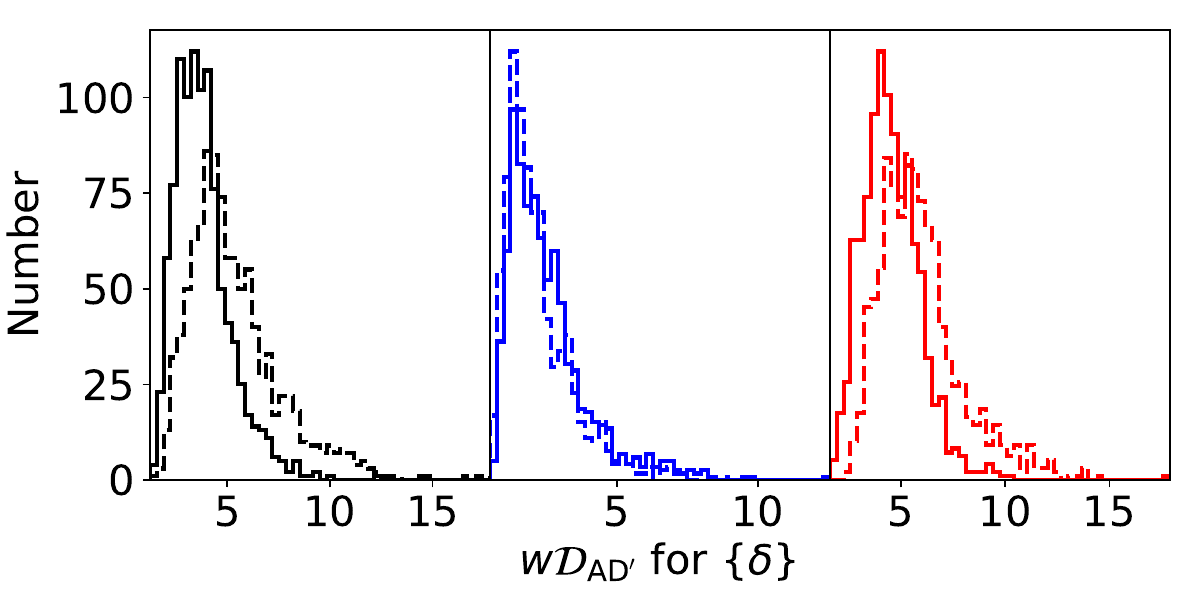} &
 \includegraphics[scale=0.28,trim={0.2cm 0.2cm 0.4cm 0.2cm},clip]{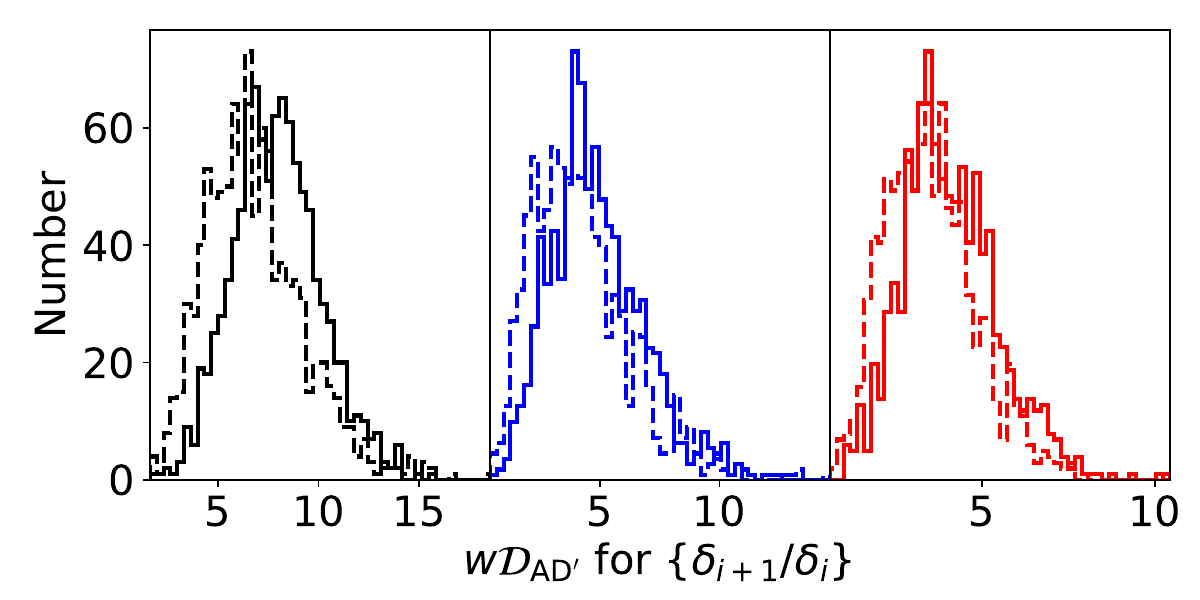} \\
 \includegraphics[scale=0.28,trim={0.2cm 0.2cm 0.4cm 0.2cm},clip]{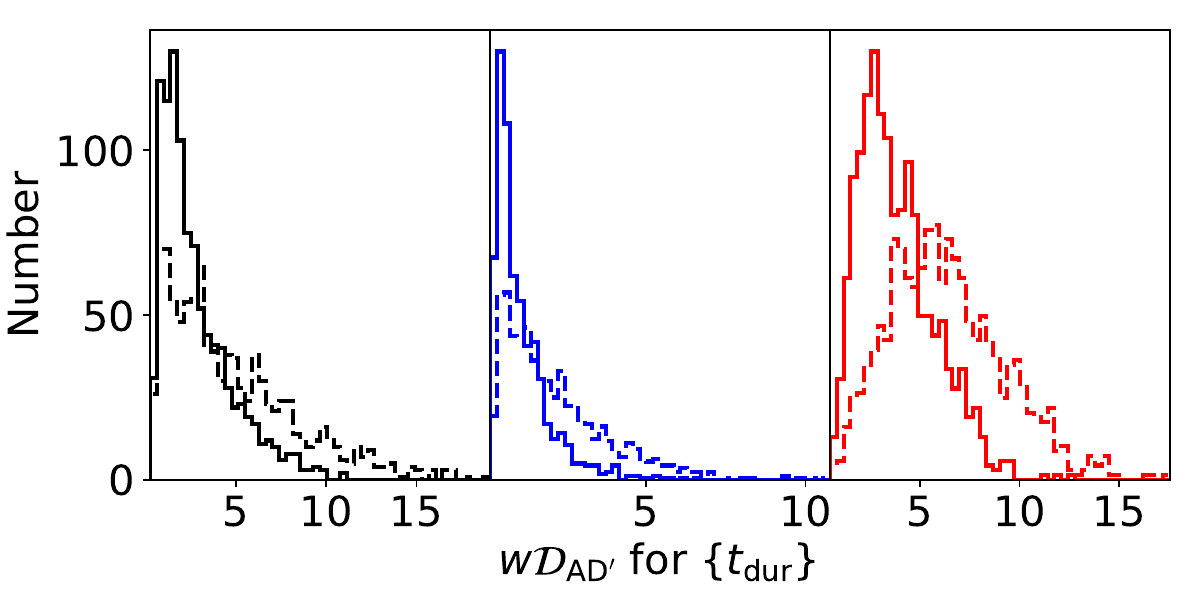} &
 \includegraphics[scale=0.28,trim={0.2cm 0.2cm 0.4cm 0.2cm},clip]{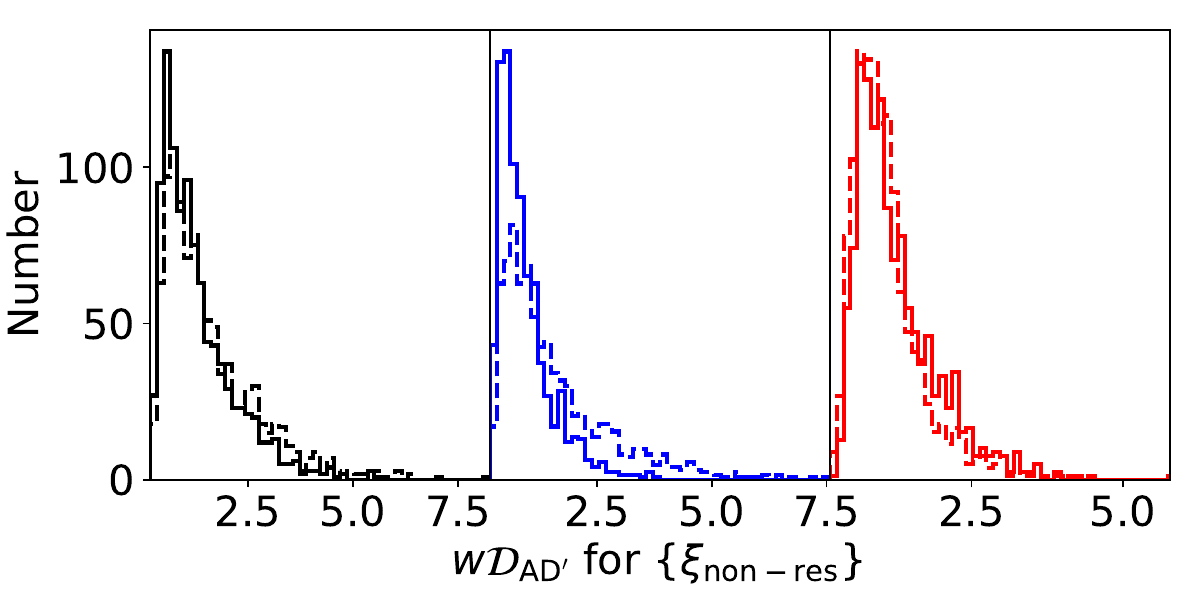} &
 \includegraphics[scale=0.28,trim={0.2cm 0.2cm 0.4cm 0.2cm},clip]{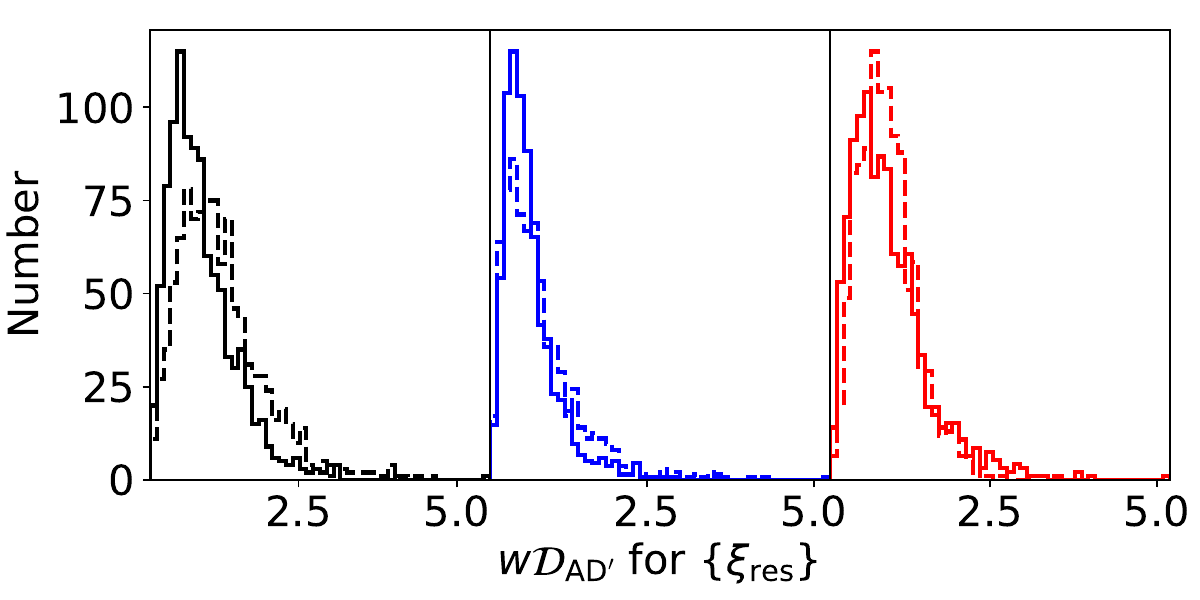} \\
\end{tabular}
\caption{Histograms of the weighted total distances using our $\mathcal{D}_{W,1}$ (AD) distance function (\textbf{top row}) and individual distances (\textbf{second row and below}), for the two-Rayleigh model (dashed lines) and the maximum AMD model (solid lines). 1000 simulated catalogs passing our distance threshold ($\mathcal{D}_{W,1} = 90$ and 80) are included for each model.}
\label{fig:dists1_AD}
\end{figure*}

\begin{figure*}
\centering
\begin{tabular}{cc}
 \includegraphics[scale=0.28,trim={1.5cm 0.2cm 0.5cm 0.2cm},clip]{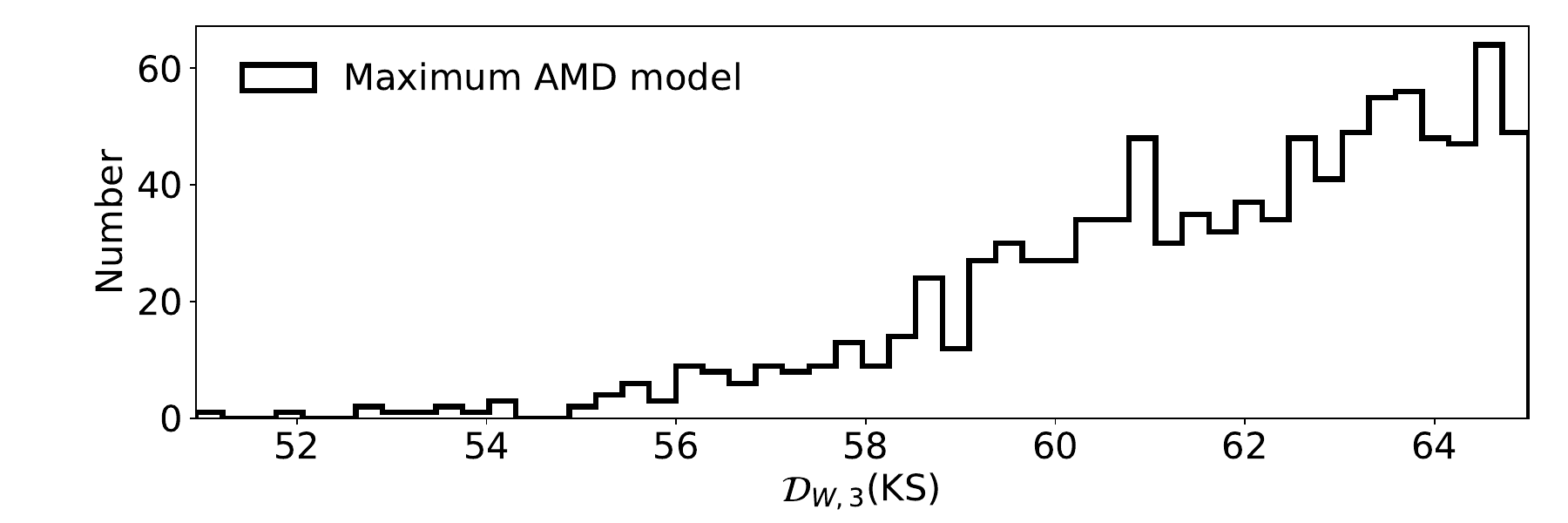} &
 \includegraphics[scale=0.28,trim={1.5cm 0.2cm 0.5cm 0.2cm},clip]{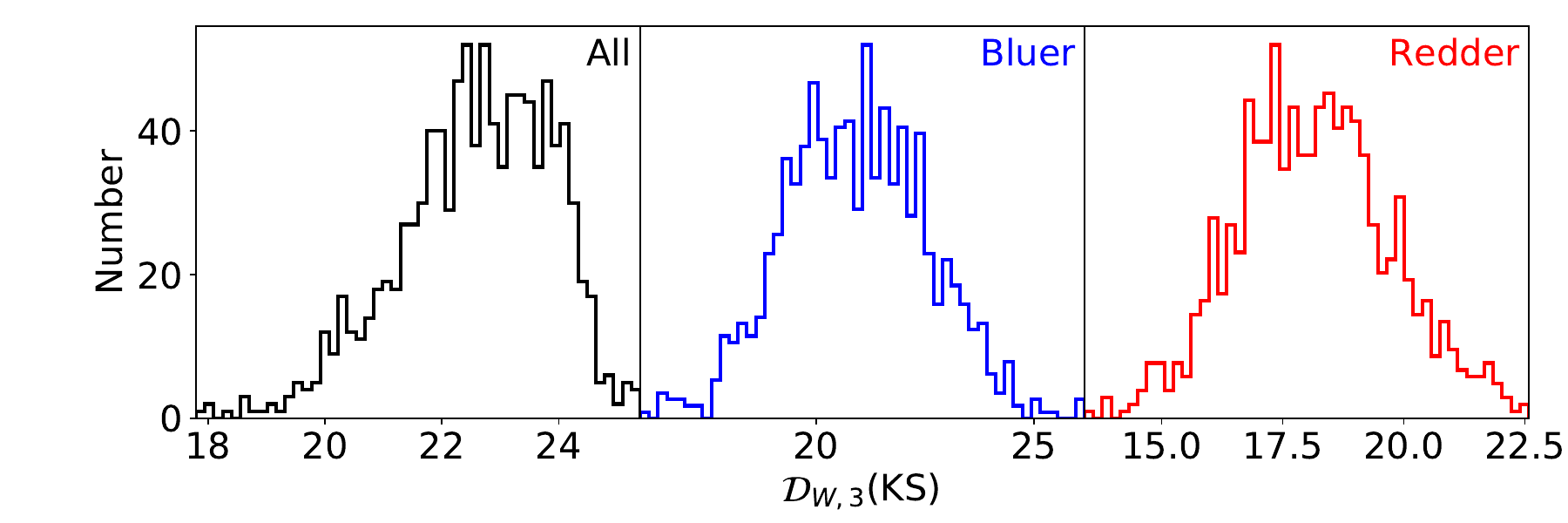} \\
 \includegraphics[scale=0.28,trim={0.6cm 0.2cm 0.4cm 0.2cm},clip]{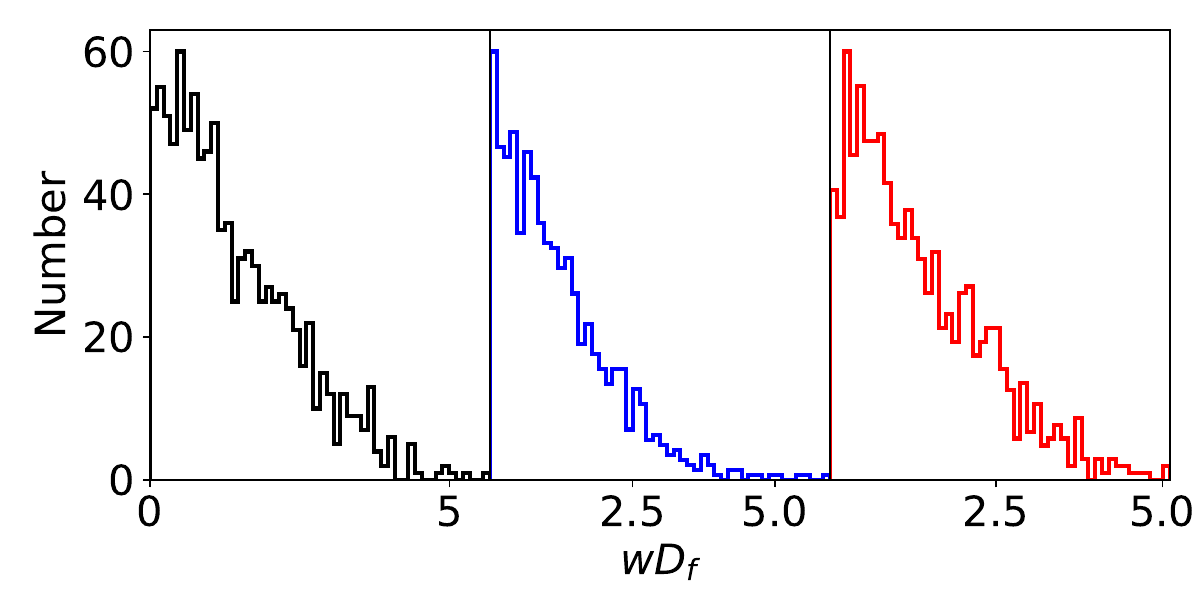} &
 \includegraphics[scale=0.28,trim={0.6cm 0.2cm 0.4cm 0.2cm},clip]{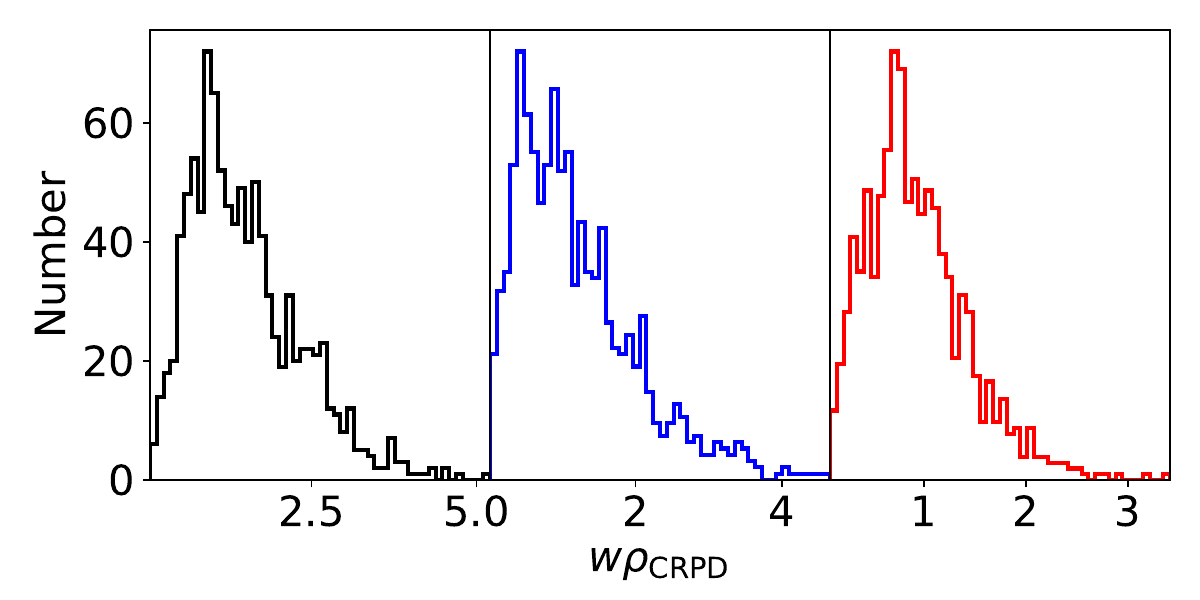} \\
\end{tabular}
\begin{tabular}{ccc}
 \includegraphics[scale=0.28,trim={0.2cm 0.2cm 0.4cm 0.2cm},clip]{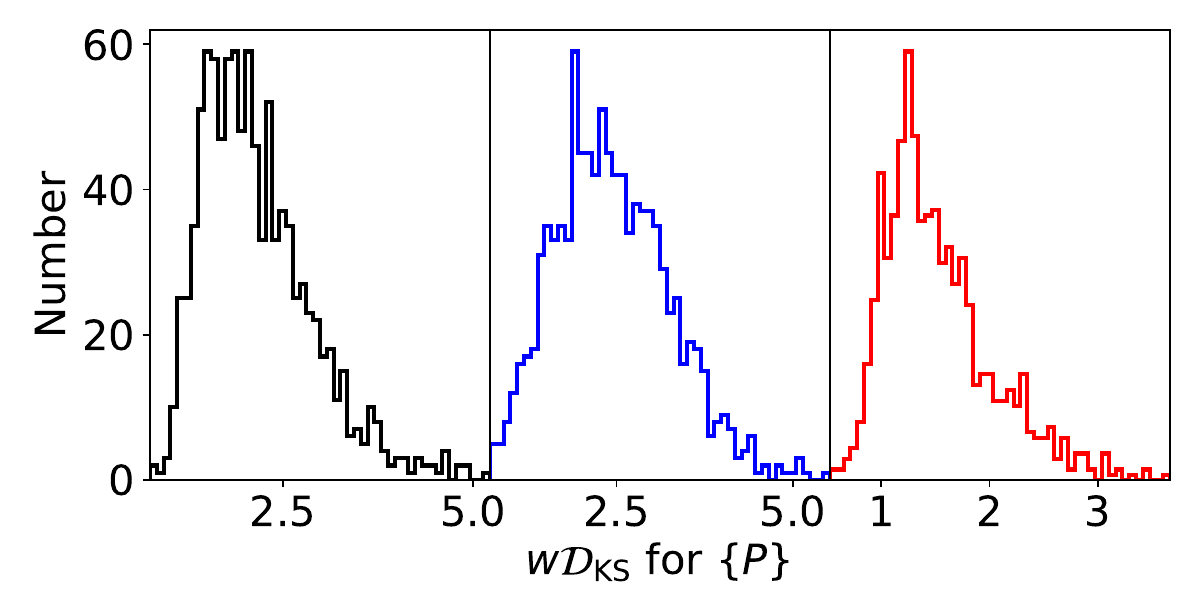} &
 \includegraphics[scale=0.28,trim={0.2cm 0.2cm 0.4cm 0.2cm},clip]{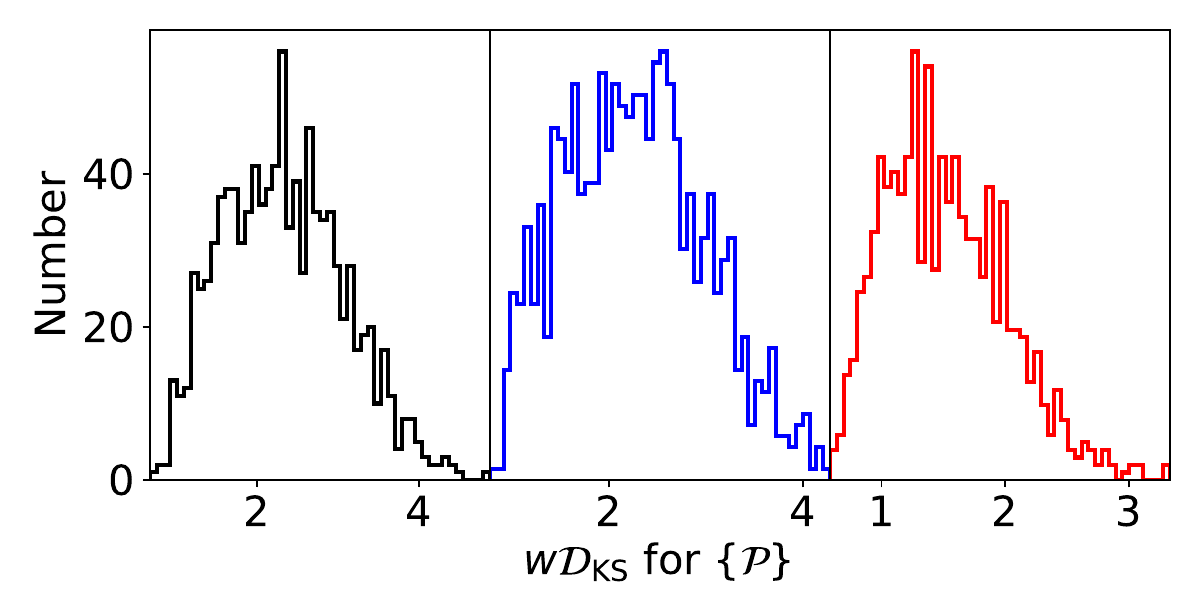} &
 \includegraphics[scale=0.28,trim={0.2cm 0.2cm 0.4cm 0.2cm},clip]{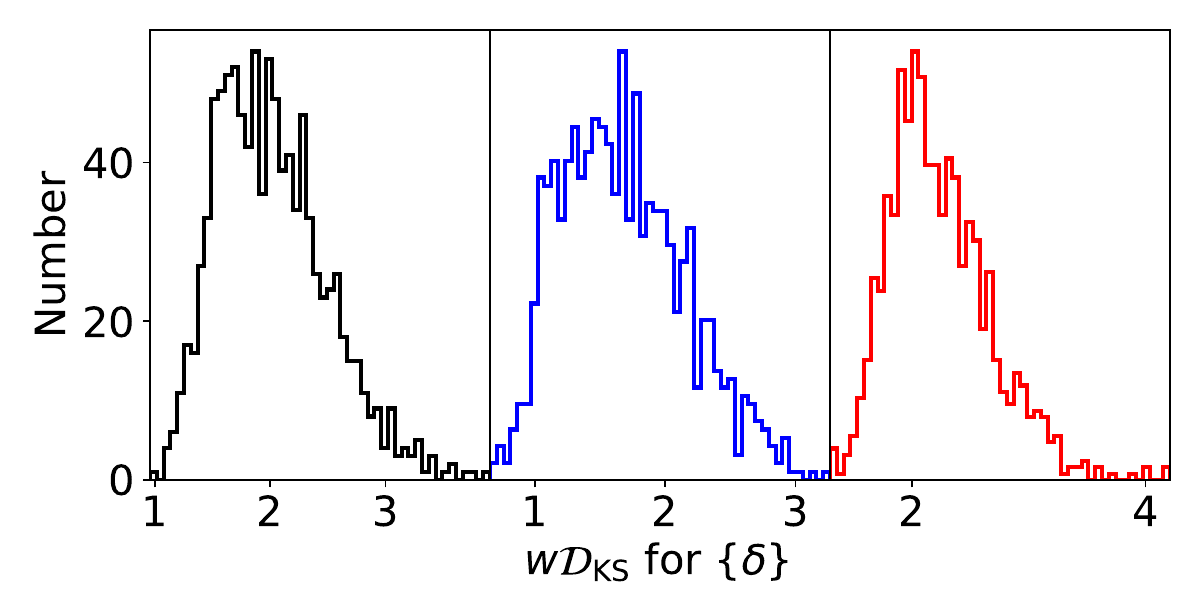} \\
 \includegraphics[scale=0.28,trim={0.2cm 0.2cm 0.4cm 0.2cm},clip]{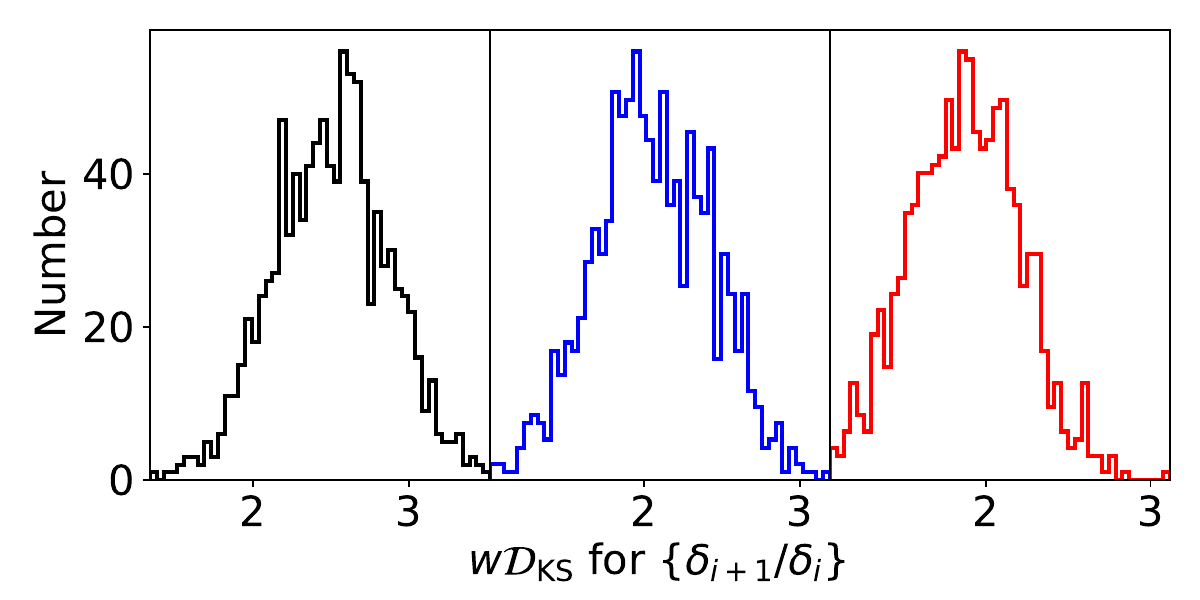} &
 \includegraphics[scale=0.28,trim={0.2cm 0.2cm 0.4cm 0.2cm},clip]{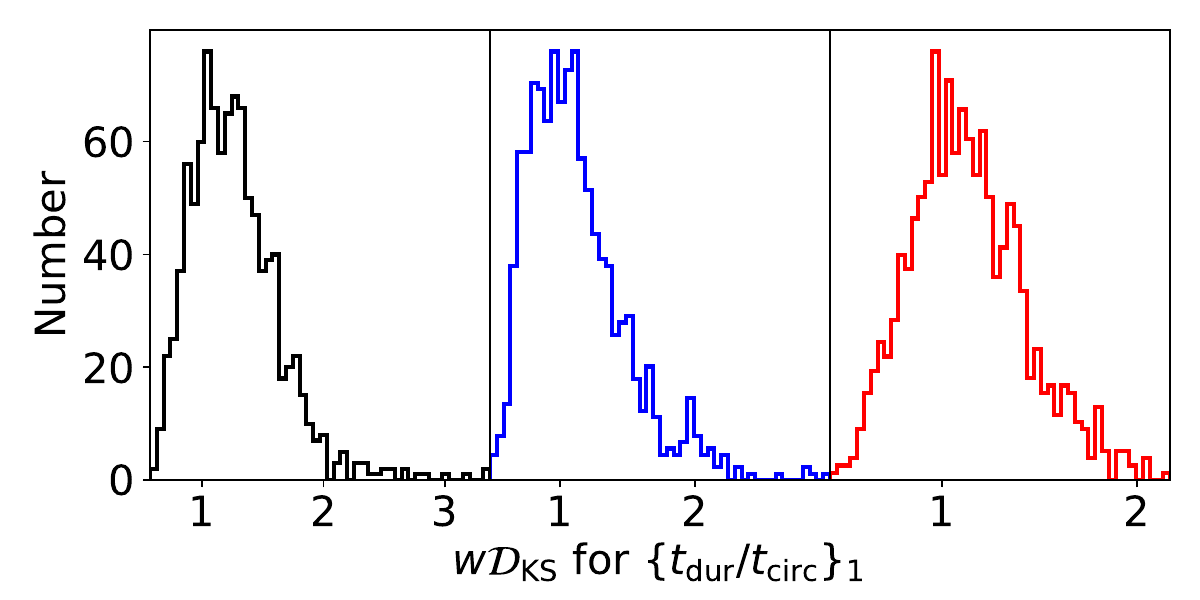} &
 \includegraphics[scale=0.28,trim={0.2cm 0.2cm 0.4cm 0.2cm},clip]{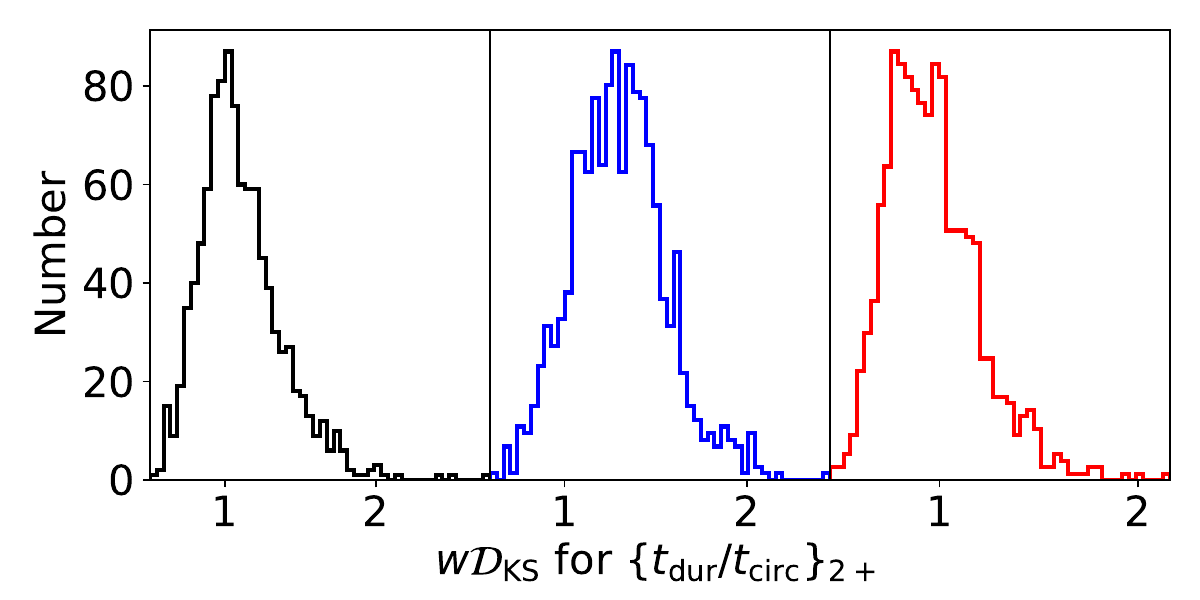} \\
 \includegraphics[scale=0.28,trim={0.2cm 0.2cm 0.4cm 0.2cm},clip]{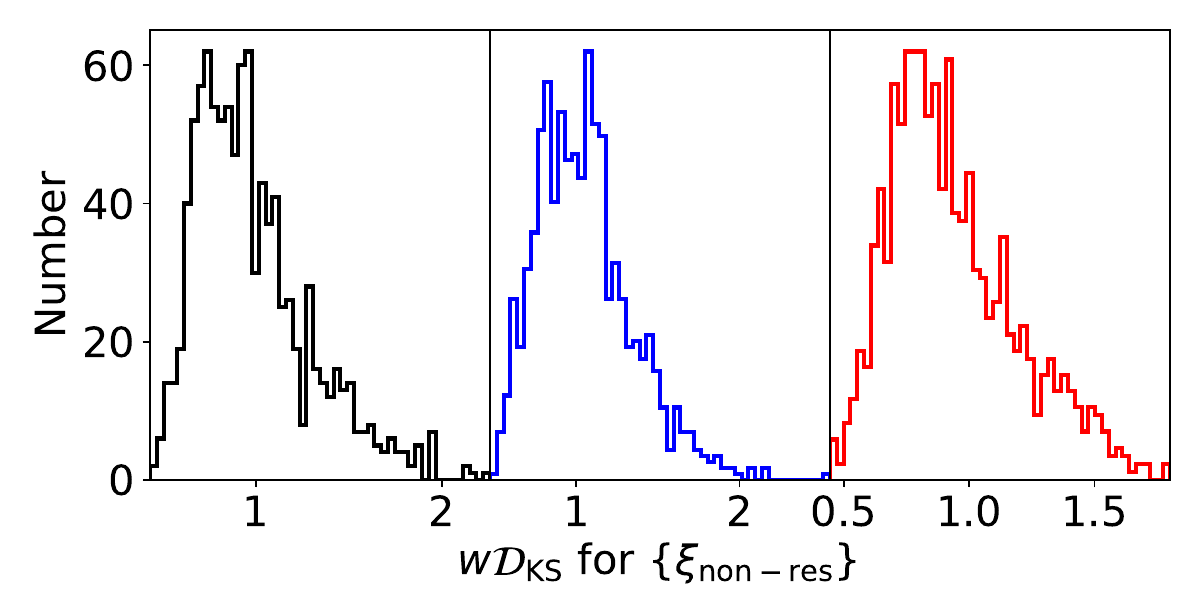} &
 \includegraphics[scale=0.28,trim={0.2cm 0.2cm 0.4cm 0.2cm},clip]{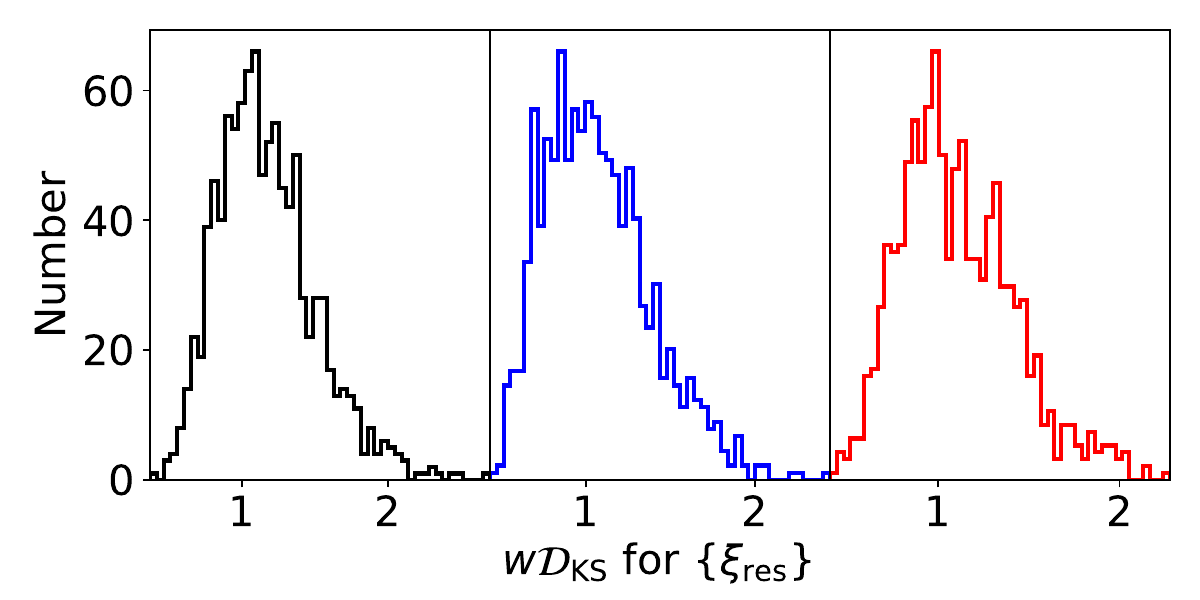} &
 \includegraphics[scale=0.28,trim={0.2cm 0.2cm 0.4cm 0.2cm},clip]{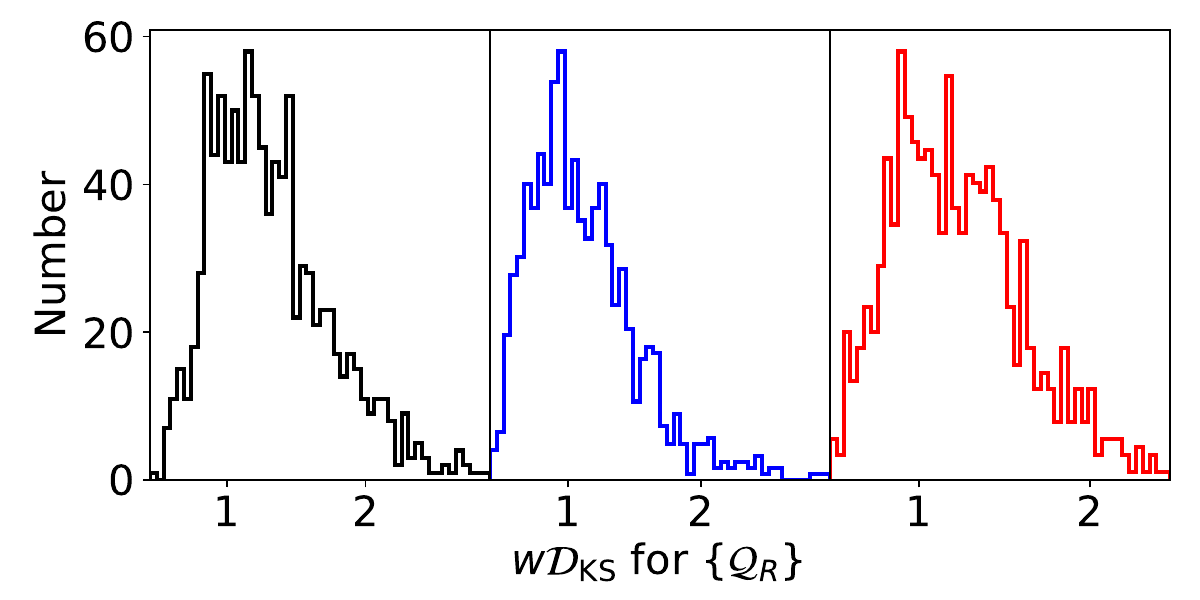} \\
 \includegraphics[scale=0.28,trim={0.2cm 0.2cm 0.4cm 0.2cm},clip]{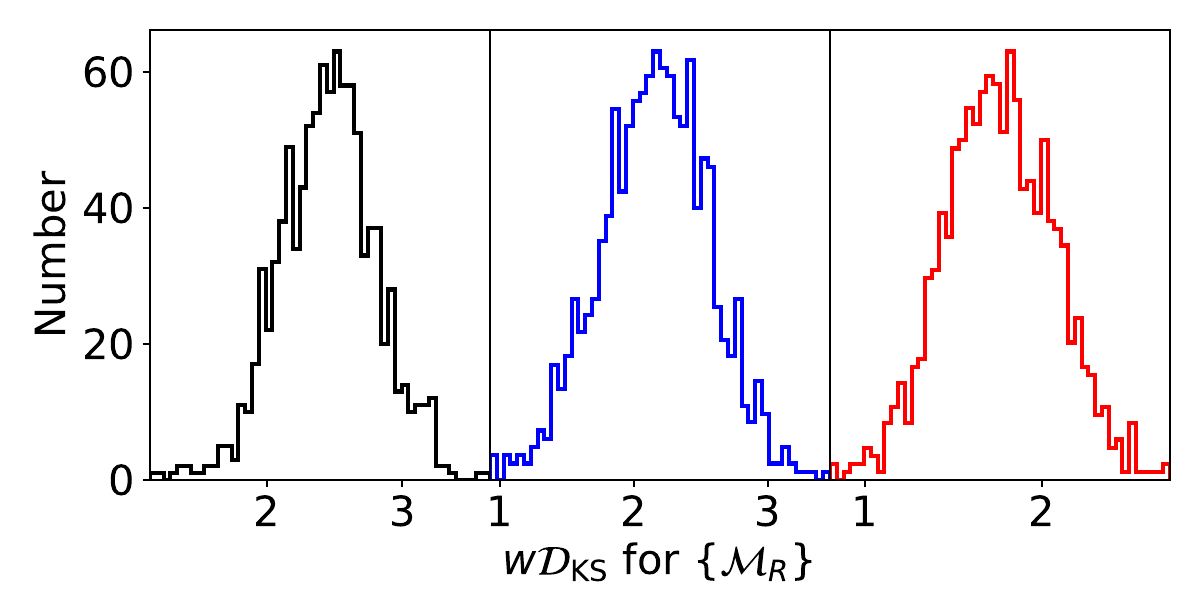} &
 \includegraphics[scale=0.28,trim={0.2cm 0.2cm 0.4cm 0.2cm},clip]{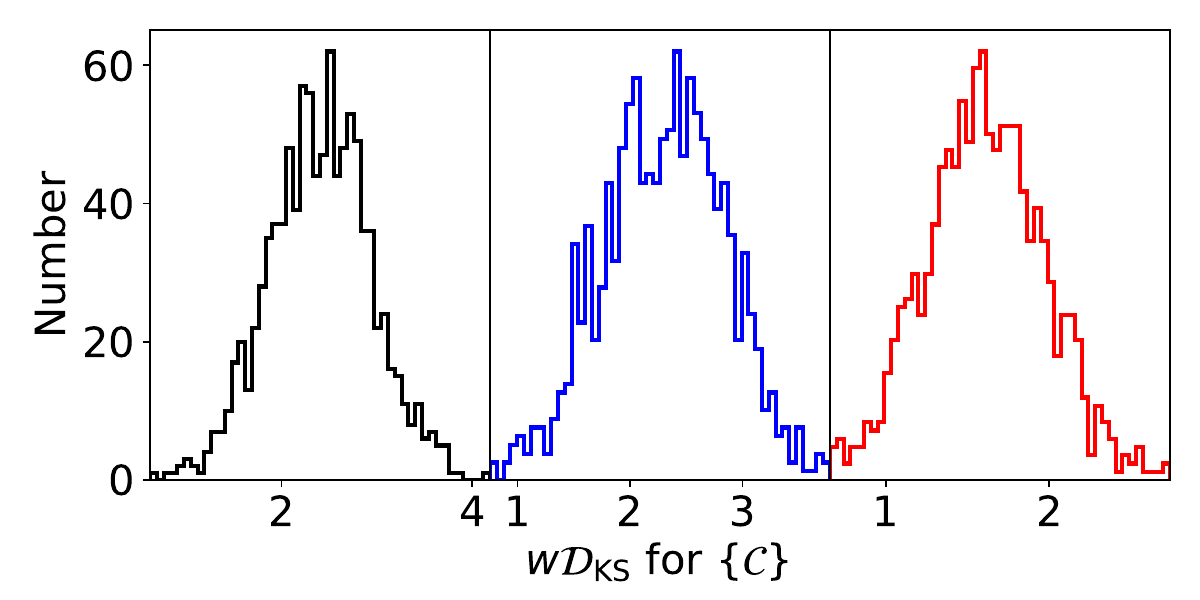} & \\
\end{tabular}
\caption{Histograms of the weighted total distances using our $\mathcal{D}_{W,3}$ (KS) distance function (\textbf{top row}) and individual distances (\textbf{second row and below}), for our maximum AMD model. 1000 simulated catalogs passing our distance threshold ($\mathcal{D}_{W,3} = 65$) are included.}
\label{fig:dists3_KS}
\end{figure*}

Figures \ref{fig:dists1_KS} and \ref{fig:dists1_AD} show the total and individual weighted distances for 1000 simulated catalogs using KS and AD terms in $\mathcal{D}_{W,1}$, respectively, for our two-Rayleigh and maximum AMD models. Likewise, Figure \ref{fig:dists3_KS} shows the weighted distances using KS terms for $\mathcal{D}_{W,3}$ for the maximum AMD model.

\begin{figure*}
\centering
\includegraphics[scale=0.29,trim={0 0 0 0},clip]{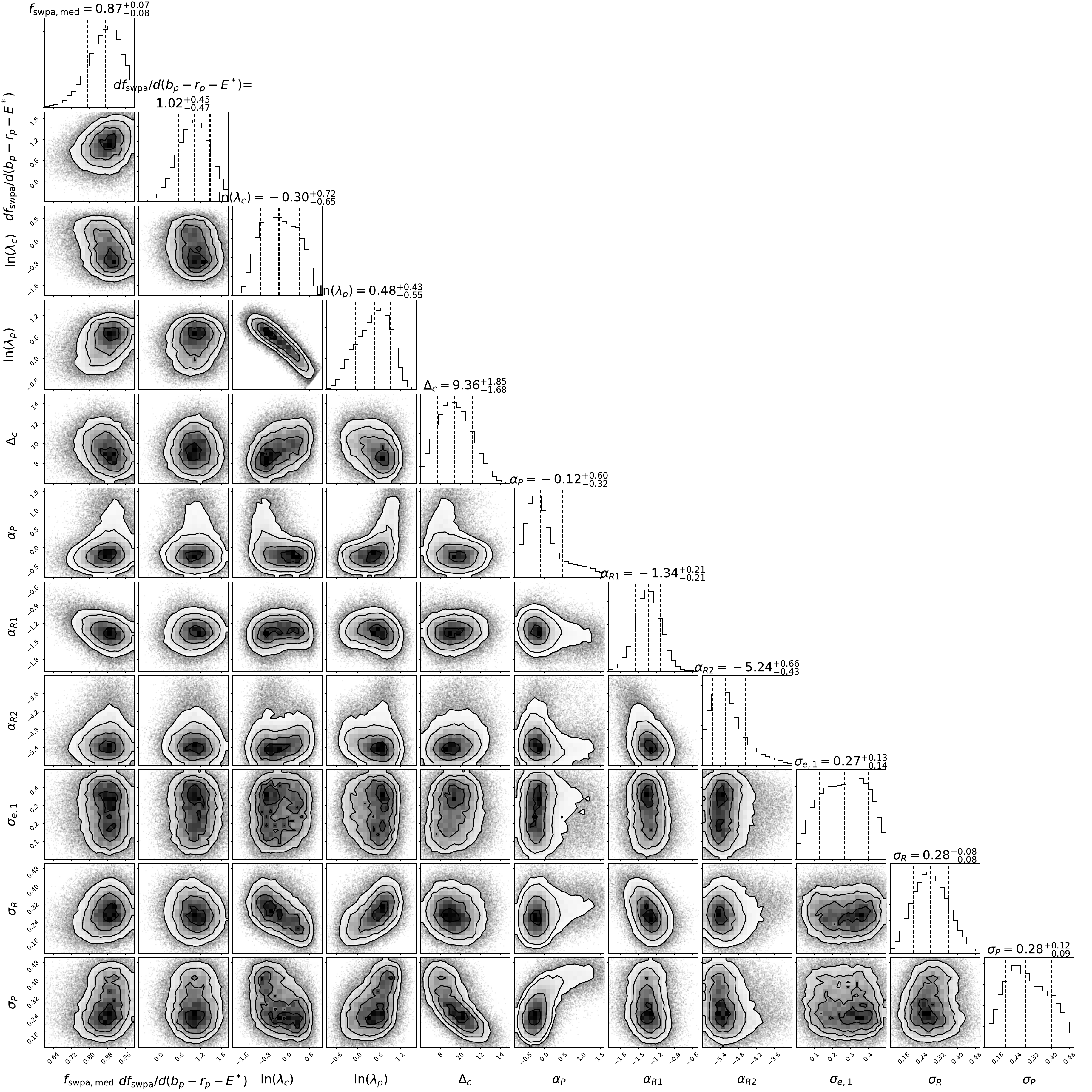}
\caption{ABC posterior distributions of the free model parameters for the maximum AMD model, using our $\mathcal{D}_{W,1}$ distance function (with KS distance terms). A total of $5 \times 10^4$ points passing a distance threshold $\mathcal{D}_{W,1} = 45$ evaluated using the GP emulator are plotted.}
\label{fig:d1_KS_corner}
\end{figure*}

\begin{figure*}
\centering
\includegraphics[scale=0.29,trim={0 0 0 0},clip]{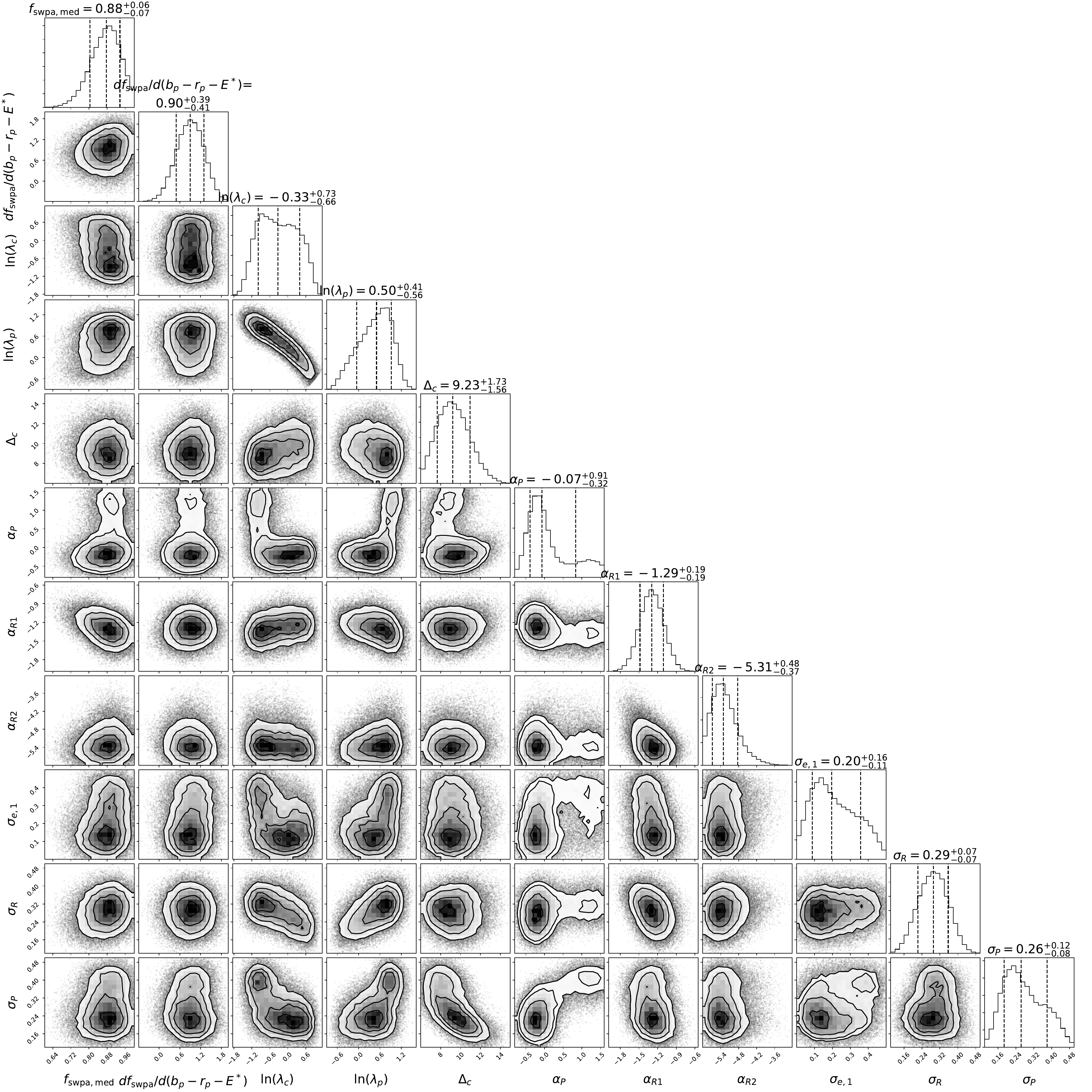}
\caption{ABC posterior distributions of the free model parameters for the maximum AMD model, using our $\mathcal{D}_{W,2}$ distance function (with KS distance terms). A total of $5 \times 10^4$ points passing a distance threshold $\mathcal{D}_{W,2} = 45$ evaluated using the GP emulator are plotted.}
\label{fig:d2_KS_corner}
\end{figure*}

\begin{figure*}
\centering
\includegraphics[scale=0.29,trim={0 0 0 0},clip]{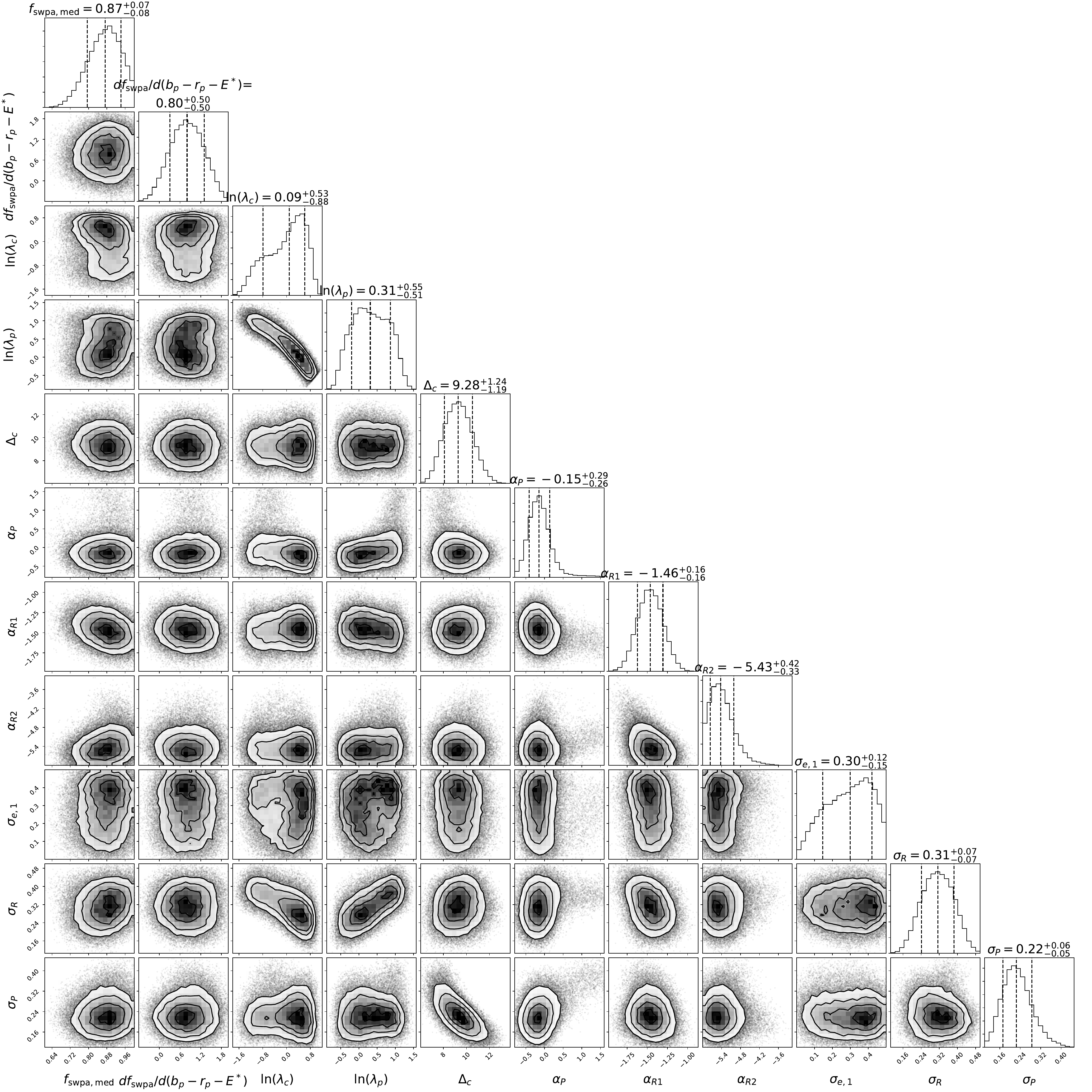}
\caption{ABC posterior distributions of the free model parameters for the maximum AMD model, using our $\mathcal{D}_{W,1}$ distance function (with AD distance terms). A total of $5 \times 10^4$ points passing a distance threshold $\mathcal{D}_{W,1} = 80$ evaluated using the GP emulator are plotted.}
\label{fig:d1_AD_corner}
\end{figure*}

\begin{figure*}
\centering
\includegraphics[scale=0.29,trim={0 0 0 0},clip]{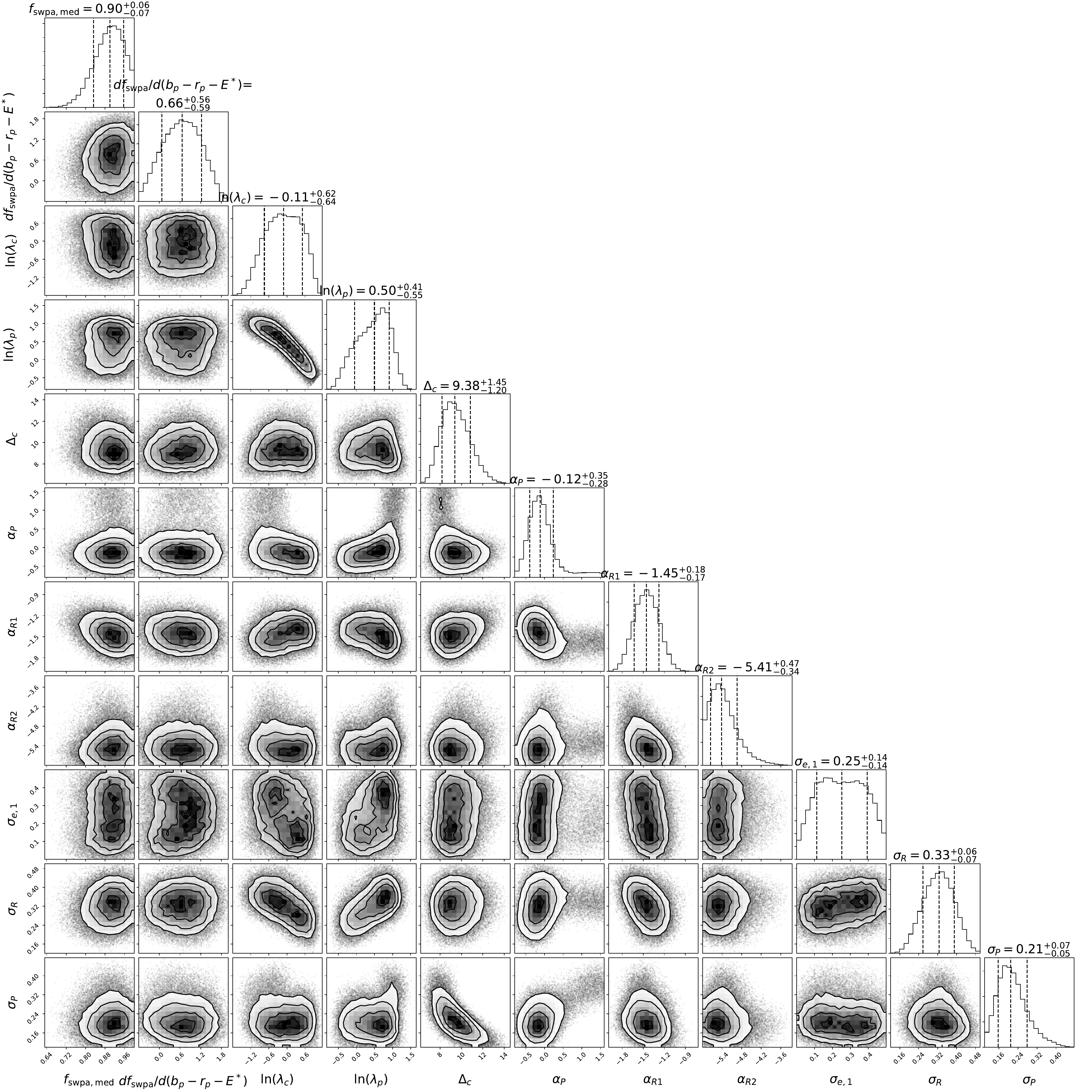}
\caption{ABC posterior distributions of the free model parameters for the maximum AMD model, using our $\mathcal{D}_{W,2}$ distance function (with AD distance terms). A total of $5 \times 10^4$ points passing a distance threshold $\mathcal{D}_{W,2} = 80$ evaluated using the GP emulator are plotted.}
\label{fig:d2_AD_corner}
\end{figure*}

\begin{figure*}
\centering
\includegraphics[scale=0.29,trim={0 0 0 0},clip]{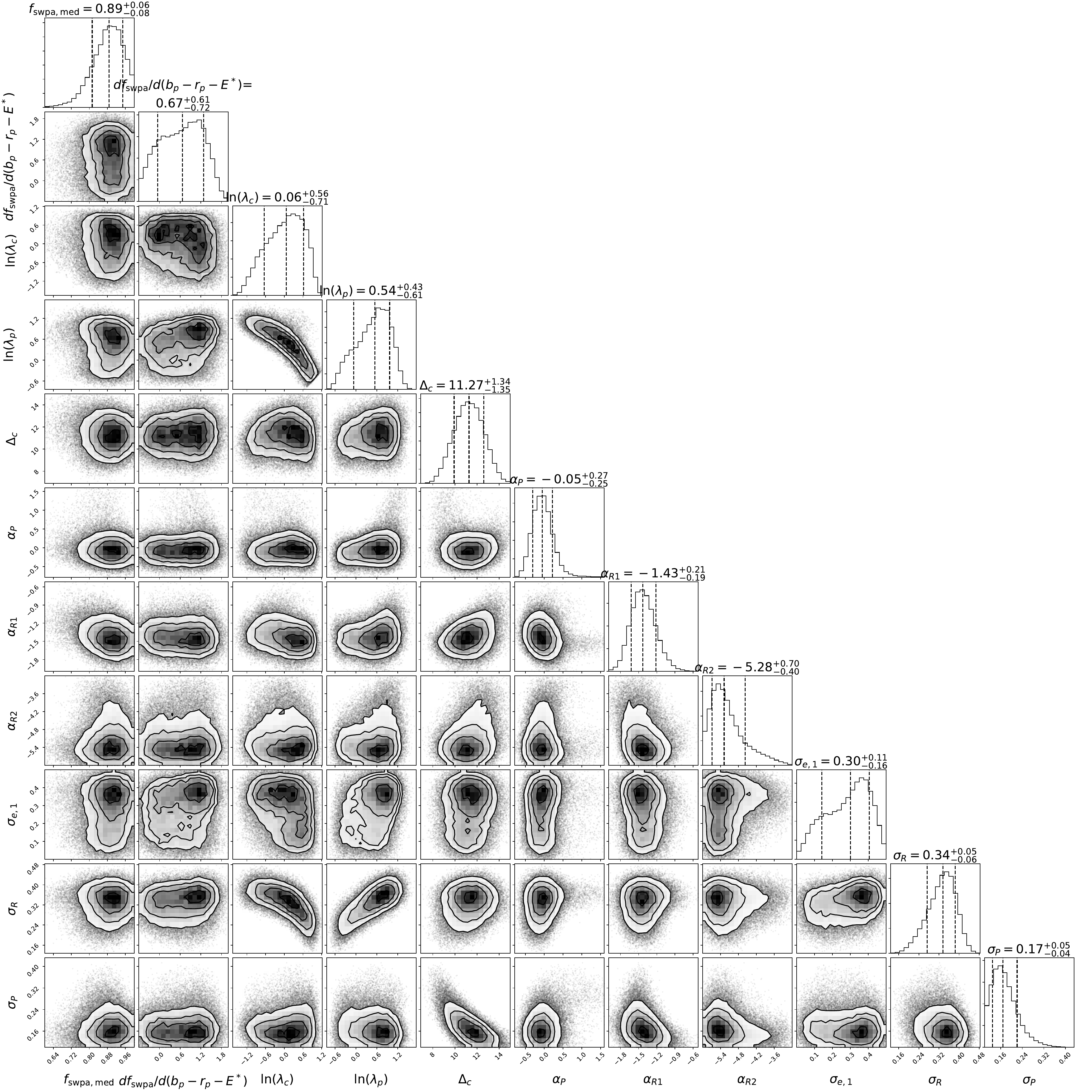}
\caption{ABC posterior distributions of the free model parameters for the maximum AMD model, using our $\mathcal{D}_{W,3}$ distance function (with AD distance terms). A total of $5 \times 10^4$ points passing a distance threshold $\mathcal{D}_{W,3} = 120$ evaluated using the GP emulator are plotted.}
\label{fig:d3_AD_corner}
\end{figure*}

Figure \ref{fig:d1_KS_corner}-\ref{fig:d3_AD_corner} show the ABC posterior distributions of the free model parameters for the maximum AMD model, for all the distance functions used in this paper, analogous to Figure \ref{fig:d3_KS_corner}.

\end{document}